\documentclass[preprint,12pt]{emulateapj}
\usepackage{epsfig}
\usepackage{natbib}
\usepackage{amssymb}
\usepackage{amsbsy}
\usepackage{natbib}
\usepackage{url}
\usepackage[mathcal]{euscript}
\bibpunct{(}{)}{,}{a}{}{,}

\newcommand{\teff}{$T_{\rm eff}$}

\newcommand{\vsini}{$v \sin i$}
\newcommand{\sini}{$\sin i$}

\pdfoutput=1

\begin{document}
 
\def\simlt{\vcenter{\hbox{$<$}\offinterlineskip\hbox{$\sim$}}}
\def\simgt{\vcenter{\hbox{$>$}\offinterlineskip\hbox{$\sim$}}}
\def\etal{et al.\ }
\def\kms{km s$^{-1}$}

\title{CSI~2264: Characterizing Young Stars in NGC~2264 with Short-Duration,
  Periodic Flux Dips in their Light Curves \footnotemark[*]}
\footnotetext[*]{Based on data from the {\em Spitzer} and {\em CoRoT}
missions, as well as the Canada France Hawaii Telescope (CFHT) MegaCam
CCD, and  the European Southern Observatory Very Large Telescope,
Paranal Chile, under program 088.C-0239. The {\em CoRoT} space mission
was developed and is  operated by the French space agency CNES, with
particpiation of ESA's RSSD and Science Programmes, Austria, Belgium,
Brazil, Germany, and Spain.  MegaCam is a joint project of CFHT and
CEA/DAPNIA, which is operated by the National Research Council (NRC)
of Canada, the Institute National des  Sciences de l'Univers of the
Centre National de la Recherche Scientifique of France, and the
University of Hawaii.}
\author{John Stauffer\altaffilmark{1}, Ann Marie Cody\altaffilmark{1,20}, 
Pauline McGinnis\altaffilmark{6},
Luisa Rebull\altaffilmark{1}, Lynne A. Hillenbrand\altaffilmark{2},  
Neal J. Turner\altaffilmark{3}, John Carpenter\altaffilmark{2},
Peter Plavchan\altaffilmark{1}, 
Sean Carey\altaffilmark{1}, 
Susan Terebey\altaffilmark{4}, Mar\'ia Morales-Calder\'on\altaffilmark{5},  
Silvia H. P. Alencar\altaffilmark{6}, 
Jerome Bouvier\altaffilmark{7}, 
Laura Venuti\altaffilmark{7}, 
Lee Hartmann\altaffilmark{8}, Nuria Calvet\altaffilmark{8}, 
Giusi Micela\altaffilmark{9}, Ettore Flaccomio\altaffilmark{9}, 
Inseok Song\altaffilmark{10}, Rob Gutermuth\altaffilmark{11},  
David Barrado\altaffilmark{5}, 
Frederick J. Vrba\altaffilmark{12}, Kevin Covey\altaffilmark{13}, 
Debbie Padgett\altaffilmark{14}, 
William Herbst\altaffilmark{15}, 
Edward Gillen\altaffilmark{16}, 
Wladimir Lyra\altaffilmark{3,18},
Marcelo Medeiros Guimaraes\altaffilmark{19},
Herve Bouy\altaffilmark{5}, 
Fabio Favata\altaffilmark{17} }
\altaffiltext{1}{Spitzer Science Center, California Institute of
Technology, Pasadena, CA 91125, USA}
\altaffiltext{2}{Astronomy Department, California Institute of
Technology, Pasadena, CA 91125, USA}
\altaffiltext{3}{Jet Propulsion Laboratory, California Institute
of Technology, Pasadena, CA 91109, USA}
\altaffiltext{4}{Department of Physics and Astronomy, 5151 State University
Drive, California State  University at Los Angeles, Los Angeles, CA 90032}
\altaffiltext{5}{Centro de Astrobiolog\'ia, Dpto. de
Astrof\'isica, INTA-CSIC, PO BOX 78, E-28691, ESAC Campus, Villanueva de
la Ca\~nada, Madrid, Spain}
\altaffiltext{6}{Departamento de F\'{\i}sica -- ICEx -- UFMG, 
Av. Ant\^onio Carlos, 6627, 30270-901, Belo Horizonte, MG, Brazil}
\altaffiltext{7}{Universit\'e de Grenoble, Institut de Plan\'etologie 
et d'Astrophysique de Grenoble (IPAG), F-38000 Grenoble, France;
CNRS, IPAG, F-38000 Grenoble, France}
\altaffiltext{8}{Department of Astronomy, University of Michigan, 
500 Church Street, Ann Arbor, MI 48105, USA}
\altaffiltext{9}{INAF - Osservatorio Astronomico di Palermo, Piazza 
del Parlamento 1, 90134, Palermo, Italy}
\altaffiltext{10}{Department of Physics and Astronomy, The University 
of Georgia, Athens, GA 30602-2451, USA}
\altaffiltext{11}{Department of Astronomy, University of Massachusetts,
Amherst, MA 01003, USA}
\altaffiltext{12}{U.S. Naval Observatory, Flagstaff Station, 10391 
West Naval Observatory Road, Flagstaff, AZ 86001, USA}
\altaffiltext{13}{Department of Physics and Astronomy (MS-9164), Western
Washington Univ., 516 High St., Bellingham, WA 98225, USA}
\altaffiltext{14}{NASA/Goddard Space Flight Center, Code 665, Greenbelt,
MD 20771, USA}
\altaffiltext{15}{Astronomy Department, Wesleyan University,
Middletown, CT 06459, USA}
\altaffiltext{16}{Department of Physics, University of Oxford, Keble Road, 
Oxford, OX1 3RH, UK}
\altaffiltext{17}{European Space Agency, 8-10 rue Mario Nikis, 
F-75738 Paris Cedex 15, France}
\altaffiltext{18}{NASA Sagan Fellow}
\altaffiltext{19}{Departamento de F\'isica e Matem\'atica - UFSJ - Rodovia MG 443, 
KM 7, 36420-000, Ouro Branco, MG, Brazil}
\altaffiltext{20}{NASA Ames Research Center, Kepler Science Office, Mountain
View, CA 94035}
\email{stauffer@ipac.caltech.edu}

\begin{abstract}
We identify nine young stellar objects (YSOs) in the NGC~2264
star-forming region with optical {\em CoRoT} light curves exhibiting
short-duration, shallow,  periodic flux dips.  All of these stars have
infrared (IR) excesses that are consistent with their having inner
disk walls near the Keplerian co-rotation radius. The repeating
photometric dips have FWHM generally less than one day,  depths almost
always less than 15\%, and periods (3 $<P<$ 11 days)  consistent with
dust near the Keplerian co-rotation period.  The flux dips vary
considerably in their depth from epoch to epoch, but usually persist
for several weeks and, in two cases, were present in data collected on
successive years.  For several of these stars, we also measure the
photospheric rotation period and find that the rotation and dip
periods are the same, as predicted by standard ``disk-locking"
models.   We attribute these flux dips to clumps of material in or 
near the inner disk wall, passing through our line of sight to the
stellar photosphere.  In some cases, these dips are also present in
simultaneous {\em Spitzer} IRAC light curves at 3.6 and 4.5 microns. 
We characterize the properties of these dips, and compare the stars
with  light curves exhibiting this behavior to other classes of YSO in
NGC~2264.   A number of physical mechanisms could locally increase the
dust scale height near the inner disk wall, and we discuss several of
those mechanisms; the most plausible mechanisms are either a disk warp
due to interaction with the stellar magnetic field or dust entrained
in funnel-flow accretion columns arising near the inner disk wall.

\end{abstract}

\keywords{open clusters and associations: individual (NGC 2264)---circumstellar matter---stars:
pre-main sequence---stars: protostars---stars: variables: T Tauri}

\section{Introduction}

Understanding how young stellar objects (YSOs) assemble themselves
and how  the accretion process works requires understanding the
structure and evolution  of the inner disk on size scales $<$ 0.1 AU. 
Even with the advent of infrared (IR) and radio interferometry,
imaging of even the nearest YSOs to detect structure on these size
scales is impractical.  It therefore becomes necessary to infer the
spatial structure and composition on these size scales indirectly
from, for example, spectral energy distributions (SEDs),
high-resolution spectra, or variability studies.

SEDs have provided some of the most important insights into the structure
and evolution of the inner disks of YSOs.  These SEDs have been used not
only to argue for the presence of an inner disk wall, located roughly at
the dust sublimation boundary, but also that the upper edge of the wall
is in most cases curved (McClure et al.\ 2013).   Some YSOs have both inner
and outer optically thick disks separated by an optically thin gap, with
the gap likely the result of gravitational sweeping by an embedded giant
planet (Quillen et al.\ 2004).  Mid-IR spectrophotometric variability seen
in a number of YSOs suggest that the height of their inner disk wall varies
by up to 20\% on timescales of days or weeks (Espaillat et al.\ 2011).

Optical or near-IR photometric variability studies offer another path
to understanding inner disk structure and to the physics of the
interaction between the inner disk and the parent star.  Such studies
date back more than half a century (Joy 1945), where variability was
used as one of the defining characteristics of the new class of young,
forming stars.  Herbst et al.\ (1994) summarized the available
information on T~Tauri star variability, identifying several different
classes dominated variously by rotational modulation of cold or hot
spots, variable accretion or variable extinction. Variable extinction
is particularly attractive as a means to study inner disk structure
because in principle it can yield information on very small size
scales, if the cadence of the time series photometry is high enough. 
Bouvier et al.\ (1999) showed that AA Tau has broad, periodic flux dips
whose amplitude and shape vary from epoch to epoch, which they ascribe
to variable extinction resulting from a warp in the star's inner disk
and to large variation in the amount of material at the inner disk rim
on $\sim$week timescales.   Photometric surveys of entire star forming
regions (e.g., Carpenter et al.\ 2001; Flaherty et al.\ 2012; Findeisen
et al.\ 2013) with modern 2-D arrays have also allowed other types of
variable extinction events -- both periodic and aperiodic -- to be
identified.  

We have recently obtained the highest cadence, highest signal-to-noise
(S/N) time domain survey of a large population of YSOs in a
star-forming region during two campaigns in 2008 and 2011.  We have
used those data for NGC~2264 for an initial study of the AA Tau-type
variables in the cluster (Alencar et al.\ 2010); to provide a
statistical basis for a new YSO light-curve morphology classification
scheme (Cody et al.\ 2014); and to identify a set of YSOs whose light
curves are dominated by short duration accretion bursts (Stauffer et
al.\ 2014).  In this paper, we use this dataset to identify a small set
of YSOs whose light curves exhibit short duration, often near-Gaussian
shaped,  periodic flux dips which must arise from dust structures near
the inner disk wall.   In a companion paper (McGinnis et al.\ 2015), we
provide a quantitative analysis of all of the other YSOs in our
NGC~2264 sample with flux dips due to variable extinction events.

\section{Observational Data }

A detailed description of the photometric data we collected in our
2008 and 2011 observing campaigns for the NGC~2264 star-forming region
is provided in Cody et al. (2014)\footnote{The {\em CoRoT} and {\em
Spitzer} light curves for all probable NGC~2264 members, as well as
our broad band photometry for these stars, are available at
\url{http://irsa.ipac.caltech.edu/data/SPITZER/CSI2264}.}.     The
subset of the data sources that we use here include several sets of 
time-series, photometric monitoring data (the first four items below)
plus additional single-epoch ancillary photometry (the last two
items):
\vskip0.05truein
\noindent
* A 2008 campaign using the {\em CoRoT} (Convection, Rotation, and
planetary Transits; Baglin et al.\ 2009) satellite, designated as
SRa01 within the {\em CoRoT} mission, with photometry being collected
from roughly MJD 54534 to 54556, a period of 23 days.  The observing
cadence was generally every 512 seconds; however, bright targets and
targets of known special interest could be done at 32 second cadence
(and the spacecraft could autonamously switch to 32 second cadence if
transit-like events were detected).  The passband is set just by the
sensitivity curve for the {\em CoRoT} CCD,  corresponding to a broad V
$+$ R filter. Only stars present in an input catalog have {\em CoRoT}
light curves.  For each cataloged star, a specific pixel mask was
designated to determine which set of pixels was queried to derive the
photometry for that star.  For the light curves discussed in this
paper and illustrated in the figures, typical RMS uncertainties in the
CoRoT relative photometry range from about 2 mmag at the bright end
(e.g. Mon-250, Figure~\ref{fig:sixctts}a) to about 6 mmag 
at the faint end (Mon-6975, Figure~\ref{fig:sixctts}f).
\vskip0.05truein
\noindent
* A 2011 campaign designated as SRa05 within the {\em CoRoT} mission,
with data collected from roughly MJD 55897 to 55936.  The noise level
in these light curves is both higher and more spatially variable due
to  increased CCD radiation damage.  In some cases, the same star was
observed with different pixel masks in the two epochs, which can also
change the derived flux.   The observing cadence and basic pipeline
reduction steps were, however, the same for the two campaigns.
\vskip0.05truein
\noindent
* {\em Spitzer} (Werner et al.\ 2004) warm mission Infrared Array
Camera (IRAC; Fazio et al.\ 2004) 3.6 and 4.5 $\mu$m data obtained
simultaneous with the {\em CoRoT} SRa05 campaign, roughly from MJD
55900 to 55930.  The cadence was approximately every 2 hours.  Stars
near the lower or upper edge of the mapped region have data in only
one of the two channels.  All of the Spitzer light curves discussed
in this paper are for stars that are fairly bright at IRAC 
wavelengths, with expected RMS uncertainties of order 1\% for their
relative photometric accuracies.
\vskip0.05truein
\noindent
* $I$-band imaging using the US Naval Observatory (USNO) 40" telescope
obtained on clear or partially clear nights during two campaigns
(winter 2010 and winter 2011), roughly MJD 55505 to 55560 and MJD
55890 to 55990.  Typically, 100-200 epochs of data were collected in
2010, and of order 900 epochs in 2011/2012.  For each campaign,
imaging was obtained for an approximately 23'x23' region, centered
on the cluster center.  Aperture photometry was performed
for all stars in all images.  The zero-points for each CCD frame were
established by identification of a set of non-variable stars in the
region surveyed, and forcing the photometry for these stars to be
constant across all images.  Each light curve was normalized to a
mean relative magnitude of zero.   For the stars in 
Table~\ref{tab:basicinformation}, the USNO relative photometry should 
have RMS uncertainties of less than 1\% at the bright end, rising to
of order 2\% for the faintest of our stars.
\vskip0.05truein
\noindent
* Single epoch $ugri$ imaging of the entire cluster,  with standard
Sloan Digital Sky Survey (SDSS) filters,  obtained using Megacam on
the Canada-France-Hawaii Telescope (CFHT)  during two observing runs
in 2010 and 2012.  These data are presented in detail in Venuti et
al.\ (2014).
\vskip0.05truein
\noindent
* {\em Spitzer} IRAC 3.6, 4.5, 5.8 and 8.0 $\mu$m and MIPS 24 $\mu$m
single epoch imaging photometry of the entire cluster obtained early
in the cryogenic mission, and reported in Sung et al.\ (2009) and
Teixeira et al.\ (2012).

\section{Overview of YSOs Whose Light Curves are Dominated by Variable Extinction Events}

If a clump of circumstellar dust passes through our line of sight to a
YSO,  the light from the star will dim by an amount roughly
proportional to the  column density of dust, modified according to the
size distribution of the grains (if the clump is optically thin), or
by an amount proportional to the fractional area of the star that is
covered (if the clump is optically thick but small in angular size
relative to the star). If the dust clump is long lived and relatively
fixed in location, the flux dips it produces should be periodic at the
Keplerian rotation period for that clump.   For some stars, a warped
inner disk has been invoked (e.g., AA Tau -- Bouvier  et al.\ 1999;
Flaherty \& Muzerolle 2010)  to explain deep flux dips that extend
over half or more of the star's  rotation period.  Dust clumps
produced by instability processes in the disk, and levitated above the
disk plane into our line of sight, could produce flux dips that do not
repeat (Turner et al.\ 2010).   Other longer timescale variable
extinction events could be associated with Keplerian rotation in the
outer disk bringing dust into our line of sight (Bouvier et al.\
2013);  with orbital motion of a binary system causing our line of
sight to intersect a circumbinary disk (Plavchan et al.\ 2008a); or
with dust entrained in a spatially or temporally variable disk wind
(Bans \& K\"onigl 2012).

In our 40+ day monitoring of the star-forming region NGC2264 with 
{\em CoRoT} and  {\em Spitzer}, we have identified three different
morphological light curve  signatures which appear to be associated
with variable extinction events.    Figure~\ref{fig:sixctts} shows
light curves for NGC~2264 Classical T Tauri stars (CTTS) which
illustrate each of these different modes. These three modes are:

\begin{figure*}
\begin{center}
\epsscale{0.80}
\plottwo{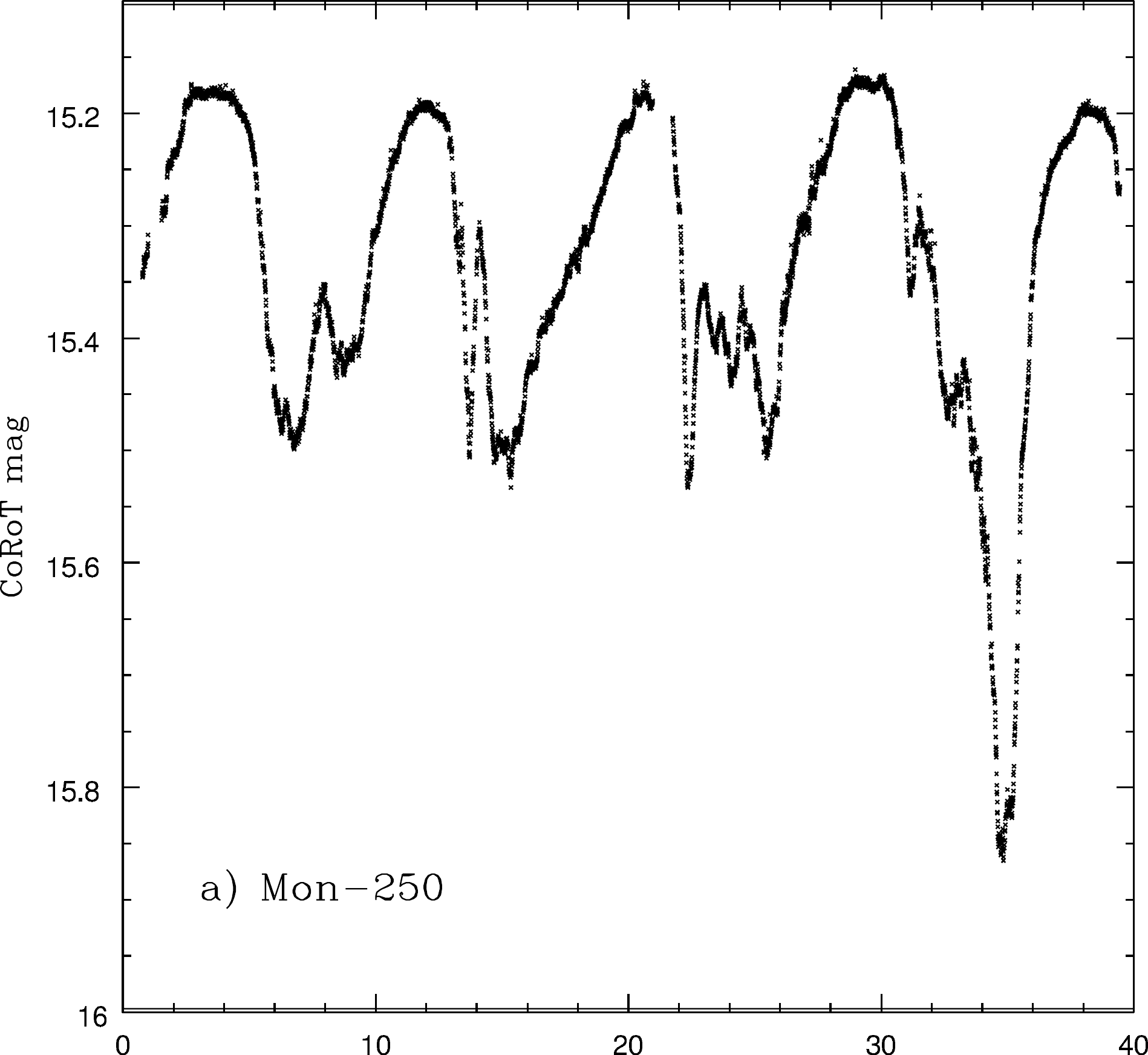}{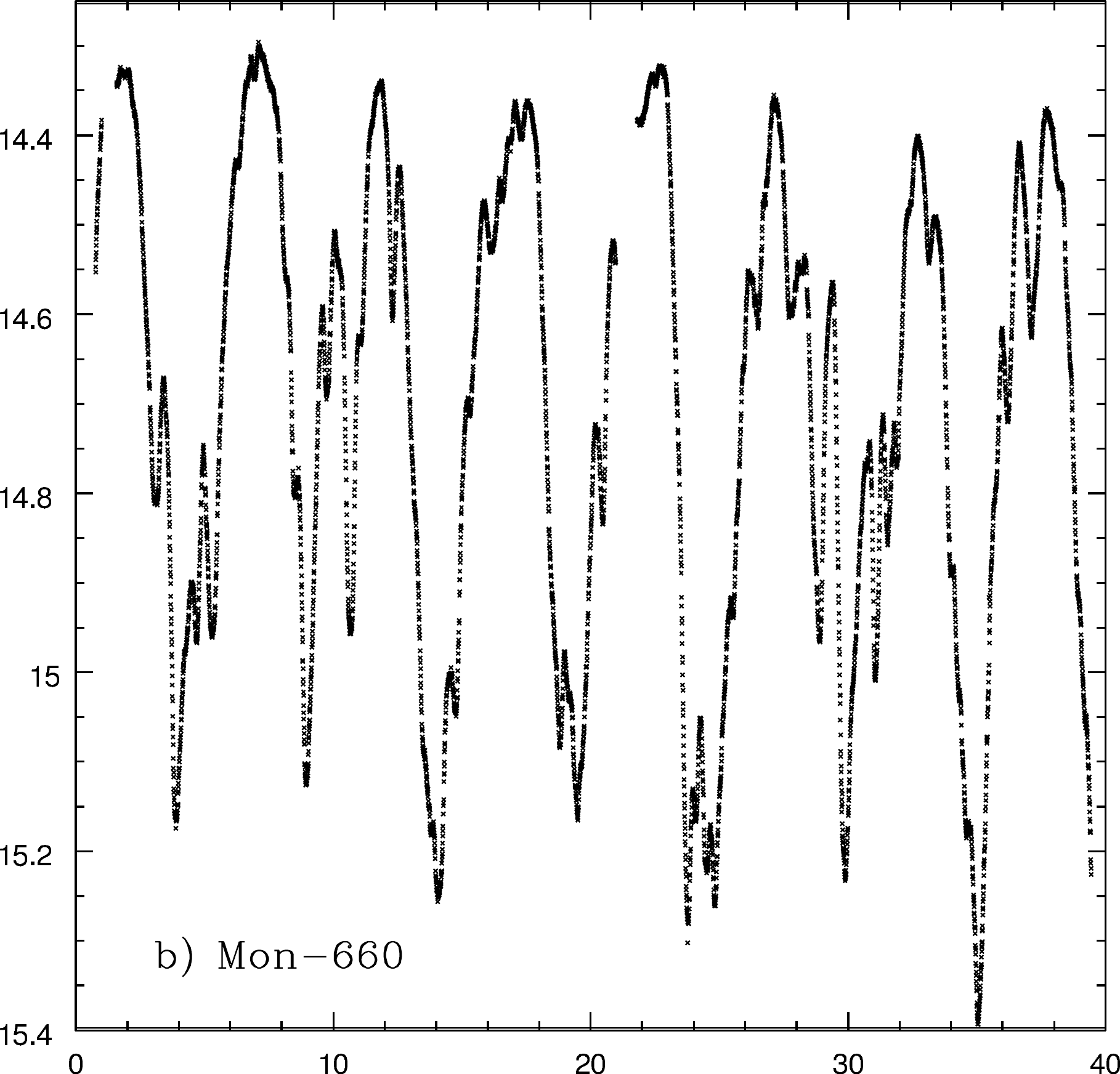}
\vspace{-0.cm}
\plottwo{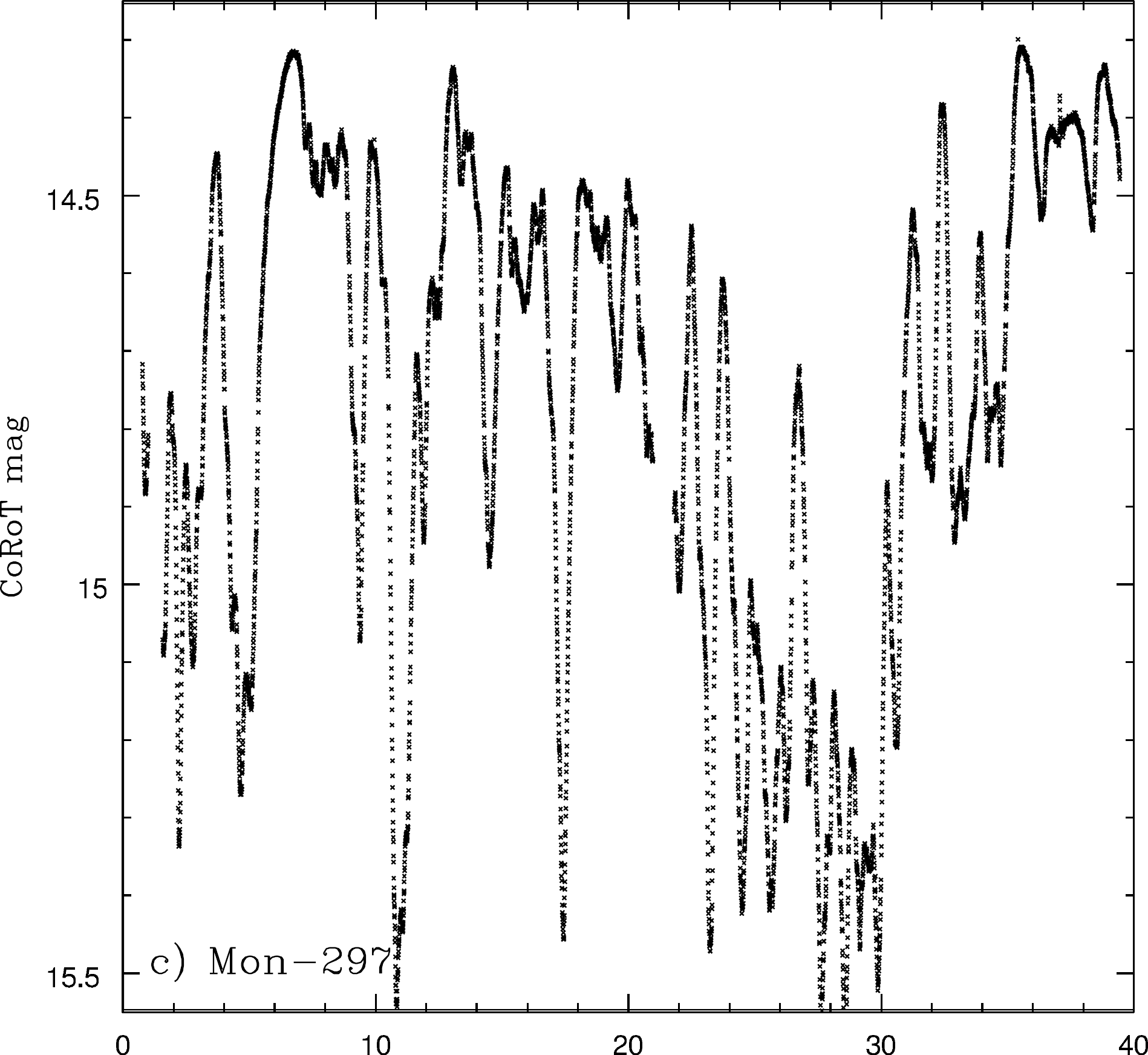}{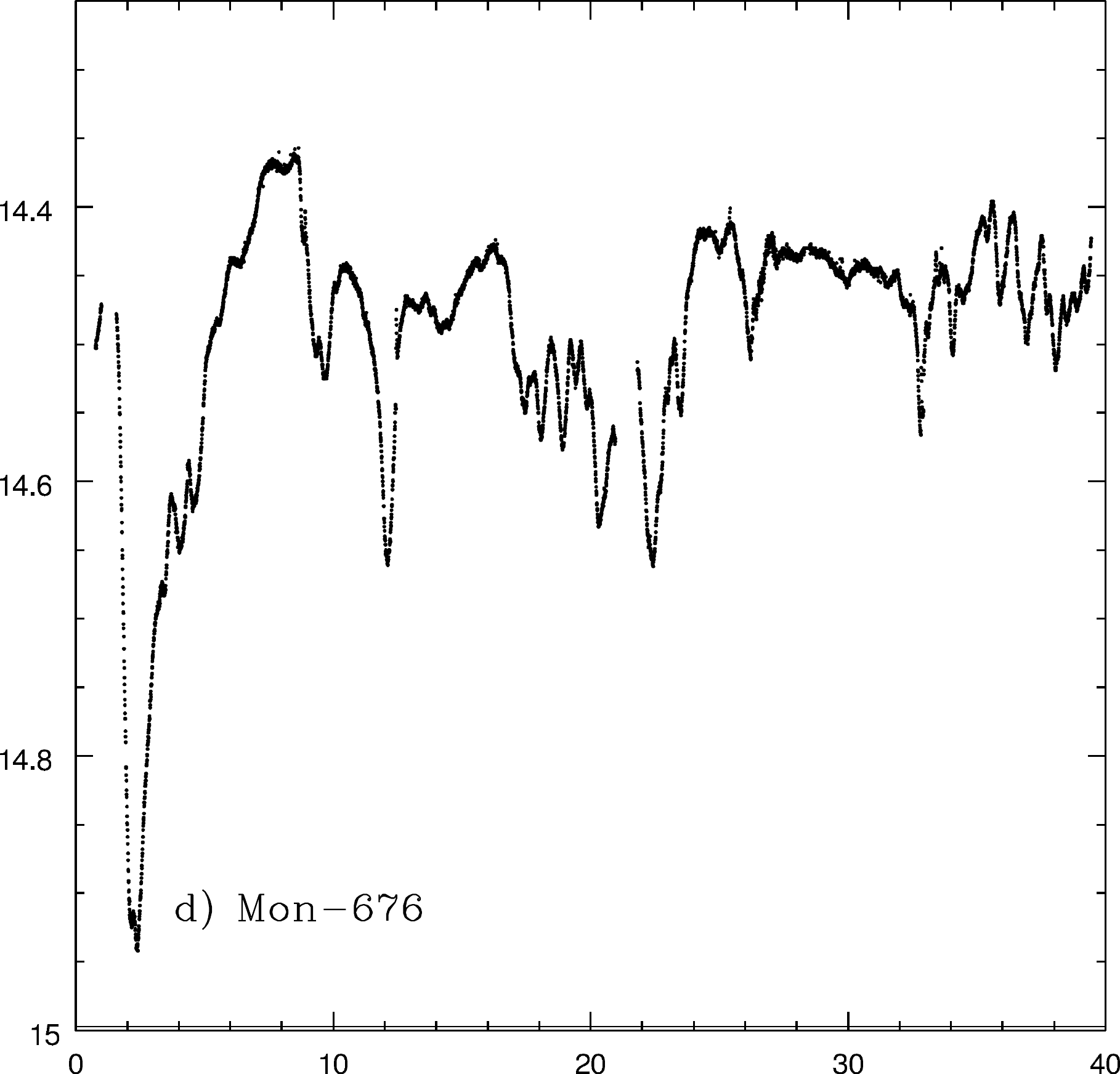}
\vspace{-0.cm}
\plottwo{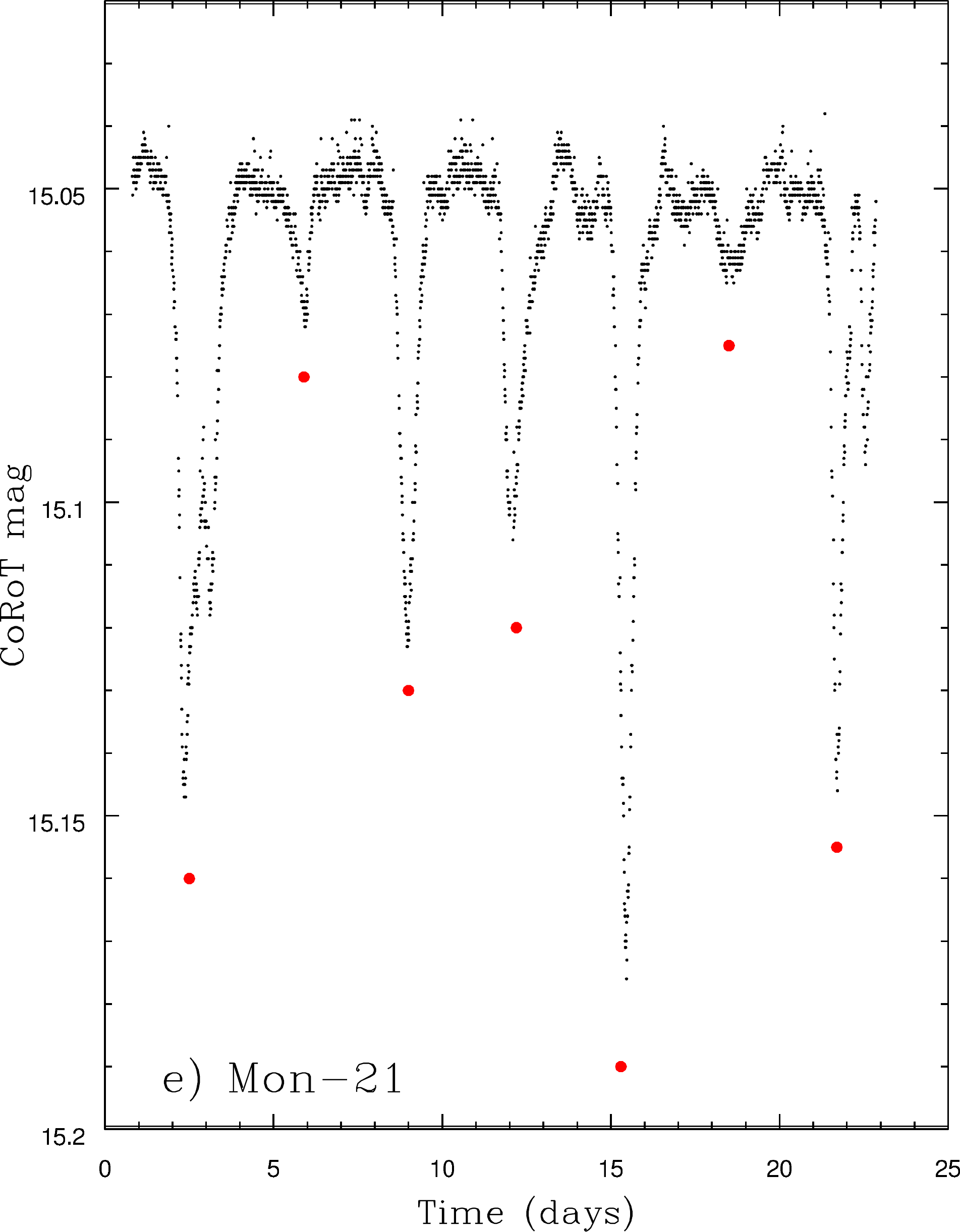}{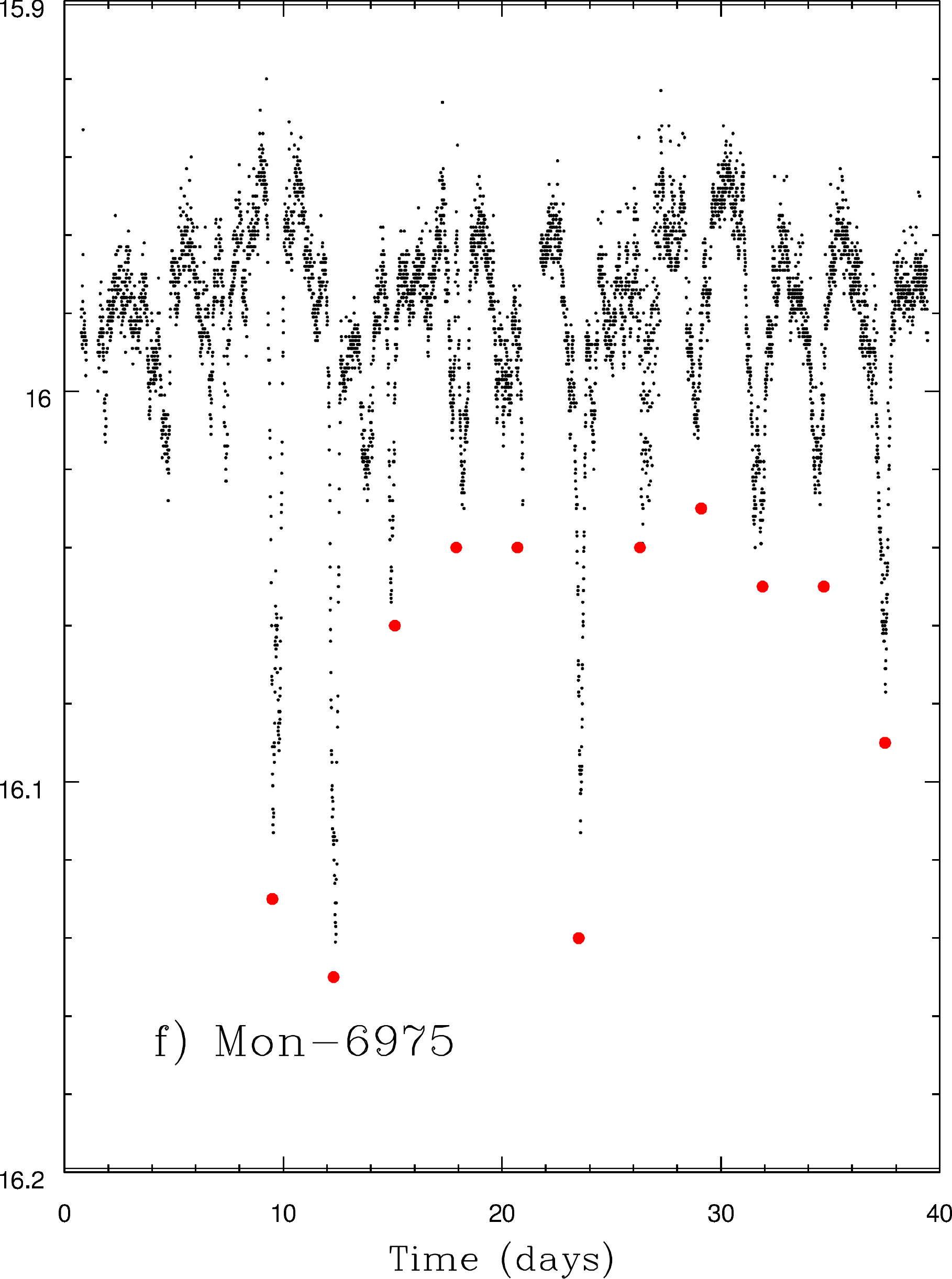}
\end{center}
\caption{Light curves for six CTTS members of NGC~2264 whose {\em CoRoT}
data are dominated by flux dips due to variable extinction.
The top two are AA Tau analogs, with broad flux dips 
most naturally explained as due to a warped inner disk intersecting
our line of sight; the middle
two light curves have aperiodic, narrow
flux dips, possibly associated with MRI (magneto-resonance instability) 
events levitating dust in
the inner disk; the bottom pair of plots exemplify the class of light
curves with periodic narrow dips which are the subject of this paper (red
dots provided to guide the eye to the periodic flux dips).
Time zero in all panels except (e) corresponds to MJD = 55896.0 days;
for panel (e) time zero corresponds to MJD= 54533.0 days.
Note, for Mon-21, the light curve shown has been processed to
remove a smoothly varying ``continuum" which we ascribe to photospheric
spots - see Section 3.1.1.  Compared to the AA Tau analogs (panels a
and b), the objects of this study (panels e and f) show, on average,
narrower, shallower and more symmetrical 
flux dips (see Figure~\ref{fig:compare_dippers}).
\label{fig:sixctts}}
\end{figure*}

\begin{itemize}
\item AA Tau analogs (examples illustrated in
     Figure~\ref{fig:sixctts}a and b.) Light curves of this type have
     relatively flat continua, punctuated by deep, broad flux dips
     occurring at the Keplerian orbital period for material near the
     inner disk rim. The generally accepted physical explanation is
     that there is a warp in the inner disk, and that a portion of
     this warp occults our line of sight (Bouvier et al. 1999);

\item Aperiodic extinctors (examples illustrated in
    Figure~\ref{fig:sixctts}c and d.) Light curves of this type have
    relatively flat continua and  deep, narrow and/or broad flux dips
    but with no obvious periodicity.  A possible physical mechanism is
    dust levitated relatively high above the disk mid-plane by
    magneto-resonance instabilities (MRI) or
    other instabilities, which then settles back towards the
    mid-plane, accretes onto the star, or escapes or sublimates on
    timescales of hours to days.  See McGinnis et al.\ (2015) for a
    discussion of these stars;

\item Short duration, periodic dippers (newly identified group -
    examples illustrated in Figure~\ref{fig:sixctts}e and f).   Light
    curves of this type have relatively shallow, relatively narrow
    flux dips with periods generally in the range 3-10 days, and with
    dip shapes often approximately Gaussian.   Sometimes these dips
    are superposed on a relatively flat continuum, but sometimes the
    dips are superposed on sine-wave light variations which we
    attribute to other physical mechanisms. A physical mechanism for
    these flux dips must also involve dust structures  in or near the
    inner disk rim that are brought into our line of sight by orbital
    motion.  We discuss options for the origin of these structures in
    \S 7. 
\end{itemize}

\begin{figure*}
\begin{center}
\epsscale{1.0}
\plottwo{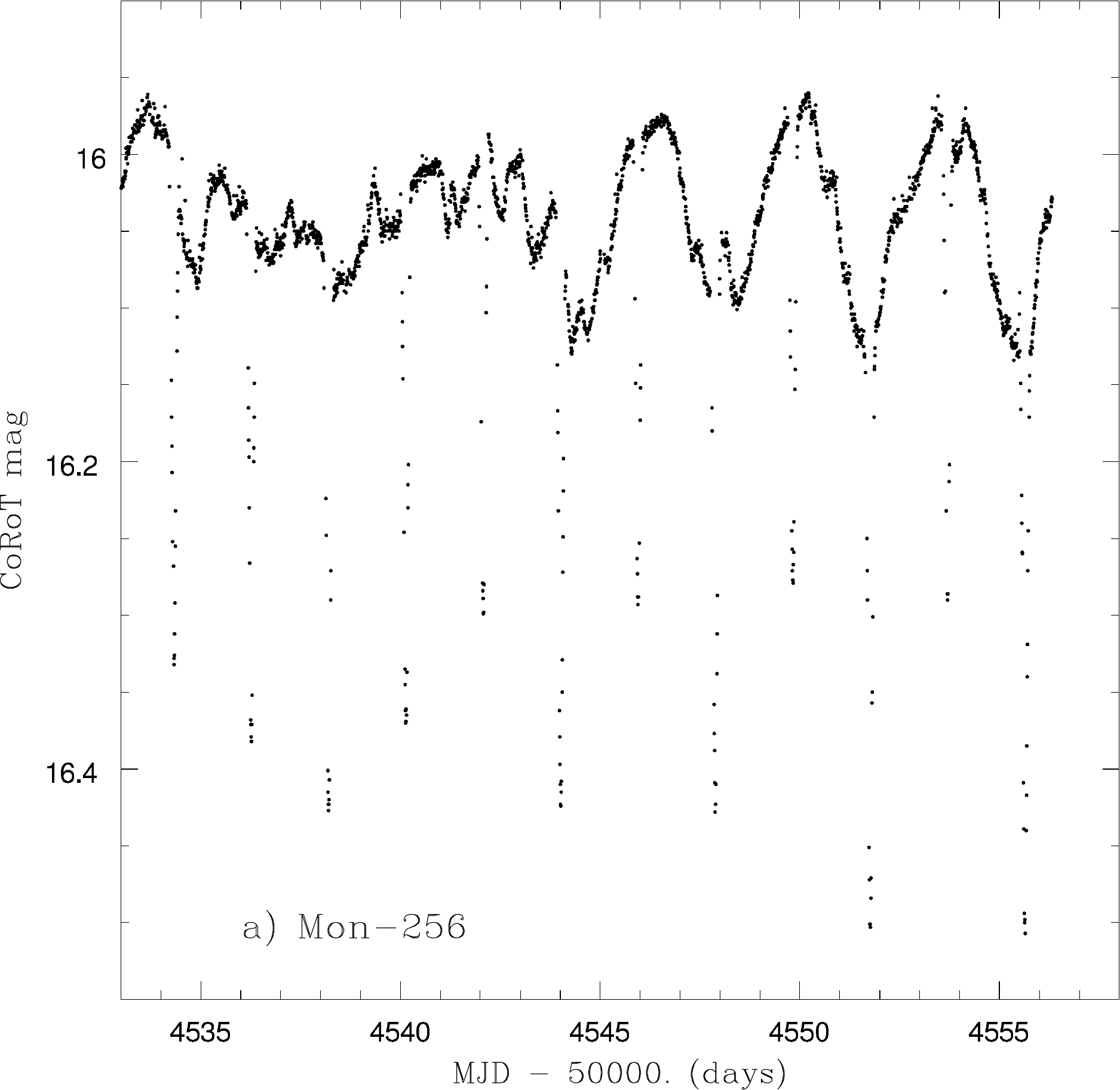}{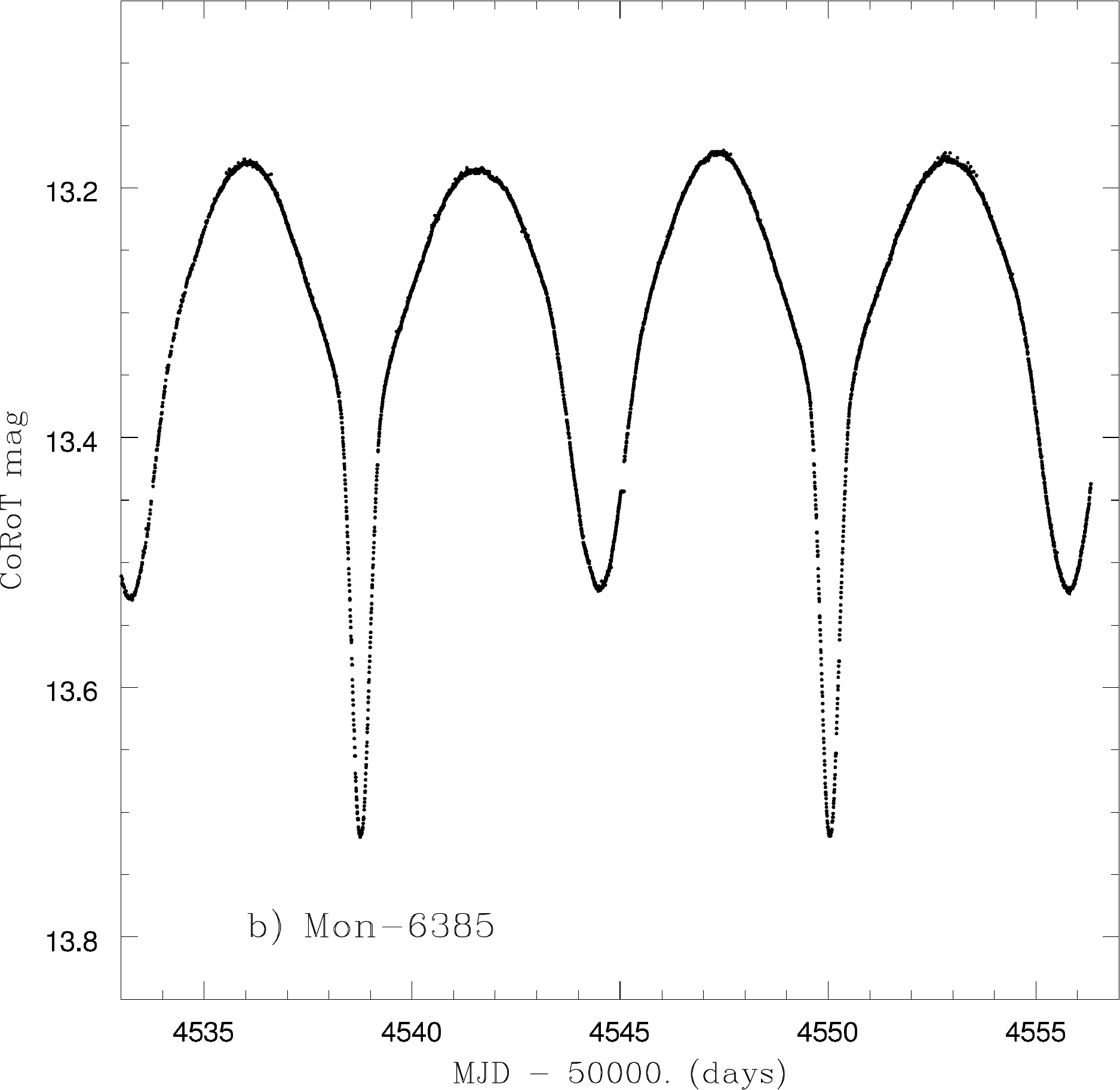}
\end{center}
\caption{{\em CoRoT} light curves for a) Mon-256 ({\em CoRoT} SRa01
223992193), an M dwarf eclipsing binary that is a member of NGC~2264; 
and  b) Mon-6385 ({\em CoRoT} SRa01 223958210), a much higher mass
field star that is a near-contact binary.   The flux dips produced by
eclipses in such systems are easily distinguished from those due to
the variable extinction events we study here.
\label{fig:EB_lightcurves}}
\end{figure*}

There are other types of variables -- primarily eclipsing binaries (or
systems with transiting exoplanets) -- which also have narrow flux
dips in their light curves, but their light curves are distinctively
different from those illustrated throughout this paper.
Figure~\ref{fig:EB_lightcurves} shows two eclipsing binary (EB) light
curves from the 2008 NGC~2264 campaign. 
Figure~\ref{fig:EB_lightcurves}a is the light curve for Mon-256, a
dM-dM  binary member of NGC~2264 (Gillen et al.\ 2014) that is a
borderline CTTS (broad and structured H$\alpha$ profile; modest IR
excess); this is the only EB in NGC~2264 with an IR excess presumably
due to  circumstellar or circumbinary dust.  For well-detached
systems, eclipses by stellar-sized bodies yield flux dips that are of
much shorter duration  (with a Gaussian fit width of $\sigma$ = 0.045
days for the eclipses of  Mon-256) than those which we study in this
paper.   The eclipse depths for Mon-256 appear to vary considerably
with time primarily due to the varying continuum level -- once that is
removed, the depths are quite stable, though slightly different at the
primary and secondary eclipse since each star has slightly different
radius and effective temperature. Figure~\ref{fig:EB_lightcurves}b
shows the light curve of 
a field star projected towards NGC~2264; its light curve is that for a
near-contact binary ($P\sim$ 11 day) where the continuum variations
are primarily due to the ellipsoidal shapes of the stars superposed on
which are narrow flux dips from the primary and secondary eclipses. 
The Gaussian width for the eclipse dip in this case ($\sigma$ = 0.21
days) is within the range of the widths for the short-duration flux
dips we discuss in this paper.  However, the regularity imposed in an
EB system yields a very stable light curve which is easily
distinguishable from light curves associated with variable
circumstellar extinction.

\section{Identification and Initial Characterization of YSOs with 
       Short-Duration, Periodic Flux Dips in their Light curves }

First results from our 2011-2012 intensive monitoring of NGC 2264
with  {\em CoRoT} and {\em Spitzer} have been published in Cody et
al.\ (2014) and Stauffer et al.\ (2014).  While working on those
papers, we noticed a small set of stars that shared some similarities
but did not seem to fit into the primary groupings identified by Cody
et al. Often, the ``narrow" flux dips in these light curves are
superposed on other, larger amplitude, types of variability -- 
primarily semi-sinusoidal light curves normally attributed to
photospheric spots --  making identification of this new light curve
signature more difficult.

\begin{figure*}
\begin{center}
\epsfxsize=.99\columnwidth
\epsfbox{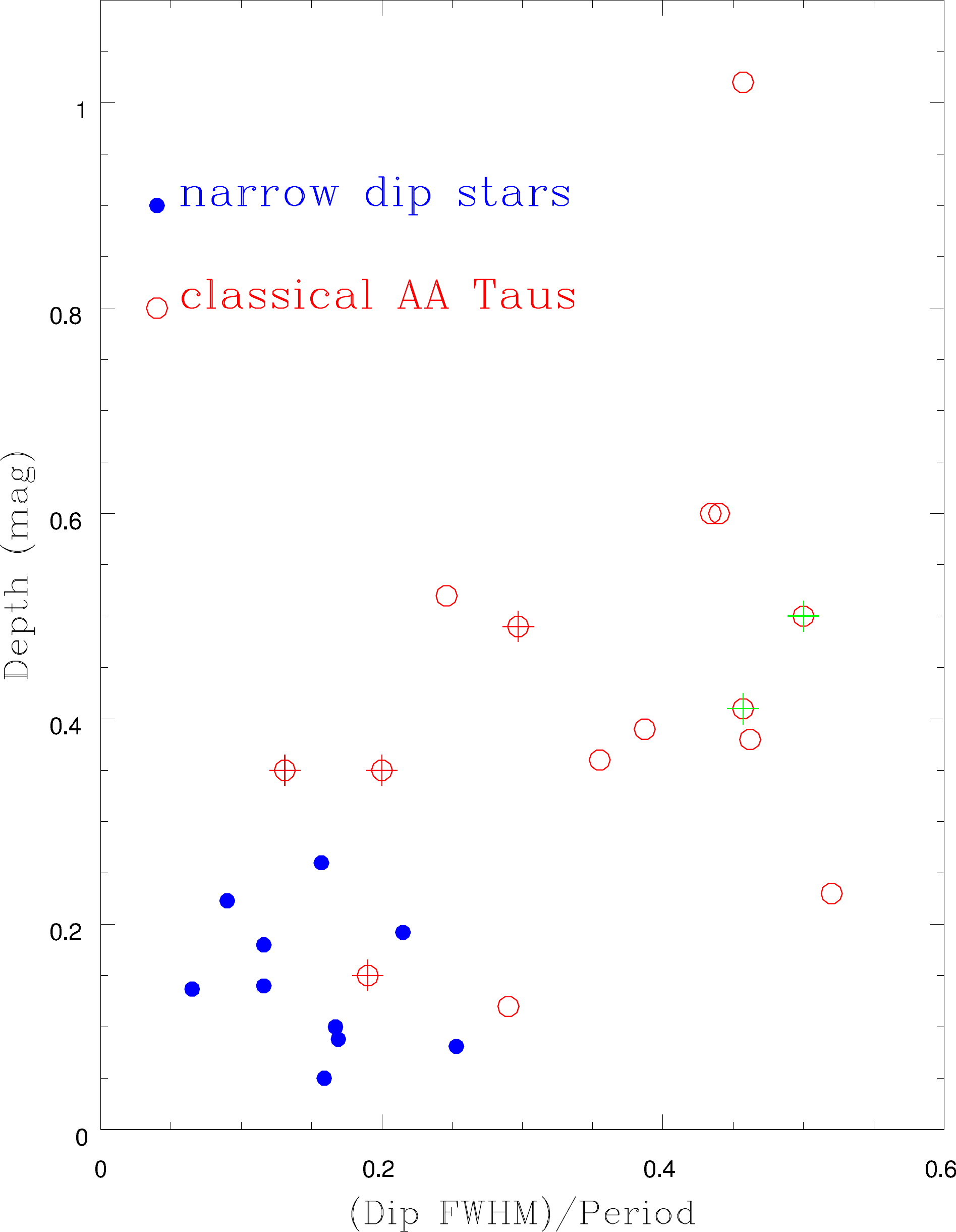}
\end{center}
\caption{Comparison of the average full depth of the flux dips to the
ratio of the dip FWHM and the dip period for {\em CoRoT} light curves
of ``classical" AA Tau stars in our data set and for the stars of
Table~\ref{tab:basicinformation}.  The stars in the newly identified
narrow-dip class  have flux dips that are both shallow and of
relatively short duration.  The stars marked with green plus signs
represent AA Tau type variables also having periodic narrow flux dips, while
the red plus signs represent stars whose {\em CoRoT} light curves
have flux dips that are
sometimes narrow and sometimes broad,
as discussed in \S 5 and illustrated in
Figure~\ref{fig:mixed_dips} and Figure~\ref{fig:composite_dips}.
\label{fig:compare_dippers}}
\end{figure*}

These narrow flux dips presumably arise because some physical
structure transits our line of sight to the stellar photosphere.   As
mentioned in the previous section, there is already a light curve
class (AA Tau analogs) with periodic, variable extinction events and
for which a standard physical model has been developed. It is possible
that this new light curve class can be fit into the same physical
model; however, we believe there are a number of reasons to consider
alternate physical models which we will discuss in \S 7.   To
emphasize the primary morphological differences between the
``classical" AA Tau analogs and our proposed new class,
Figure~\ref{fig:compare_dippers} compares the average widths and depths of the
flux dips for the new group of stars to those of the AA Tau analogs in
NGC~2264 identified in the companion paper  by McGinnis et
al.\footnote{Primarily because our definition of the AA Tau class
requires that the optical light curves satisfy two quantitative tests
(Cody et al.\ 2014),  our (Stauffer and Cody) list of NGC~2264 AA Tau
analogs is slightly smaller than that of McGinnis et al.\ (2015).  
Specifically, we do not include Mon-358, Mon-654, Mon-1054, and
Mon-1167;  we also do not include Mon-56 and Mon-1131 because we
instead place them in our new category of stars with narrow, periodic
flux dips.} While the distribution of points in
Figure~\ref{fig:compare_dippers} could be regarded as simply one
population with a range of properties, it could also  be viewed as two
populations whose properties only slightly overlap.  The goal of this
paper is to better characterize the light curves with these narrow
flux dips, thereby determining if they should be considered as a new
class of event or if they should simply be regarded as an extension of
the AA Tau class.  Throughout the paper, we use the terms ``narrow"
and ``short-duration" interchangeably to describe the flux-dips of this
set of stars, and ``broad" (or long-duration) to describe the flux
dips for the classic AA Tau light curves, with the dividing line being
approximately FWHM/Period = 0.25 (see Figure~\ref{fig:compare_dippers}).

In Table~\ref{tab:basicinformation} and
Table~\ref{tab:basicphotometry}, we provide basic photometric and
spectroscopic data for the nine stars which we believe fall into this
``narrow-dip"  group, compared to a total of 159 NGC~2264 CTTS with
{\em CoRoT} light curves (McGinnis et al. 2015). All but one of these
stars are previously identified members of NGC~2264 that had been
classified as CTTS based on either H$\alpha$\ emission or SEDs.  
Spectral types are available for all of these objects, and all are low
mass stars with spectral types of K5 or later.   In the remainder of
this section, we provide a brief description of each star  and its
light curve properties.  In Table~\ref{tab:quant_information},  we
provide the  fitted times, widths and depths of all the flux dips for
these stars where those quantities could be determined accurately,
and
we give a label for each of these dips - the dip ID column of the
table - primarily as a means for us to refer to specific dips in
the text.

\begin{deluxetable*}{lccccccccccc}
\tabletypesize{\scriptsize}
\tablecolumns{12}
\tablewidth{0pt}
\tablecaption{Basic Information for YSOs with Short Duration Flux Dips\label{tab:basicinformation}}
\tablehead{
\colhead{Mon ID\tablenotemark{a}}  & \colhead{2MASS ID} &
\colhead{{\em CoRoT}\tablenotemark{b}} & \colhead{{\em CoRoT}\tablenotemark{c}} &
\colhead{SpT\tablenotemark{d}} & \colhead{H$\alpha$ EW
\tablenotemark{d}} & \colhead{FR4.5\tablenotemark{e}} &
\colhead{Period\tablenotemark{f}} 
& \colhead{L$_{*}$\tablenotemark{g}} & \colhead{R$_*$\tablenotemark{g}} 
& \colhead{R$_{co}$\tablenotemark{h}} & \colhead{R$_d$\tablenotemark{i}} \\
\colhead{} & \colhead{} & \colhead{} & \colhead{} & \colhead{} & 
\colhead{(\AA)} & \colhead{} & \colhead{(days)} & \colhead{(L$_{\odot}$)} & \colhead{(R$_{\odot}$)} 
& \colhead{(R$_{\odot}$)} & \colhead{(R$_{\odot}$)} 
}
\startdata
CSIMon-000021 & 06405944+0959454 & 223980412 & 223980412 & K5 & -7.4 & 1.6 &  3.2 & 0.96 & 1.69 & 9.7 & 7.3/14.5 \\
CSIMon-000056 & 06415315+0950474 & 223994760 & 223994760 & K5 & -1.8 & 1.0 & 5.83 & 0.77 & 1.51 & 14.3 & 6.5/13.0  \\
CSIMon-000378 & 06405292+0944544 & \nodata   & 616872605 & K5.5 & -8.5 & 2.1 &  11.1 & 0.90 & 1.71 & 21.3 & 7.0/14.0 \\
CSIMon-000566 & 06405275+0943004 & 400007955 & 400007955 & M3.5 & -19.4 & 0.9 & 4.5 & 0.16 & 1.23 & 7.3 & 2.9/5.9 \\
CSIMon-001076 & 06403819+0929524 & 400007702 & 605424384 & M1 & -2.6 & 0.6 & 5.85 & 0.26 & 1.25 & 10.4 & 3.8/7.5 \\
CSIMon-001131 & 06393441+0954512 & 223957455 & 223957455 & M1.5: & -31.1  & 1.4 & 5.2  & 0.55 & 2.02 & 8.9 & 5.5/11.0 \\
CSIMon-001165 & 06404192+0943551 & \nodata   & \nodata   & M3: &  -7.9  & 1.3 & 5.5 & 0.13 & 1.15 & 8.5 & 2.7/5.3 \\
CSIMon-001580 & 06410285+1007119 & 223981250 & \nodata   & M1: &  -17.2  & 0.6 & 7.1 & 0.27 & 1.28 & 11.9 & 3.8/7.7 \\
CSIMon-006975 & 06395318+1001137 & \nodata   & 616803898 &  M2.5:  &  -15.  & 0.2 & 2.8 & 3.42 & 3.44 & 5.8 & 9.3/18.6 \\
\enddata
\tablenotetext{a}{The CSIMon IDs are our internal naming scheme for stars in
the field of NGC~2264 -- see Cody et al.\ (2014) and 
\url{http://irsa.ipac.caltech.edu/data/SPITZER/CSI2264 for RAs and Decs and
alternative IDs}.  
In the main text of this paper, we omit the ``CSI" and the
leading zeros in the object names for brevity. }
\tablenotetext{b}{{\em CoRoT} identification number in 2008 SRa01 campaign.  }
\tablenotetext{c}{{\em CoRoT} identification number in 2011 SRa05 campaign. }
\tablenotetext{d}{See Cody et al.\ (2014) and the Appendix for the sources of the
spectral type and H$\alpha$ equivalent width (EW) data.  Here we use the convention
that negative values are in emission, which is opposite to the convention in
Cody et al.  }
\tablenotetext{e}{FR4.5 is an estimate of the ratio of the flux from the disk to the
flux from the stellar photosphere at 4.5 $\mu$m, based on all the available broad-band
photometry and the published spectral types.  Because the SED models do not include
veiling, these values are likely lower limits.}
\tablenotetext{f}{Period determined in this paper, see Section 3.  }
\tablenotetext{g}{Stellar luminosity and radius estimates from Venuti et al. (2014). }
\tablenotetext{h}{Co-rotation radius; see formula in Bouvier et al. (1999). }
\tablenotetext{i}{Dust sublimation radius; see formula in Monnier \& Millan-Gabet (2002). We adopt
  values of 1500 K for the dust sublimation temperature, and minimum and maximum values of their
  Q$_R$ parameter of 1.0 and 4.0, based on their Figure 2.}
\end{deluxetable*}

\begin{deluxetable*}{ccccccccccccc}
\tabletypesize{\scriptsize}
\tablecolumns{10}
\tablewidth{0pt}
\tablecaption{Photometric Data for the Stars of
Table~\ref{tab:basicinformation}\label{tab:basicphotometry}}
\tablehead{
\colhead{Mon ID}  & \colhead{$u$} & \colhead{$g$} & \colhead{$r$} &
\colhead{$i$} & \colhead{$J$} & \colhead{$H$} & \colhead{$K_s$} &
\colhead{[3.6]} &
\colhead{[4.5]} & \colhead{[5.8]} & \colhead{[8.0]} & \colhead{[24]}
}
\startdata
CSIMon-000021 & 18.174 & 15.835 & 14.810 & 14.452 & 12.781 & 12.093 & 11.685 & 11.248 & 10.936 & 10.707 & 10.052 & 7.556 \\
CSIMon-000056 & 18.785 & 16.205 & 15.056 & 14.475 & 12.959 & 12.302 & 12.003 & 11.717 & 11.390 & 11.038 & 10.290 & 7.866 \\
CSIMon-000378 & 18.277 & 16.142 & 14.943 & 14.239 & 12.699 & 11.862 & 11.267 & 10.882 & 10.540 & 10.091 &  9.371 & 6.270 \\
CSIMon-000566 & 20.988 & 19.109 & 17.747 & 16.418 & 14.318 & 13.588 & 13.279 & 12.823 & 12.576 & 12.140 & 11.132 & \nodata \\
CSIMon-001076 & 20.754 & 18.336 & 16.896 & 16.012 & 13.931 & 13.194 & 12.891 & 12.485 & 12.289 & 12.025 & 11.278 & 8.566 \\
CSIMon-001131 & 18.769 & 17.180 & 15.832 & 14.856 & 12.991 & 12.215 & 11.932 & 11.492  & 11.152  & 10.893 & 10.195 & 6.998 \\
CSIMon-001165 & 21.924 & 19.528 & 17.983 & 16.427 & 14.679 & 13.845 & 13.582 & 13.043 & 12.789 & 12.380 & 11.571 & \nodata \\
CSIMon-001580 & 20.489 & 17.976 & 16.553 & 15.678 & 13.978 & 13.218 & 12.874 & 12.624 & 12.455 & 12.223 & 11.244 & 8.013 \\ 
CSIMon-006975 & \nodata & 20.027 & 17.974 & 16.626 & 13.476 & 12.576 & 12.151 & 11.666 & 11.56 & \nodata & 11.10 & \nodata \\

\enddata 
\tablecomments{Broad-band photometry for the stars from
Table~\ref{tab:basicinformation}, in AB magnitudes for $ugri$ but in
Vega magnitudes for longer wavelengths. The $ugri$ data are from CFHT,
as reported in Venuti et al.\ (2014); the $JHK_s$ data are from the
on-line 2MASS all-sky point source catalog; the IRAC data are from
Sung et al.\ (2009), or from our own analysis of archival IRAC
imaging.  Typical photon-noise and calibration uncertainties for these
magnitudes are of order 0.02 mag; however, because all of these stars
are photometric variables with amplitudes up to several tenths of a
magnitude, these single-epoch data could differ from absolute,
time-averaged values by 0.1 mag or more.}
\end{deluxetable*}

\subsection{Mon-21}

Mon-21, also known as V602 Mon, is a relatively poorly studied member of NGC~2264 of spectral
type K5 (Rebull et al.\ 2002), with weak H$\alpha$\ emission (Furesz
et al.\ 2006) and a relatively modest IR excess.  It has no
discernible ultraviolet (UV) excess (Venuti et al.\ 2014). We obtained
{\em CoRoT} light curves for it in both 2008 and 2011, and an IRAC
light curve for it in 2011.

\begin{deluxetable*}{lccccc}
\tabletypesize{\scriptsize}
\tablecolumns{6}
\tablewidth{0pt}
\tablecaption{Flux Dip Gaussian Fit Parameters for Stars in 
Table~\ref{tab:basicinformation} \label{tab:quant_information}}
\tablehead{
\colhead{Mon ID}  & \colhead{Year} &
\colhead{Dip ID} & \colhead{Center (days)} &
\colhead{Width (days)} & \colhead{Percent Depth} \\
\colhead{} & \colhead{} & \colhead{} & MJD-50000. & \colhead{} & \colhead{}
}
\startdata
CSIMon-000021 & 2008 & A1 & 4535.39 & 0.17 & 10\%  \\
CSIMon-000021 & 2008 & A2 & 4536.15 & 0.17 & 7\%  \\
CSIMon-000021 & 2008 & B & 4541.99 & 0.18 & 7\%  \\
CSIMon-000021 & 2008 & C & 4545.07 & 0.18 & 5\%  \\
CSIMon-000021 & 2008 & D & 4548.47 & 0.17 & 12\%  \\
CSIMon-000021 & 2008 & E1 & 4554.73 & 0.16 & 10\%  \\
CSIMon-000021 & 2008 & E2 & 4555.60 & 0.15 & 4\%  \\
\nodata    & \nodata & \nodata & \nodata & \nodata & \nodata \\
CSIMon-000021 & 2011 & A & 5897.65 & 0.22 & 10\% \\
CSIMon-000021 & 2011 & B1 & 5900.35 & 0.35 & 2.5\% \\
CSIMon-000021 & 2011 & B2 & 5900.77 & 0.29 & 4.5\% \\
CSIMon-000021 & 2011 & C & 5910.29 & 0.26 & 8\% \\
CSIMon-000021 & 2011 & D1 & 5913.10 & 0.33 & 3.5\% \\
CSIMon-000021 & 2011 & D2 & 5913.30 & 0.11 &  3.5\% \\
CSIMon-000021 & 2011 & D3 & 5913.64 & 0.18 &  6\% \\
CSIMon-000021 & 2011 & E & 5916.31 & 0.41 &  4\% \\
CSIMon-000021 & 2011 & F1 & 5919.23 & 0.38 &  4.5\% \\
CSIMon-000021 & 2011 & F2 & 5919.73 & 0.25 &  6.5\% \\
CSIMon-000021 & 2011 & G1 & 5922.51 & 0.25 &  6.0\% \\
CSIMon-000021 & 2011 & G2 & 5922.96 & 0.20 & 8.2\% \\
CSIMon-000021 & 2011 & H1 & 5925.04 & 0.34 &  1.5\% \\
CSIMon-000021 & 2011 & H2 & 5925.86 & 0.29 &  10\% \\
CSIMon-000021 & 2011 & I1 & 5928.60 & 0.35 &  5.5\% \\
CSIMon-000021 & 2011 & I2 & 5929.20 & 0.27 &  10\% \\
CSIMon-000021 & 2011 & J1 & 5931.79 & 0.39 &  4\% \\
CSIMon-000021 & 2011 & J2 & 5932.37 & 0.23 &  12\% \\
\nodata    & \nodata & \nodata & \nodata & \nodata & \nodata \\
CSIMon-000056 & 2011 & A & 5912.63 & 0.32 & 12.5\% \\
CSIMon-000056 & 2011 & B & 5919.00 & 0.25 & 5.5\% \\
CSIMon-000056 & 2011 & C1 & 5924.30 & 0.25 & 8.3\% \\
CSIMon-000056 & 2011 & C2 & 5925.08 & 0.27 & 7.7\% \\
\nodata    & \nodata & \nodata & \nodata & \nodata & \nodata \\
CSIMon-001165 & 2011 & A & 5911.87 & 0.20 & 25\% \\
CSIMon-001165 & 2011 & B & 5922.83 & 0.15 & 14\% \\
\nodata    & \nodata & \nodata & \nodata & \nodata & \nodata \\
CSIMon-006975 & 2011 & A1 & 5905.54 & 0.10 & 15\% \\
CSIMon-006975 & 2011 & A2 & 5905.82 & 0.10 & 12\% \\
CSIMon-006975 & 2011 & B & 5908.36 & 0.13 & 14\% \\
CSIMon-006975 & 2011 & C & 5910.94 & 0.10 & 7\% \\
CSIMon-006975 & 2011 & D & 5919.59 & 0.11 & 11\% \\
\nodata    & \nodata & \nodata & \nodata & \nodata & \nodata \\
CSIMon-001131 & 2011 & A & 5901.06 & 0.30 & 24\% \\
CSIMon-001131 & 2011 & B & 5906.14 & 0.30 & 23\% \\
CSIMon-001131 & 2011 & C1 & 5910.66 & 0.27 & 12\% \\
CSIMon-001131 & 2011 & C2 & 5911.03 & 0.18 & 10\% \\
CSIMon-001131 & 2011 & C3 & 5911.56 & 0.36 & 11\% \\
CSIMon-001131 & 2011 & D & 5926.75 & 0.75 & 13\% \\
CSIMon-001131 & 2011 & E1 & 5931.66 & 0.34 & 6.5\% \\
CSIMon-001131 & 2011 & E2 & 5932.04 & 0.14 & 10\% \\
CSIMon-001131 & 2011 & E3 & 5932.31 & 0.15 & 12\% \\
\nodata    & \nodata & \nodata & \nodata & \nodata & \nodata \\
CSIMon-001580 & 2008 & A1 & 4541.73 & 0.18 & 13\% \\
CSIMon-001580 & 2008 & A2 & 4542.15 & 0.24 & 9\% \\
CSIMon-001580 & 2008 & B1 & 4548.53 & 0.15 & 6\% \\
CSIMon-001580 & 2008 & B2 & 4548.7  & 0.06 & 5\% \\
CSIMon-001580 & 2008 & B3 & 4548.98 & 0.17 & 8\% \\
CSIMon-001580 & 2008 & C & 4554.83 & 0.20 & 25\% \\
\nodata    & \nodata & \nodata & \nodata & \nodata & \nodata \\
CSIMon-000378 & 2011 & A & 5918.64 & 0.55 & 14\% \\
\nodata    & \nodata & \nodata & \nodata & \nodata & \nodata \\
CSIMon-001076 & 2008 & A & 4538.65 & 0.17 & 15.7\% \\
CSIMon-001076 & 2008 & B & 4550.54 & 0.16 & 13\% \\
CSIMon-001076 & 2008 & C & 4556.20 & 0.15 & 12.5\% \\
\nodata    & \nodata & \nodata & \nodata & \nodata & \nodata \\
CSIMon-000566 & 2011 & A & 5909.33 & 0.32 & 5\% \\
\enddata
\end{deluxetable*}

\begin{figure*}
\begin{center}
\epsscale{0.75}
\plottwo{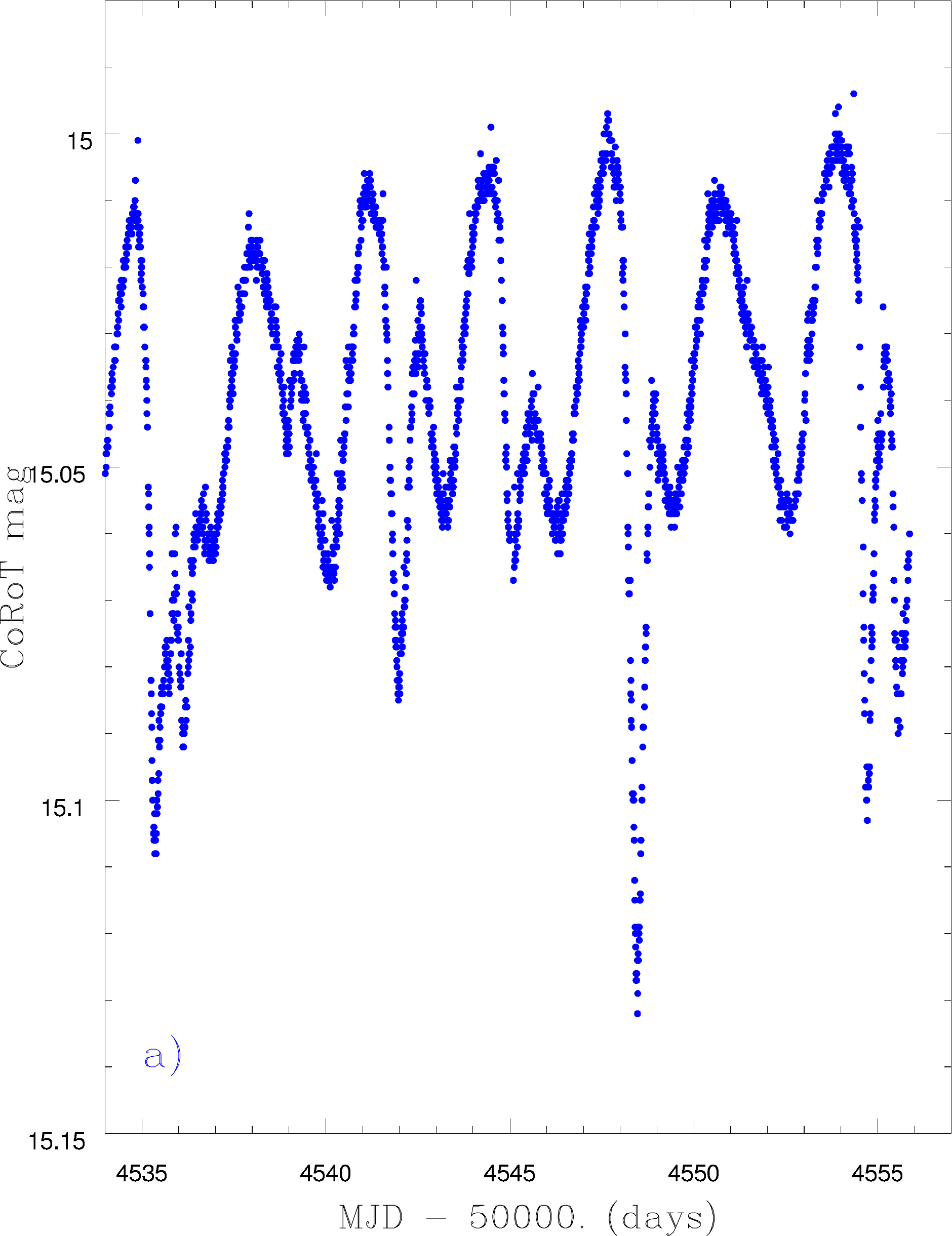}{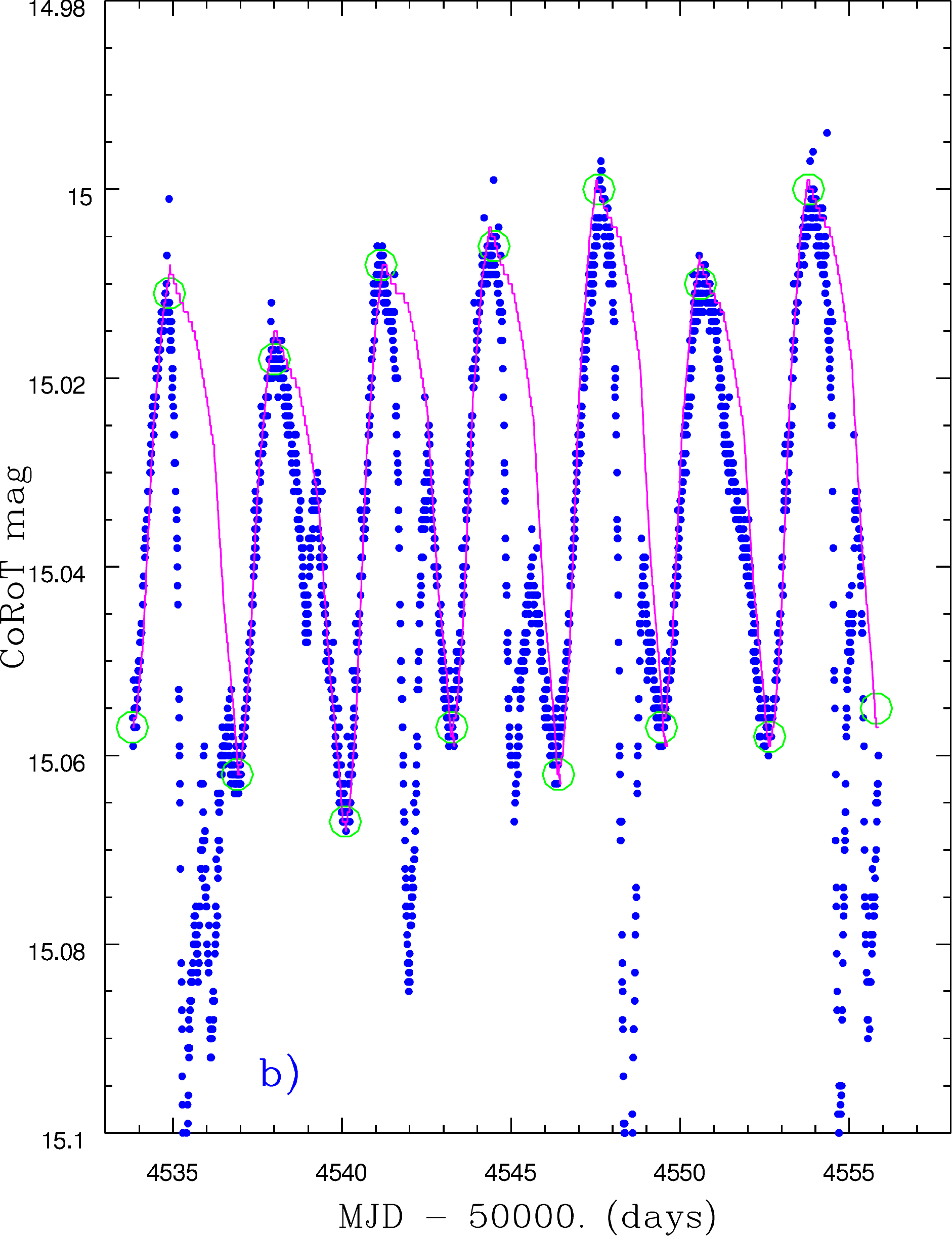}
\vspace{-0.0cm}
\plottwo{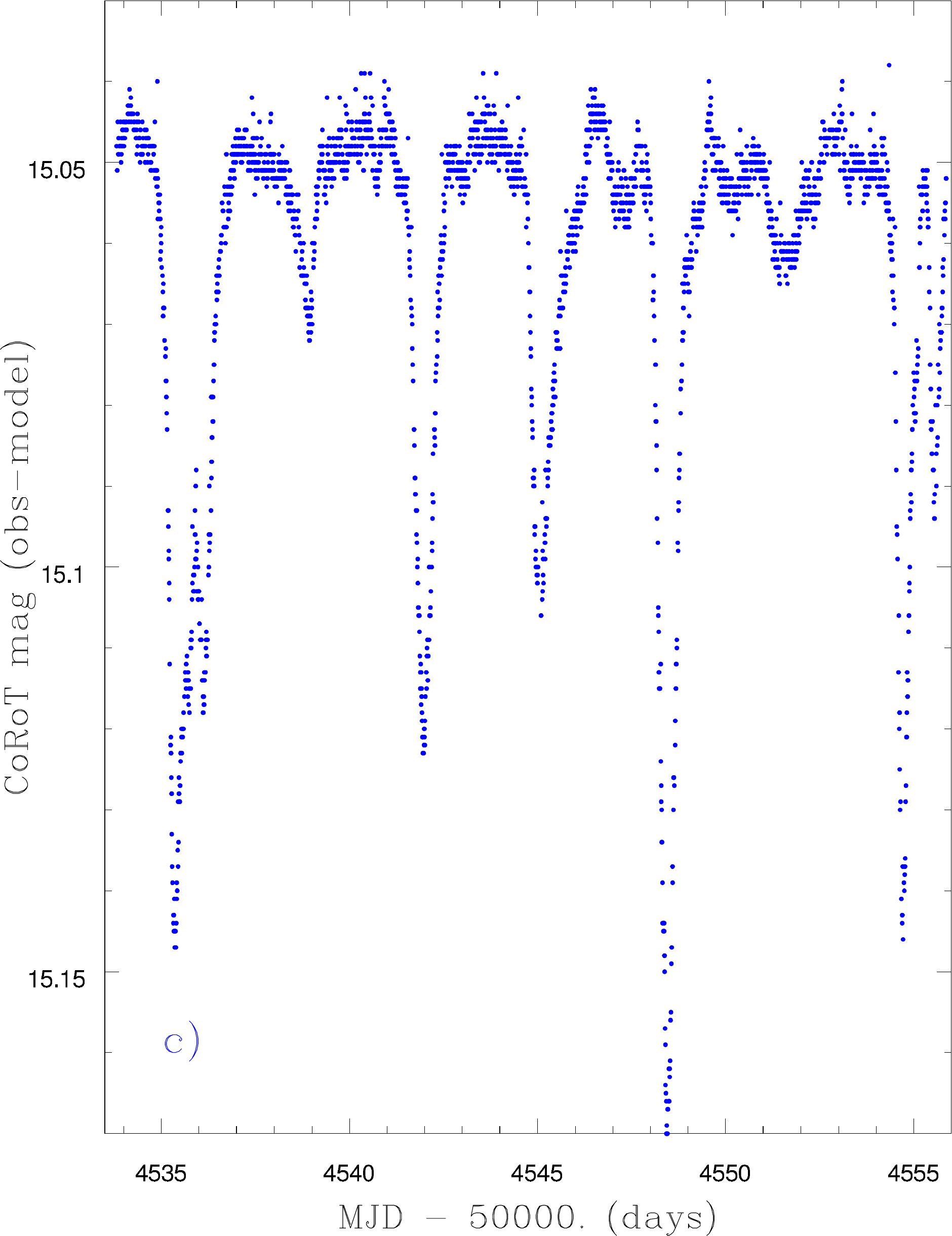}{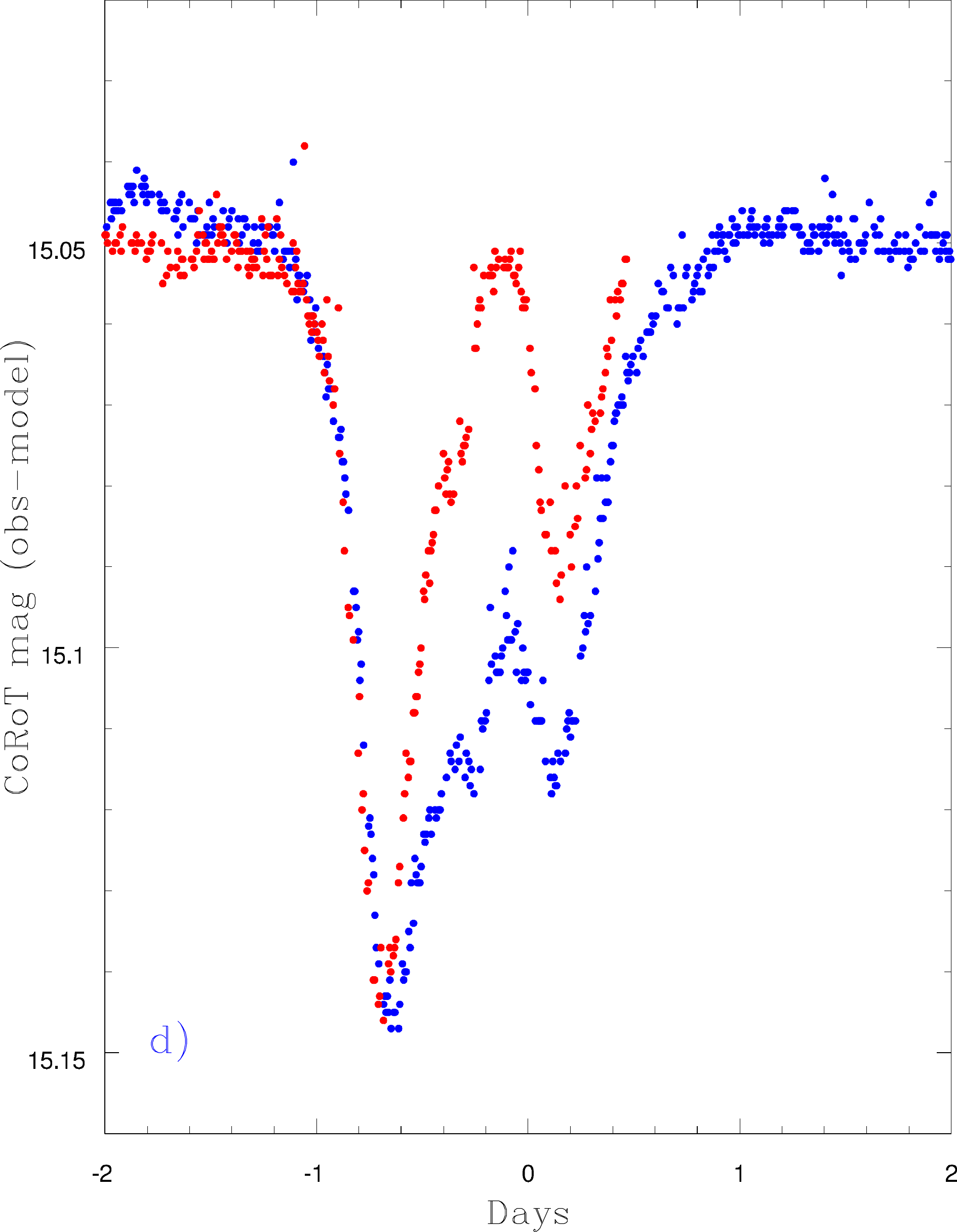}
\vspace{-0.0cm}
\plottwo{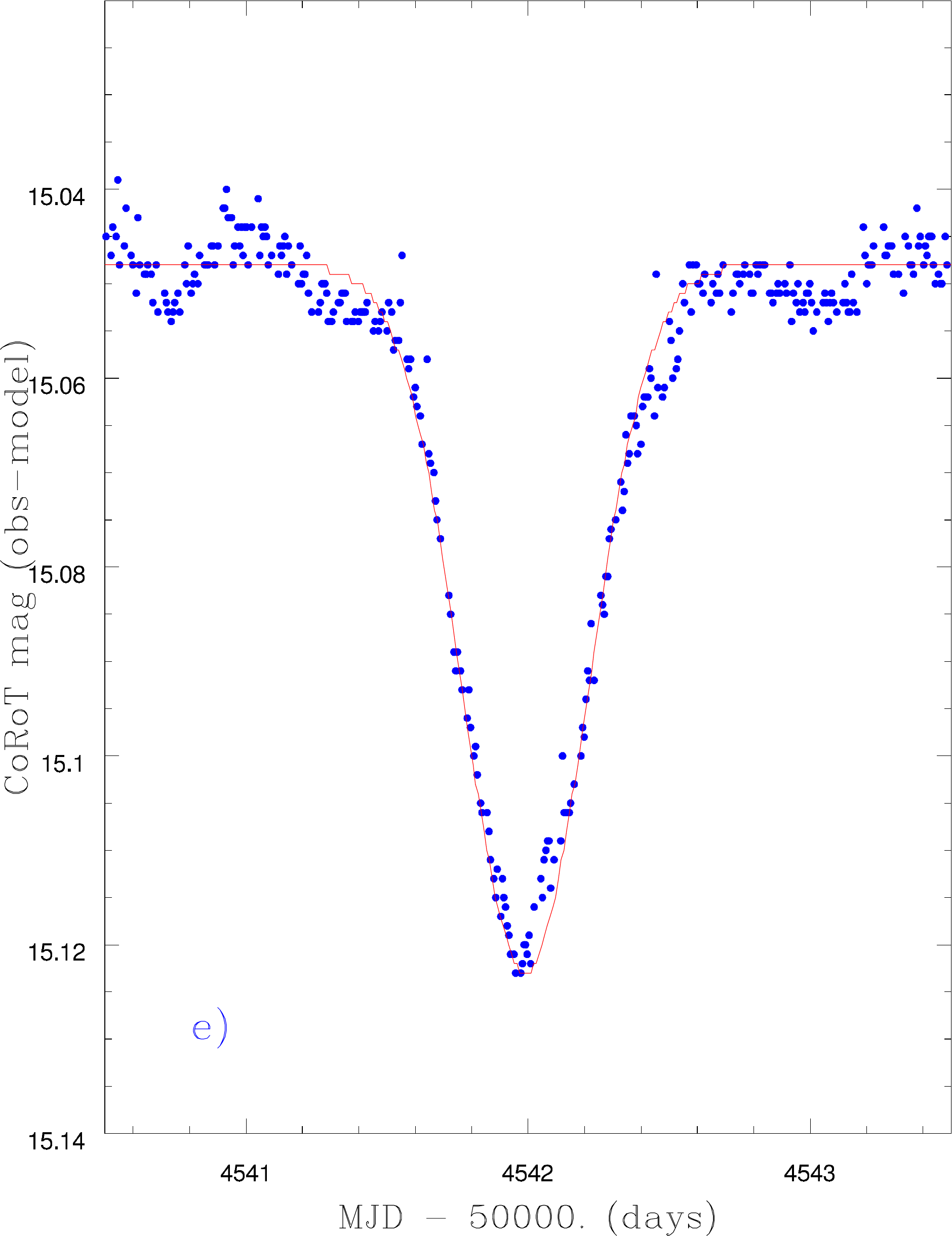}{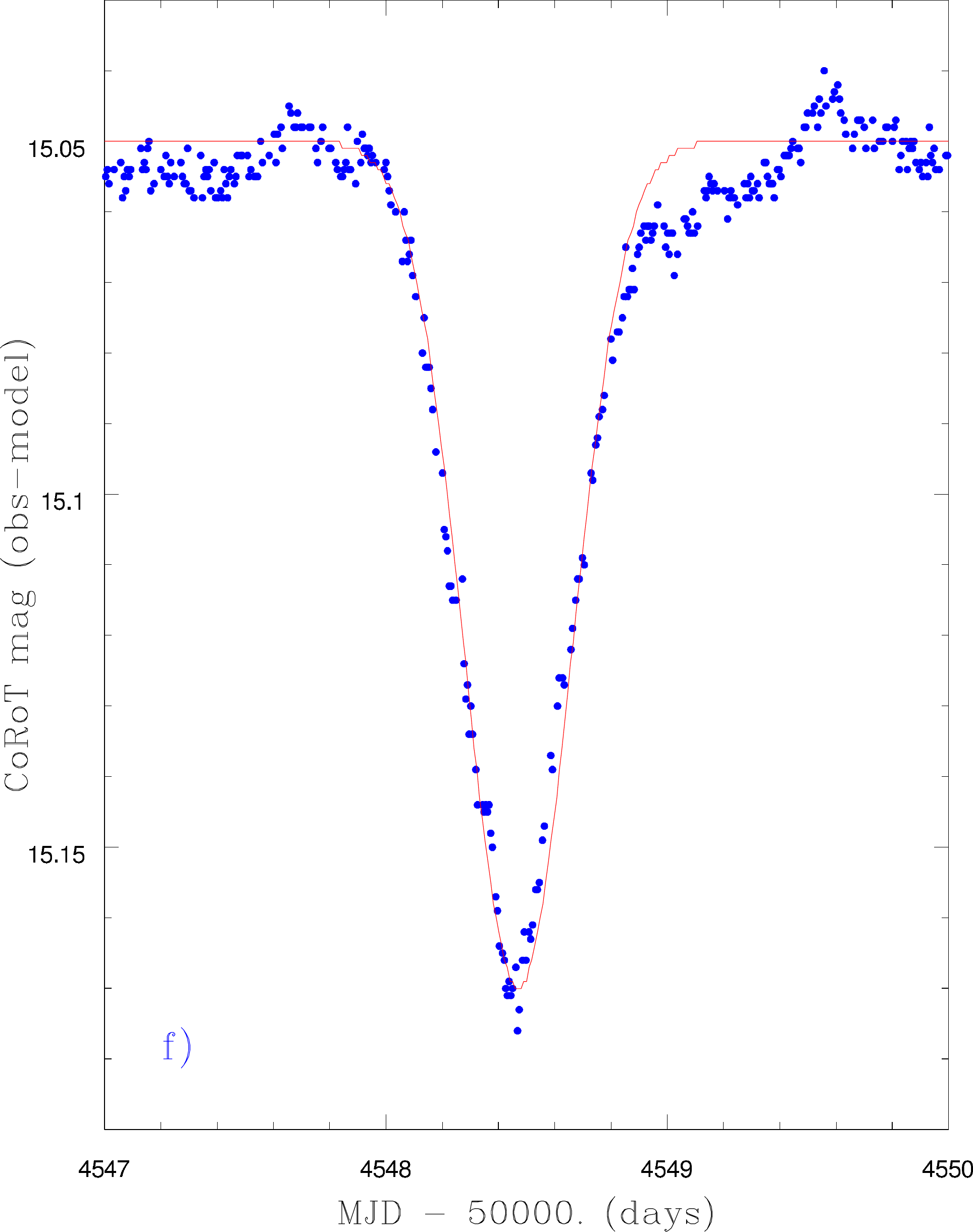}
\end{center}
\caption{a) As observed {\em CoRoT} light curve for Mon-21; b) same as
(a), except now overplotted in red by a model light curve designed to
approximate the light curve shape attributable to spots on the star's
photosphere; c) residual light curve produced by subtracting the
spotted star model from the observed light curve for Mon-21; d)
superposition of the flux dips approximately centered at MJD 54536 
(in blue, dips A1 and A2 in Table~\ref{tab:quant_information}
and 54555 (dips
E1 and E2, in red), highlighting their double-peak nature and their
qualitative similarity; e and f) illustration that some of the narrow
flux dips for Mon-21 are nearly Gaussian in shape (dips B and D
in Table~\ref{tab:quant_information},
respectively).  The period corresponding to the spotted-star waveform -
the red curve in panel (b) - is about 3.15 days; the period corresponding
to the average dip interval is about 3.22 days (see text).  \label{fig:Mon21_modela}}
\end{figure*}

\subsubsection{Mon-21 (2008 {\em CoRoT} campaign)}

The as-observed {\em CoRoT} light curve for Mon-21 in  2008
(Figure~\ref{fig:Mon21_modela}a) appears relatively complex because,
we believe, it is a superposition of two distinctly different
waveforms arising from two different physical mechanisms. 
Specifically,  we ascribe the broad, semi-sinusoidal variation to
rotational modulation due to the presence of a large, high latitude
spot group on the  star's photosphere.  The narrow flux dips that are
also present are the component we wish to isolate and quantify.  The
two waveforms appear to have nearly identical periods, and to produce
features of very similar amplitude.

In order to measure accurately the properties of the flux dips, it is
necessary to remove the comparatively smoothly varying,
semi-sinusoidal pattern.   We construct a model light curve to emulate
the sinusoidal variation by taking the {\em CoRoT} data from day MJD
4537.4 to 4540.6, the portion of the light curve least affected by
narrow flux dips, and use those data to produce  a sine-wave-like
template covering one full period (3.22 days). (The light curve
interval from $\sim$4550 to 4553 is also relatively clean, but we felt
that the inflection in the curve presumably due to a narrow dip was
harder to eliminate unambiguously compared to the earlier time
interval.)   From examination of the light curve, we mark beginning
points (flux and time) for each cycle, which are indicated as the
lower green circles in Figure~\ref{fig:Mon21_modela}b. A linear
correction term is added to the flux for each replicated template so
that its ending flux matches the next cycle's beginning flux. Finally,
we slightly adjust the amplitude of each waveform (by multiplying all
the measurements in that replicated waveform by a constant) so that
the maximum brightness matches the observed maximum at that epoch
(upper green circles in Figure~\ref{fig:Mon21_modela}b).  The model
light curve is shown in red in Figure~\ref{fig:Mon21_modela}b;  the
result of subtracting this model from the observed light curve -- and
hence our representation of the light curve solely due to variable
extinction events -- is shown in Figure~\ref{fig:Mon21_modela}c.   We
made no further adjustments to the model to better fit the observed
data other than the deterministic steps described above.  We note that
the first and last of the narrow dips are in fact double-dips, with
the second dip being shallower and offset by of order 0.8 days in both
cases (see Figure~\ref{fig:Mon21_modela}d).

We have ascribed the sinusoidal variations to non-axisymmetrically
distributed  star spots on the stellar photosphere -- probably cold
spots.  Could it instead be ascribed to variable extinction due to a
warped disk intersecting our line of sight?  We offer two arguments in
favor of our hypothesis and counter to the alternative.  First, we
know what the light curves of YSOs with typical  warped disks look
like because there are many of them in NGC~2264 (see the companion
paper to this, McGinnis et al.\ 2015).  Those light curves look
nothing like the semi-sinusoidal pattern shown in red in
Figure~\ref{fig:Mon21_modela}b; the warped disk light curves are much
more irregularly shaped and much more time variable.  Second, we also
know that light curves dominated by non-axisymmetrically distributed,
high-latitude spots are common amongst YSOs (Herbst 1986; Bouvier \&
Bertout 1989) and their light curves closely resemble the red curve in
Figure~\ref{fig:Mon21_modela}b.  In fact, we have identified one CTTS
in NGC~2264 -- Mon-103 -- whose {\em CoRoT} light curve is nearly
identical, except for minor scale factors, to the red curve in
Figure~\ref{fig:Mon21_modela}b.  We provide a quantitative comparison
of the 2008 Mon-21 light curve and that for Mon-103 in the Appendix.

Figures~\ref{fig:Mon21_modela}e and f compare a Gaussian fit to the
observations for the dips that are narrowest and best fit by such a
model (dips B and D - see Table~\ref{tab:quant_information}).
The derived times, Gaussian widths and dip depths (as a
percentage of the ``continuum" flux) are provided in 
Table~\ref{tab:quant_information}.  Most of the dips are well fit, and
most have widths in the narrow range from 0.15 to 0.18 days.

The semi-sinusoidal pattern and the narrow flux dips both have
apparent periodicities of about 3.2 days.   However, are the dips
exactly periodic? And, are the periods for the semi-sinusoidal
waveform and the dips precisely the same?  The answer to the first
question is no, as can be seen from  examination of the data in
Table~\ref{tab:quant_information}.  The centroids of each dip can be
measured to an accuracy of a few hundredths of a day, yet dip B and C
are separated by 3.08 days, while dips C and D are separated by 3.40
days.  Applying the period-finding algorithm of Plavchan  et al.\
(2008b) to the light curve for just the flux dips
(Figure~\ref{fig:Mon21_modela}c) yields a period of 3.224 days; a
similar estimate of 3.223 days is obtained by  subtracting the
measured times for dip E1 and A1 and assuming they are separated by
six periods. The Lomb-Scargle algorithm yields a period of 3.15 days
for the semi-sinusoidal pattern (represented either just by the red
curve in Figure~\ref{fig:Mon21_modela}b or by the residual remaining
after subtracting the Gaussian fits from the original light curve). 
Given the real variations in the timing of the flux dips and the
comparative shortness of our {\em CoRoT} campaign, we cannot conclude
the two periods are significantly different.

\subsubsection{Mon-21 (2011 {\em CoRoT} campaign)}

\begin{figure*}
\begin{center}
\epsscale{1.0}
\plottwo{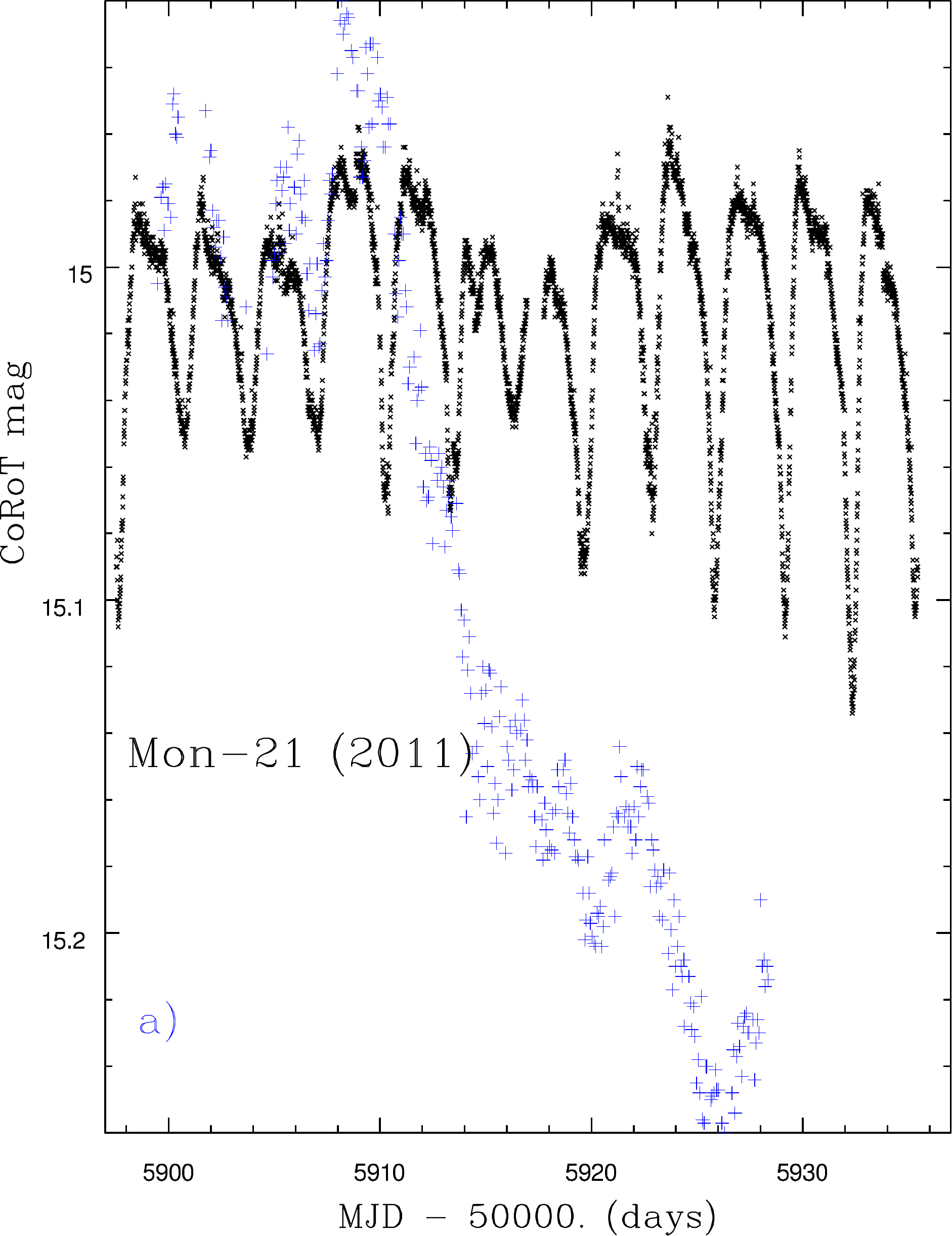}{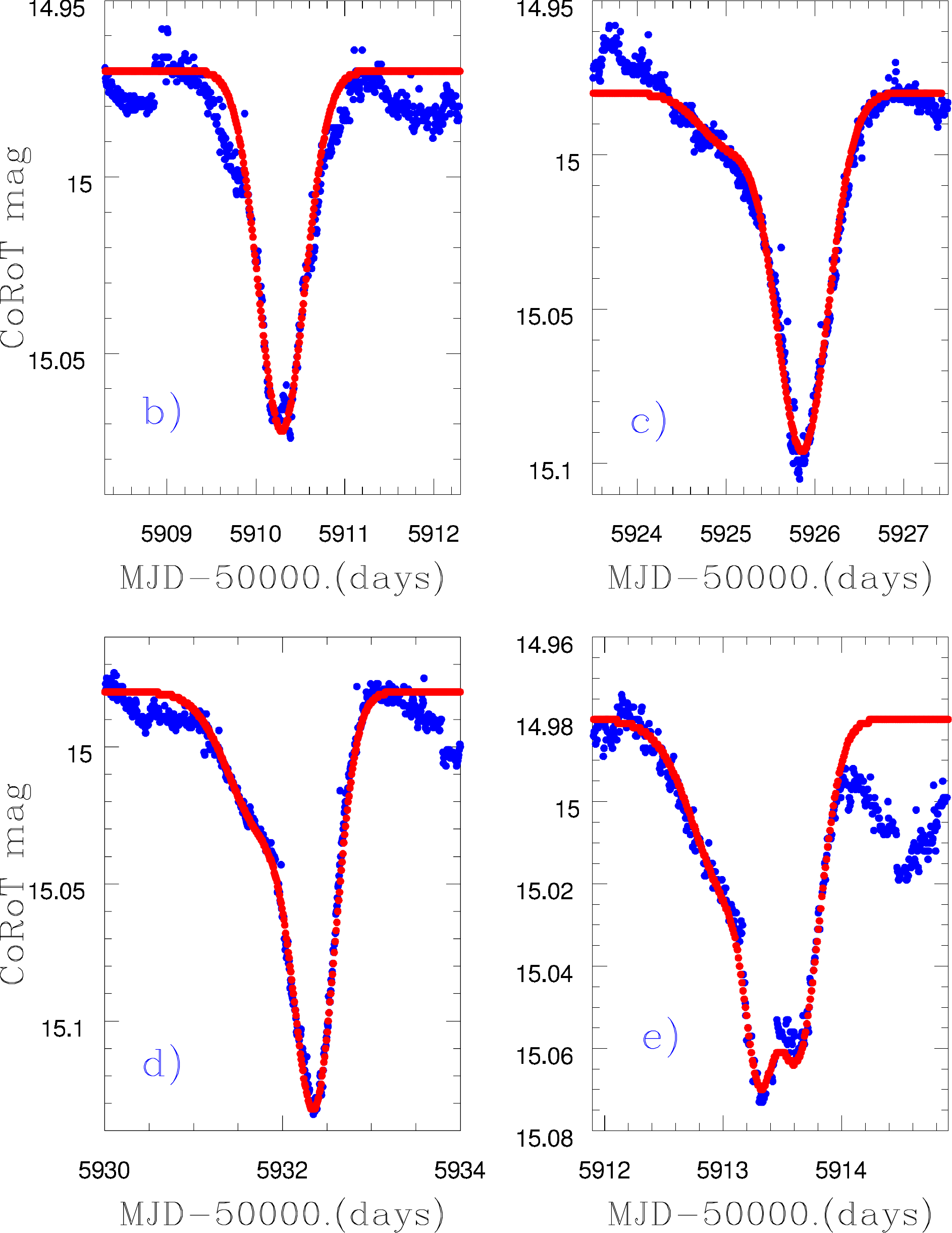}
\end{center}
\caption{a) {\em CoRoT} light curve for Mon-21 in 2011 (in black).  
The blue crosses show the 2011 IRAC light curve for Mon-21, shifted in
zero point to overlap the {\em CoRoT} data.  The two light curves
track each other fairly well at the start of the campaign and some of
the later {\em CoRoT} flux dips have possible IRAC counterparts, but
the 0.3 magnitude dimming at IRAC wavelengths beginning at MJD 55910
is not seen at optical wavelengths;  b) flux dip C (see Table~\ref{tab:quant_information})
which is
reasonably well fit by a single Gaussian; c) flux dip H, where
a shoulder to the flux dip requires the presence of a
second narrow flux dip; d) flux dip J, also requiring a two component
fit; and  e) flux dip D, which appears to be well fit by a three Gaussian
model. \label{fig:mon21_corot}}
\end{figure*}

By 2011, the {\em CoRoT} light curve for Mon-21 changed morphology, as
illustrated in Figure~\ref{fig:mon21_corot}a. The semi-sinusoidal
pattern is either essentially absent or much reduced in amplitude. The
flux dips appear somewhat broader and in many cases asymmetric.   One
possibility is that we are still seeing a blend of a spotted waveform
and narrow flux dips from variable extinction, with the spot being
responsible for the weaker feature causing the apparent asymmetry in
each observed waveform; another possibility is that spots no longer
contribute  in any significant way and the observed waveforms are due
to the blend of two narrow dips.   In order to document the full light
curve shape,  we have chosen to simply model the observed features
with Gaussian fits, usually with two components;  the fits  to a
number of the dips are shown in Figure~\ref{fig:mon21_corot}. On
average, the primary dip at each epoch has a Gaussian sigma width of
0.26$\pm$0.02 days, about 50\% broader than the dips in 2008.   The
weaker secondary dips  always precede the deeper dip.  See
Table~\ref{tab:quant_information} for the fit parameters. The average
period inferred from the dip centroids is 3.16$\pm$0.02 days, 
somewhat different from the dip period in 2008, but essentially equal
to the period for the sinusoidal waveform we attribute to spots on the
stellar surface.

\begin{figure*}
\begin{center}
\epsscale{1.0}
\plottwo{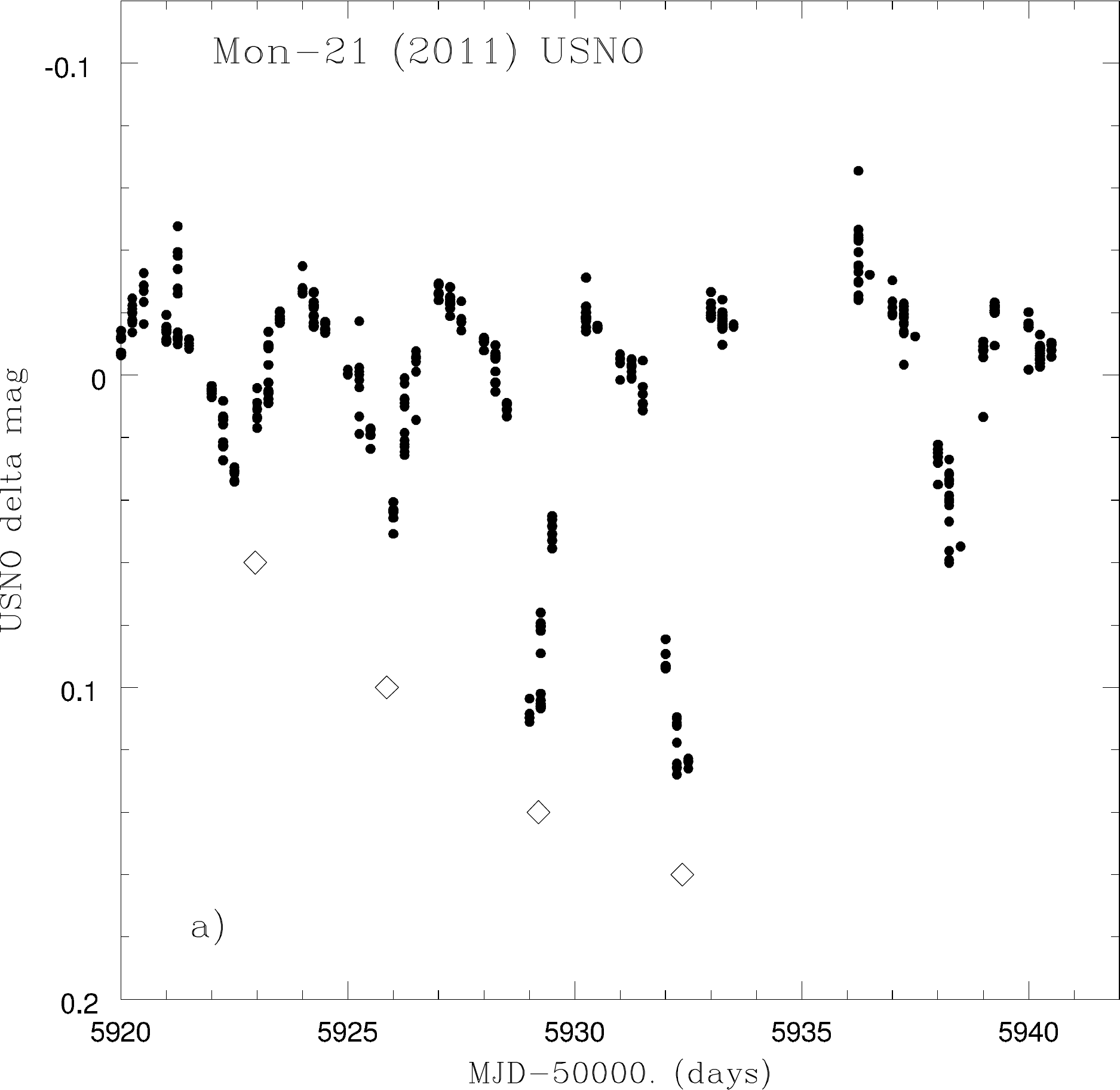}{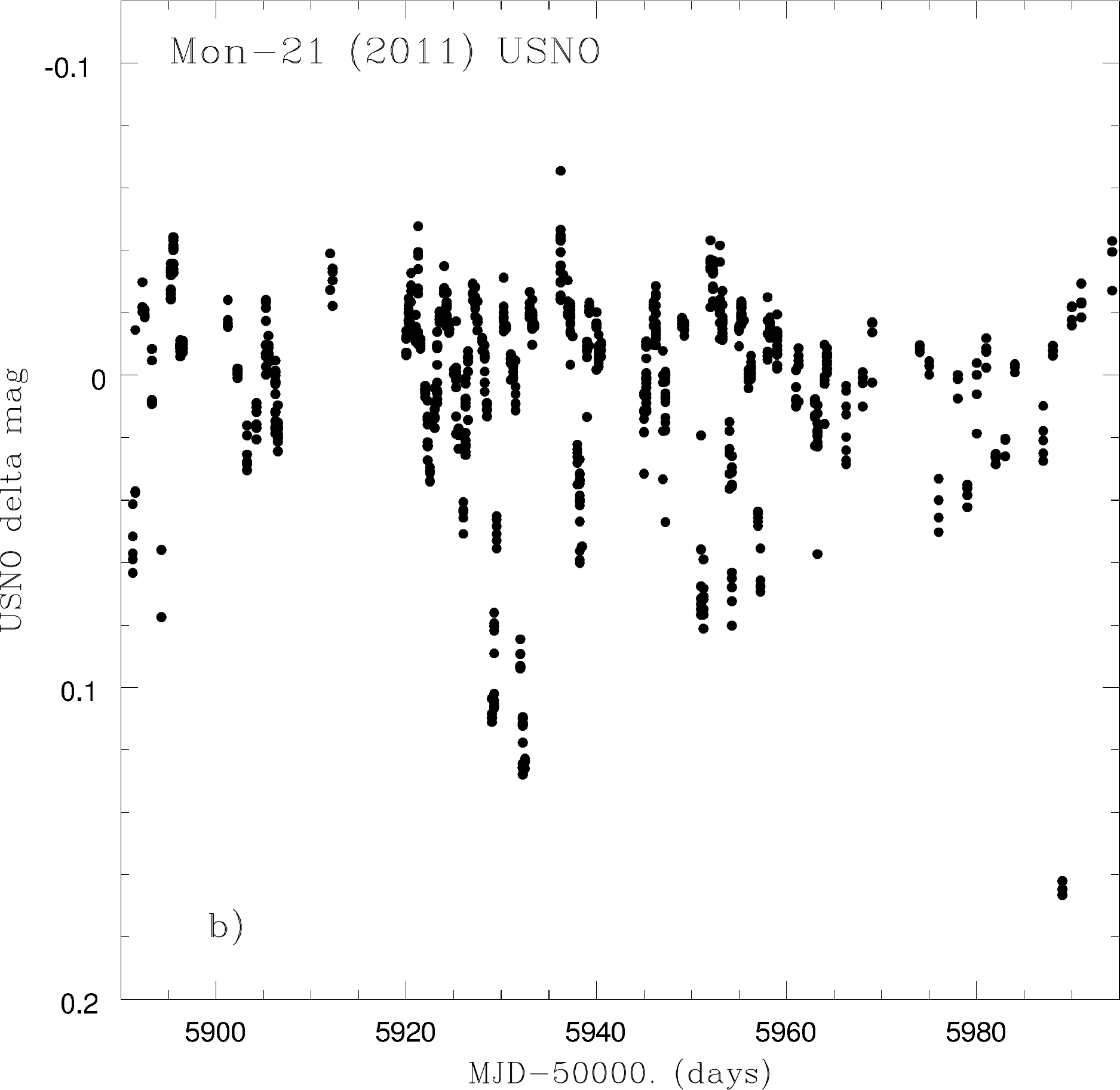}
\end{center}
\caption{a) A portion of the {\em USNO} $I$-band light curve for
Mon-21, illustrating generally good agreement between the USNO and
{\em CoRoT} data.  The diamond symbols are plotted at the times of the
deepest flux dips in the {\em CoRoT} data. b) The entire USNO $I$-band
light curve.  Additional flux dips appear to be present in the date
range 55950 to 55960.
\label{fig:mon21_usno}}
\end{figure*}

A ground-based USNO $I$-band light curve for Mon-21 is available that
overlaps in time with the {\em CoRoT} light curve.  For a star of
Mon-21's I magnitude, the expected RMS uncertainty in the relative
USNO photometry is about 0.02 mag.   Because the {\em
CoRoT} data have significantly lower noise and higher cadence, they
provide the best measure of the shape of the flux dips.   However, the
USNO data extend beyond the {\em CoRoT} campaign by nearly two months,
and therefore provide valuable information as to the persistence and
amplitude variability of the flux dips.  Figure~\ref{fig:mon21_usno}a
shows an expanded view of the region of best overlap between the {\em
CoRoT} and USNO data; Figure~\ref{fig:mon21_usno}b shows the entire
USNO light curve.  These data illustrate that the same flux dip
structure is detected in the ground-based data, and that it persists
for at least twenty days after the end of the {\em CoRoT}  campaign. 
Because the ground-observing only provided data for about 5-6 hours
per night, the window function for detecting the flux dips in the USNO
data is not very good and it is not possible to deduce whether similar
depth flux dips persisted beyond MJD 55960.

We also obtained a {\em Spitzer}/IRAC [3.6] light curve for Mon-21  in
2011.  As shown in Cody et al.\ (2014), for a star of Mon-21's brighness,
the expected RMS uncertainty in the relative IRAC photometry should be
of order 1\%.   The most obvious feature of the IRAC light curve is a gradual
flux decrease of 0.25 mag beginning around MJD 55910.  On small
timescales, there are possible correspondences between the {\em CoRoT}
flux dips and features in the IRAC data, but there are also places
where there is no correspondence. Given the shallowness of the dips in
the {\em CoRoT} data, and that the disk contributes equal or more
light at [3.6], the lack of distinct dips in the IR is not unexpected.

\subsection{Mon-56}

\begin{figure*}
\begin{center}
\epsscale{0.80}
\plottwo{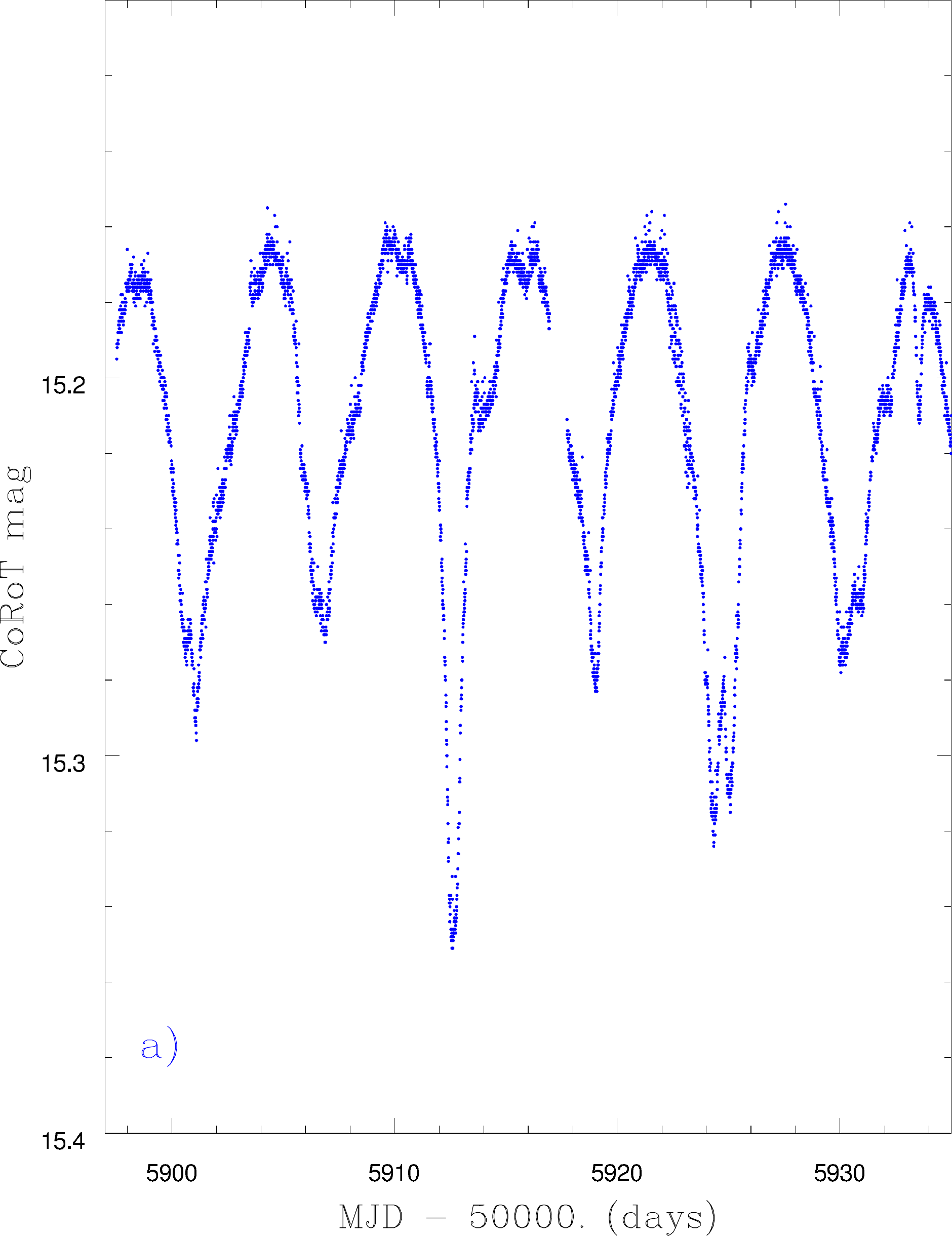}{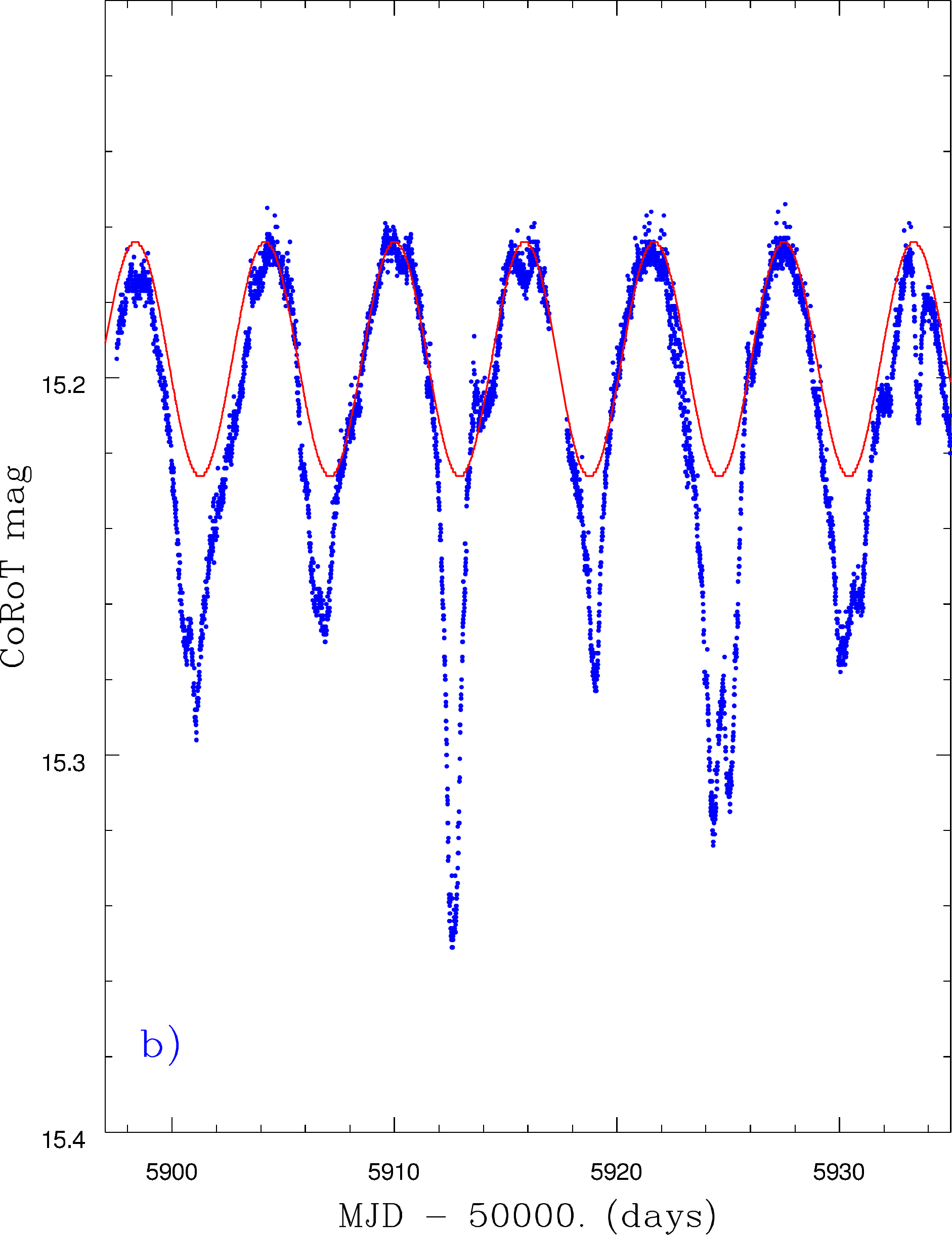}
\vspace{-0.cm}
\plottwo{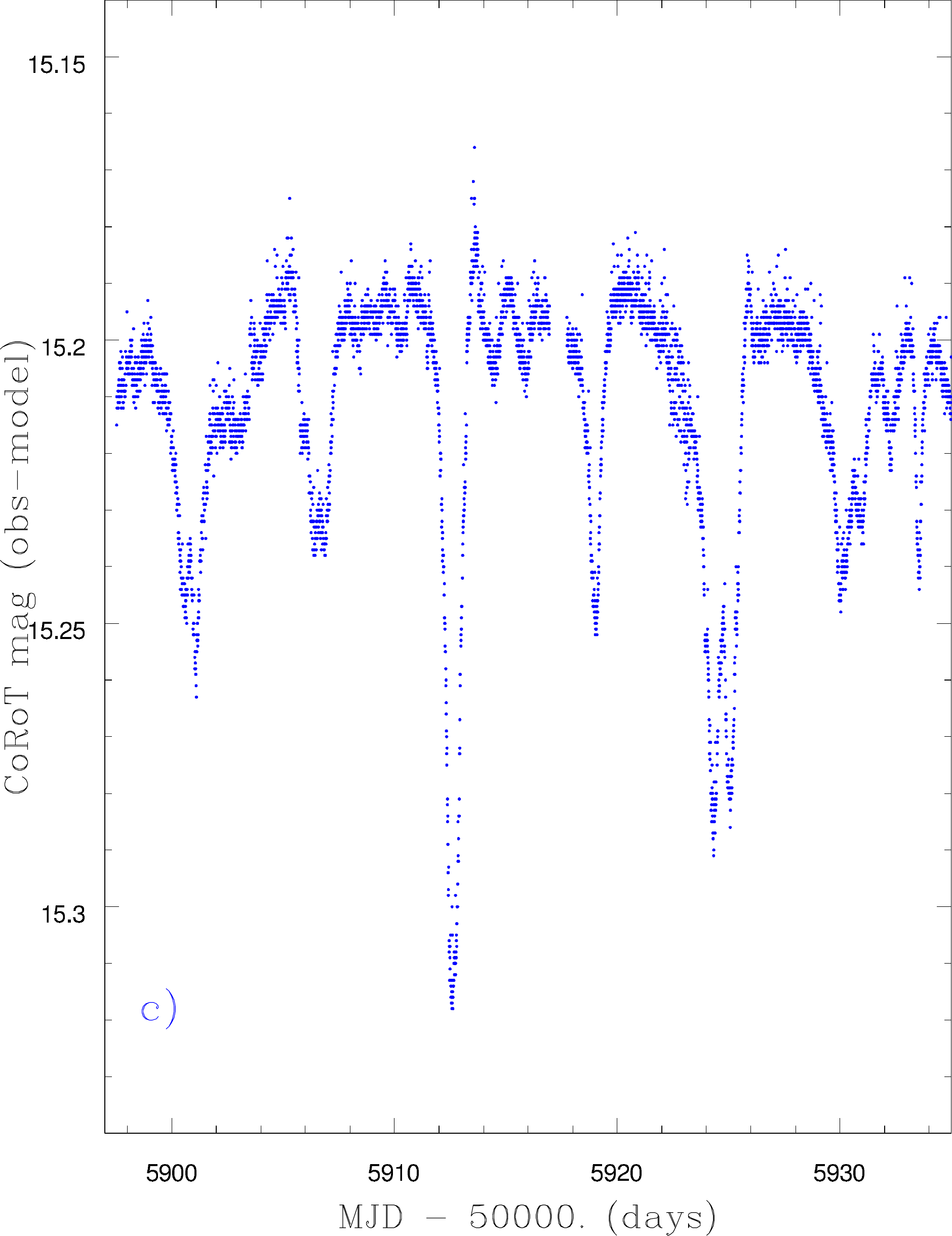}{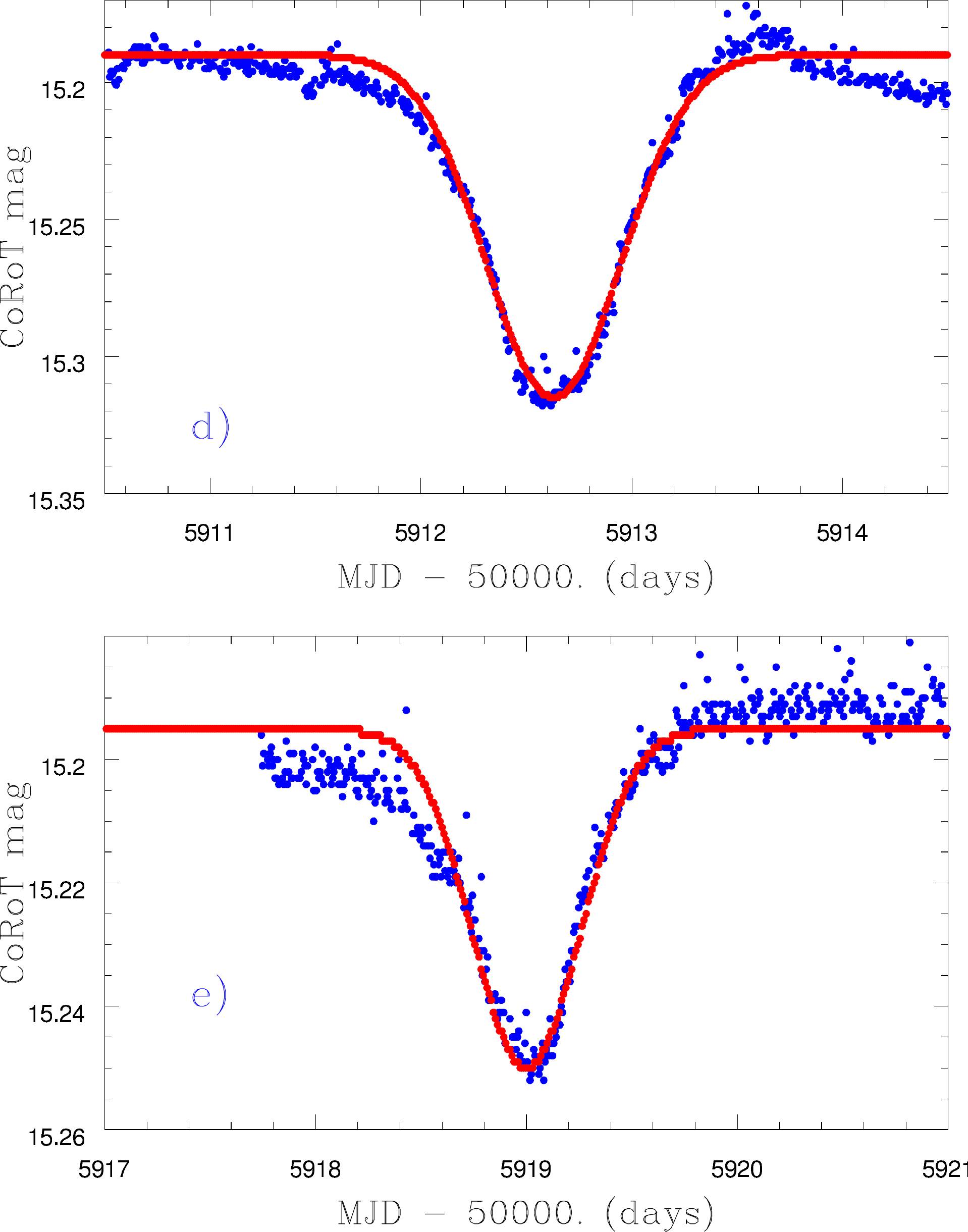}
\end{center}
\caption{a) {\em CoRoT} light curve for Mon-56 in 2011; b) {\em CoRoT}
light curve  for Mon-56, with a model light curve designed to
approximate the light curve shape attributable to spots on the star's
photosphere overplotted in red; c) Residual light curve for Mon-56,
produced by subtracting the spotted star model from the observed light
curve; d and e) {\em CoRoT} light curves for  two of the prominent flux dips
in Mon-56 (dips A and B in Table~\ref{tab:quant_information}).  The adopted
perod for both the spotted-star waveform and the flux dips is about 5.8 days.
\label{fig:mon56_corot}}
\end{figure*}

In many respects, Mon-56 is a near twin to Mon-21.  It also has a
published spectral type of K5 (Dahm \& Simon 2005), H$\alpha$\
emission with an inverse P-Cygni profile but quite small equivalent
width (Furesz et al.\ 2006), essentially no UV excess and IR colors
placing it near the center of the Class~II population in the IRAC
color-color diagram (Allen et al.\ 2004). Similar to Mon-21,  we
interpret its {\em CoRoT} light curve in 2011 as also being the
superposition of two components with essentially identical periods -
a smoothly varying, semi-sinusoidal component, possibly due to
star-spots and a system of periodic, shallow, short duration flux dips which are
sometimes single and sometimes double.  Figure~\ref{fig:mon56_corot}a
shows the as-observed {\em CoRoT} light curve.

Compared to Mon-21, however, the Mon-56 light curve is more difficult
to deconvolve because the flux dips are generally stronger and broader
and in phase with the flux minima of the sinusoidal waveform, thereby
making it harder to construct a high quality model for the spotted
light curve (or to be certain that this component is indeed due to
spots as opposed to a warped-disk model).   We have therefore resorted
to a lower fidelity but simpler model for the smoothly varying
component of the light  curve -- specifically, we adopt a simple
sine-wave, whose phase, amplitude and period we set to yield the
best fit (i.e. to yield a residual light curve with a nearly flat
continuum level interspersed with narrow dips).
Figure~\ref{fig:mon56_corot}b shows our final model for the smoothly
varying component of the light curve superposed on the observed Mon-56
light curve, and Figure~\ref{fig:mon56_corot}c shows the result of
subtracting this model from the observations, yielding our best
visualization of the short-duration flux dips in this star.   The full
amplitude of the adopted sinusoid  is 0.06 mag (about 6\%), and the
period is 5.83 days (the same as determined from ground-based
monitoring by Lamm et al.\ 2004).   Because of the approximate nature
of the deconvolution, we attempt to derive flux dip fits only for the
three strongest features in Figure~\ref{fig:mon56_corot}c. Those
parameters are provided in Table~\ref{tab:quant_information} and the
fits are illustrated for the apparently single flux dips in
Figure~\ref{fig:mon56_corot}d and e.

We note that the Mon-56 {\em CoRoT} data are interpreted differently
in McGinnis et al.\ (2015). McGinnis et al.\ attribute all of the
variability seen in the  system to variable extinction, as opposed to
a superposition of spot  and extinction variability.  Regardless of
these differences, both models agree that the narrow, periodic flux
dips are best attributed to extinction  and we therefore believe
Mon-56 fits within the class of variability discussed in this paper.

\subsection{Mon-1165}

Mon-1165 has spectral type M3 (see appendix).  In an IRAC color-color
diagram, it falls within the box defining the locus for Class~II YSOs
(Allen et al.\ 2004).   However, as for the previous two YSOs, it has
essentially no UV excess, and only a weak H$\alpha$\ emission line.

\begin{figure*}
\begin{center}
\epsscale{0.80}
\plottwo{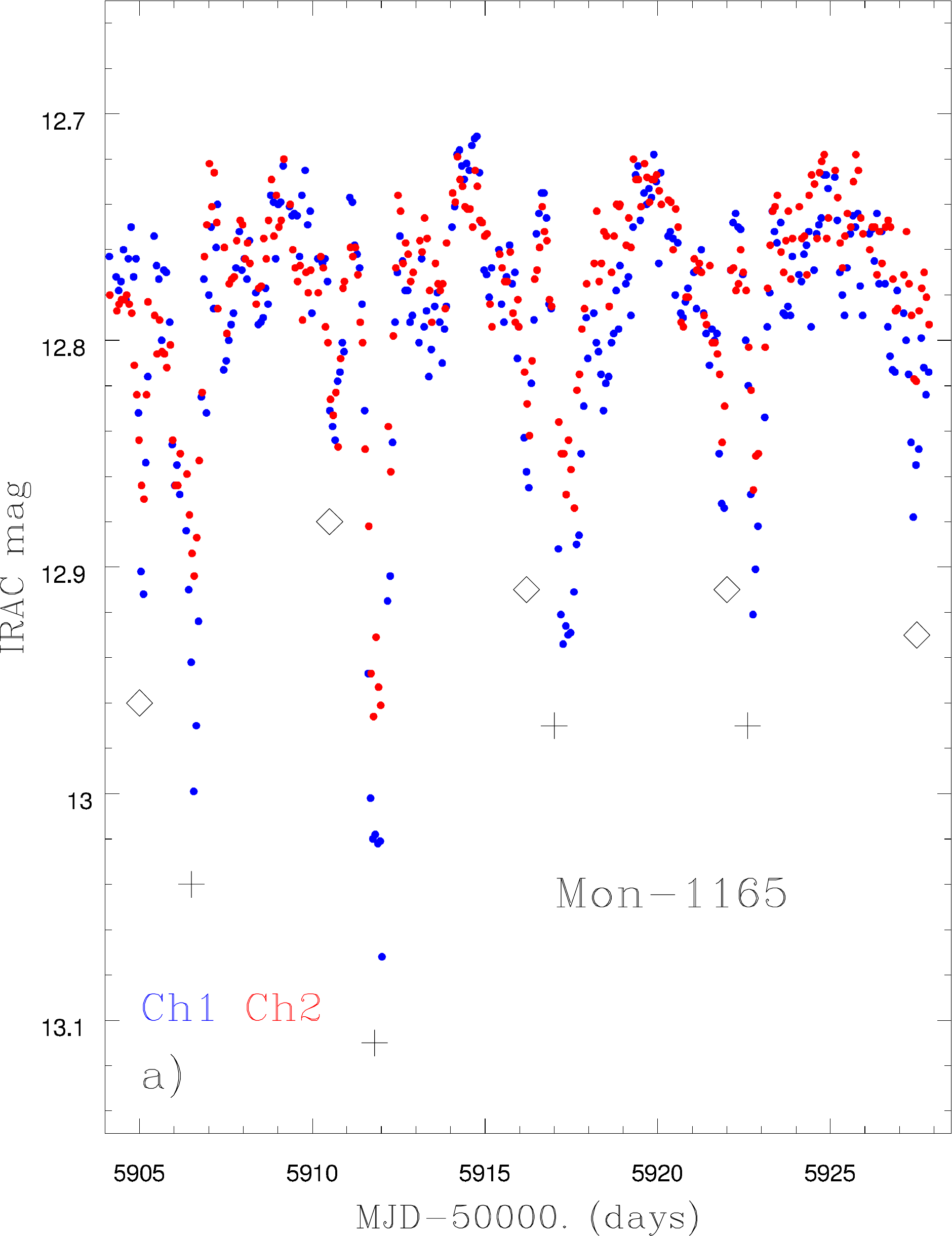}{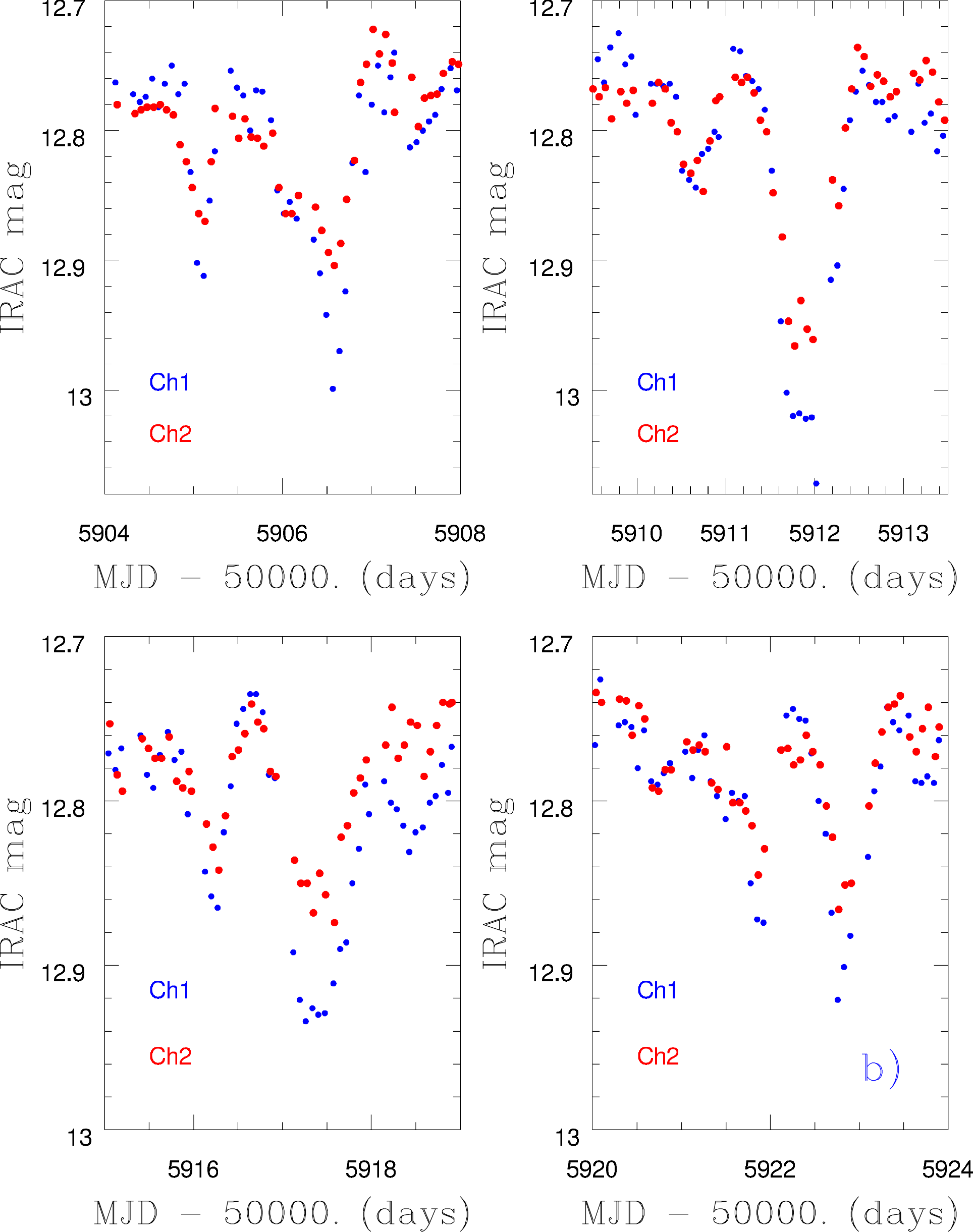}
\vspace{-0.cm}
\plottwo{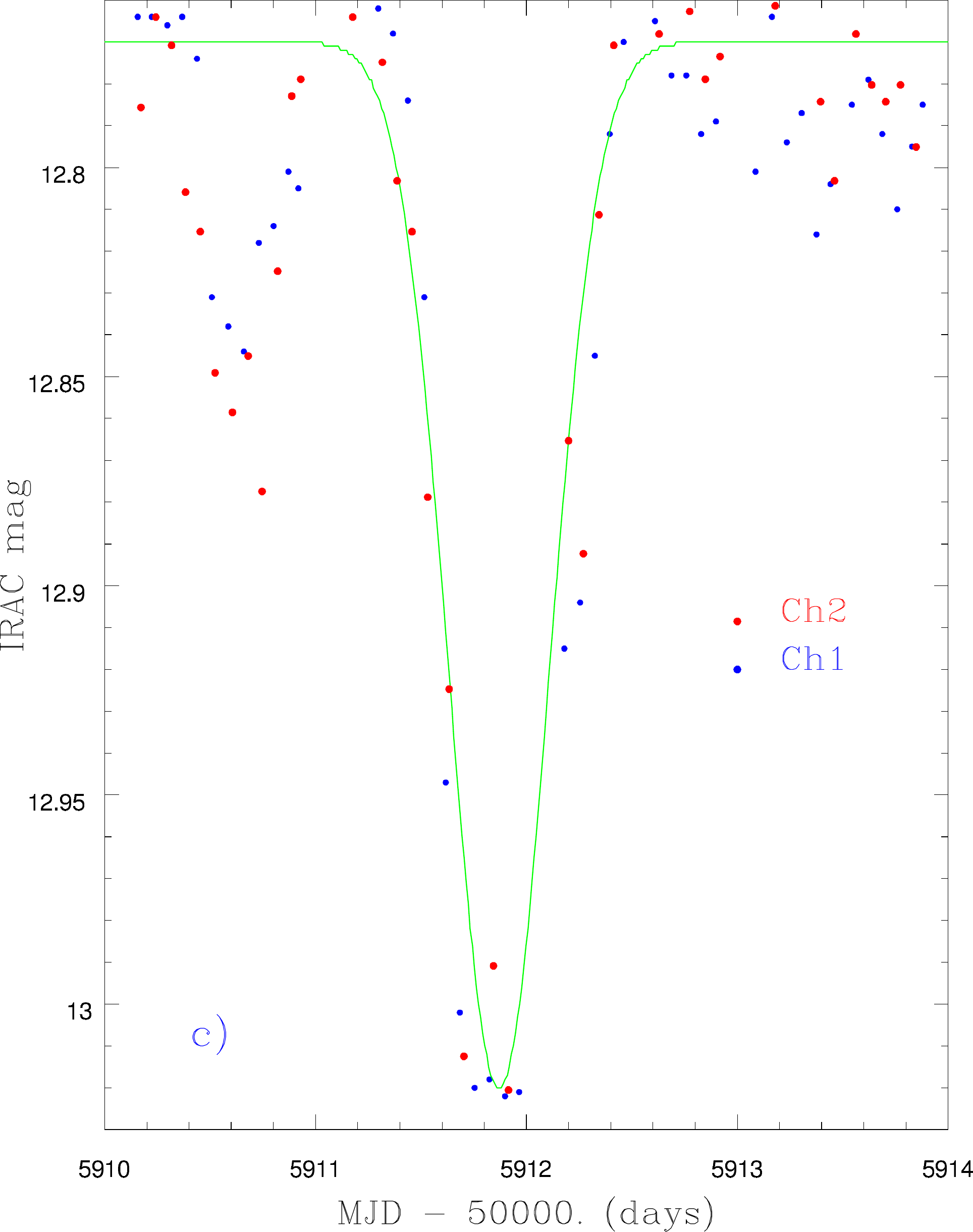}{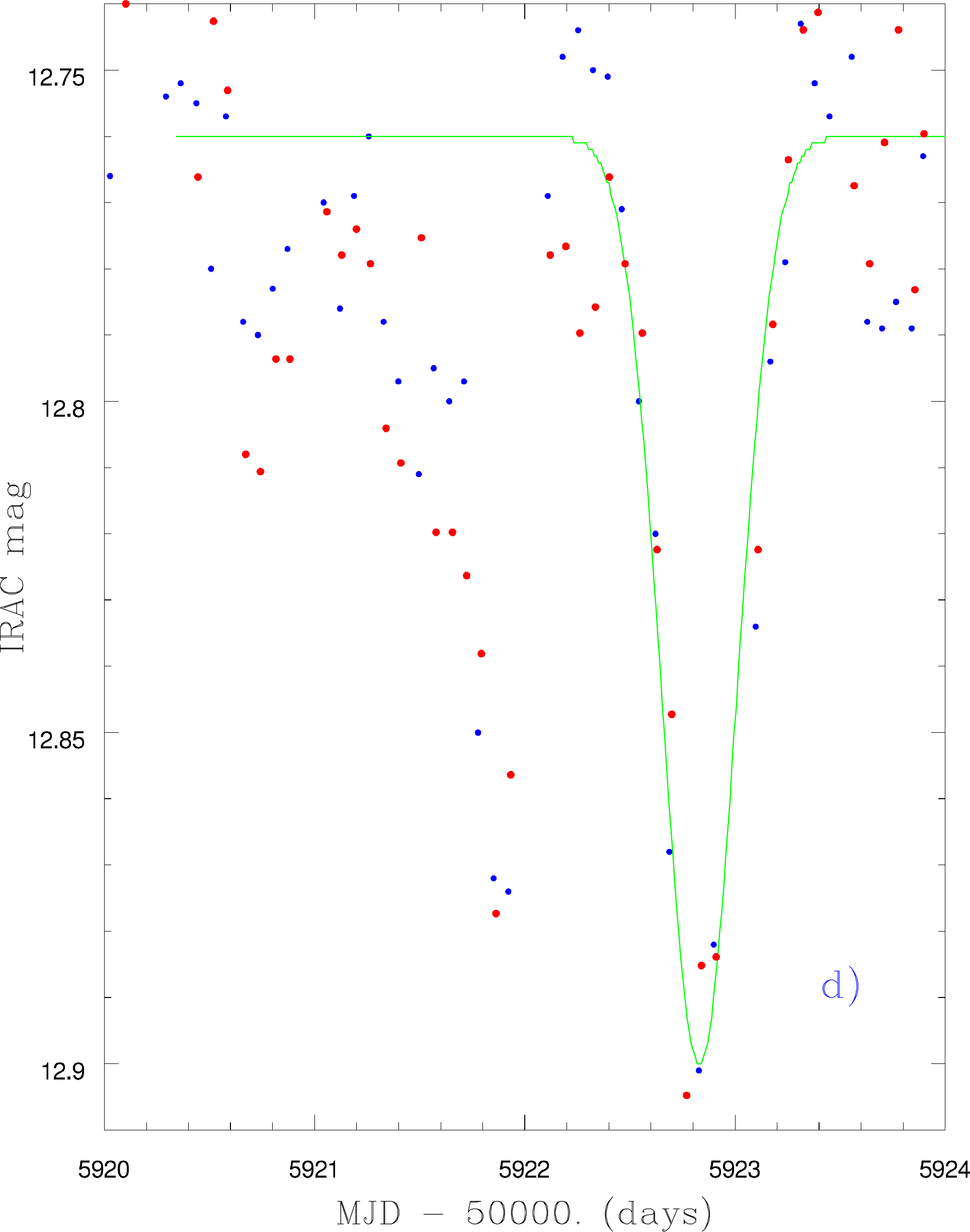}
\end{center}
\caption{a) IRAC light curve for Mon-1165; 0.3 mag has been added to
the 3.6 $\mu$m (Ch. 1) photometry in order for the 3.6 and 
4.5 $\mu$m (Ch. 2) light
curves to approximately align vertically.   The plus signs and
diamonds mark the two components of each dip pair; b) Cutouts from the
IRAC  light curve -- with the 3.6 and 4.5 micron points shifted
vertically to align with each other -- at the times of the four
well-sampled flux dip pairs; c and d) Illustration that  the narrow
flux dips for Mon-1165 are reasonably Gaussian in shape.   In this
case, the $y$-axis for [4.5] has been both shifted and stretched, so
that the dip has the same apparent amplitude in both channels.
\label{fig:mon1165_irac}}
\end{figure*}

We do not have a {\em CoRoT} light curve for Mon-1165, but the IRAC
light curve shows distinct, narrow, periodic, short-duration, paired
flux dips (see Figure~\ref{fig:mon1165_irac}a).   An auto-correlation
analysis of the light curve yields a period for the flux dips of 5.5
days.  The lower amplitude dip always  precedes the main dip by about
1.2 days (see Figure~\ref{fig:mon1165_irac}b).   The flux dips are
generally deeper at 3.6 $\mu$m than at 4.5 $\mu$m by, on average,
about 50\% but this is poorly determined due to the shallowness and
narrowness  of the dips and the relatively low cadence of the
observations. Only two of the dips are well-enough defined to
accurately measure their shape and width -- those fits are shown in
Figure~\ref{fig:mon1165_irac}c and d;  the Gaussian ${\sigma}$ values
for the two fits are 0.15 and 0.20 days.

\begin{figure*}
\begin{center}
\epsscale{1.0}
\plottwo{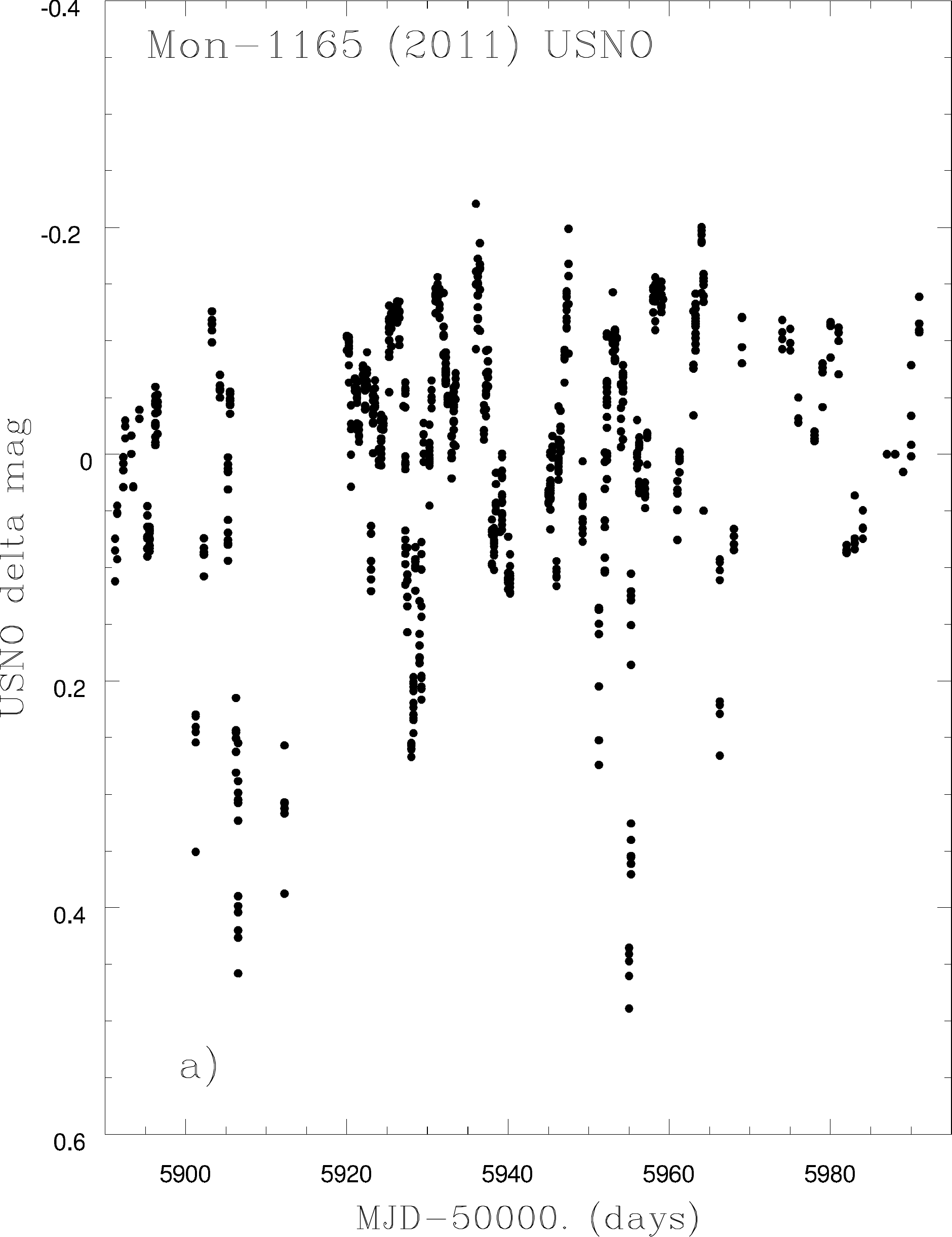}{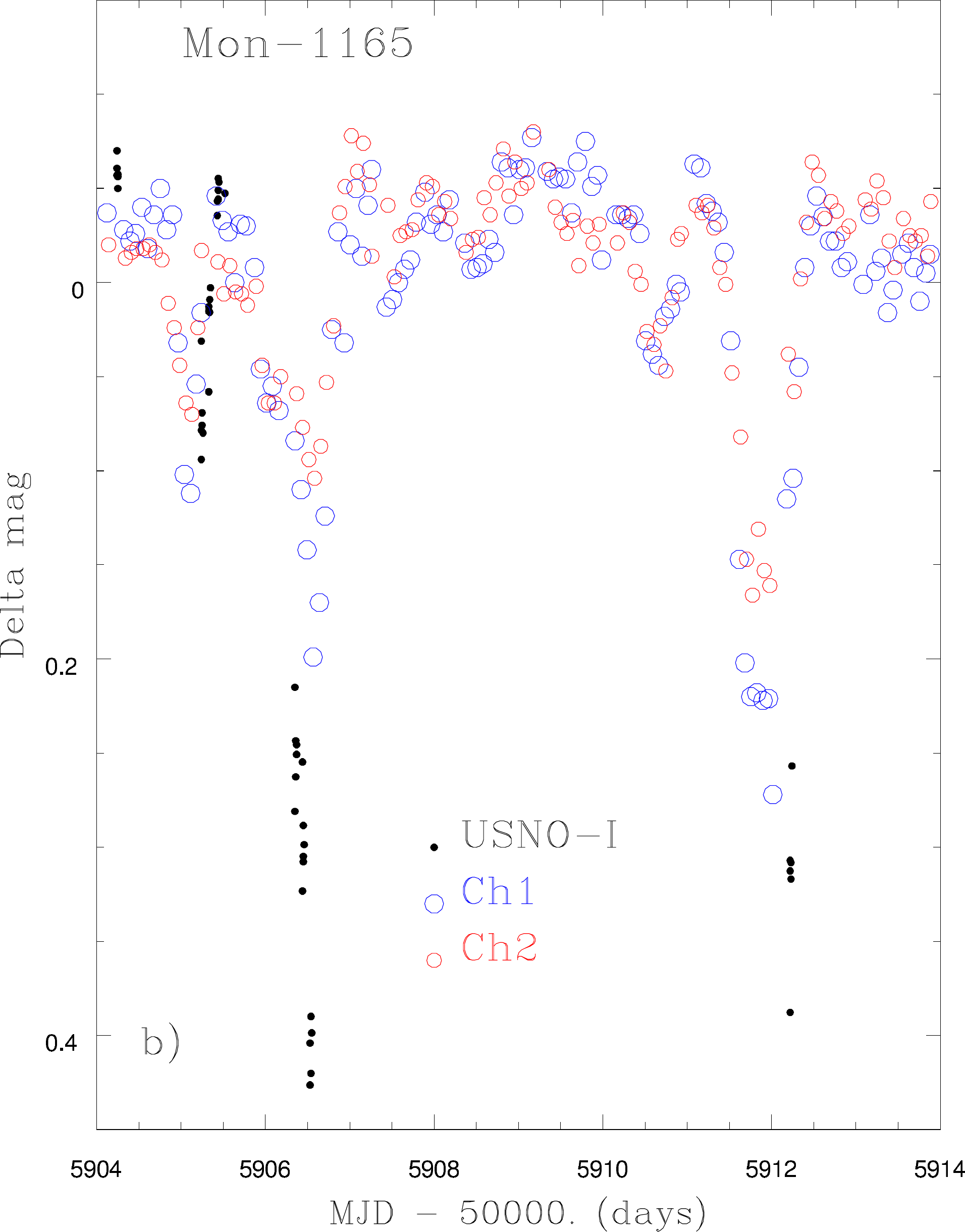}
\end{center}
\caption{a) {\em USNO} $I$-band light curve for Mon-1165; flux dips
with depths up to about 0.4 mag appear to be present.  b) Overlay of
the IRAC and USNO light curve for Mon-1165, with relatively good data
in all three bands for the flux dip near MJD 55907.
\label{fig:mon1165_usno}}
\end{figure*}

We have ground-based, optical, synoptic photometry which only overlap
with a small portion of the {\em Spitzer} data due to bad weather in
Arizona in December 2011.  Figure~\ref{fig:mon1165_usno}a shows the
entire USNO light curve, illustrating that at least a few flux dips
are present beyond the end of the {\em CoRoT} campaign.  
Figure~\ref{fig:mon1165_usno}b overplots the IRAC and USNO data for
the time window where we have some optical data that falls at or near
the center of one of the strong IR flux dips.  The ratio of the dip
depth at 4.5 $\mu$m to the depth in the I band --  $A_{4.5}/A_{I}$ --
is about 0.3.  In order to use this ratio to help constrain the
occultation geometry, we must make a correction for the fraction of
the IR light that comes from the disk (assuming the disk flux is not
significantly occulted).  Based on a fit to the star's SED, our
estimate of  f$_{4.5}$(disk)/f$_{4.5}$(star) is 1.3; hence the
corrected dip depth ratio becomes $A_{4.5}/A_{I} \sim$ 0.7.   Given
the uncertainties, this is consistent with equal dip depths at both
wavelengths, as one would expect from dust structures that are
optically thick but cover only a small part of the star.  If instead
the dust structures are optically thin, this ratio would indicate an
extreme deficit of small particles compared to the standard ISM
mixture where the ratio should be $\sim$0.10 (Indebetouw et al.\
2005).

\subsection{Mon-6975}

Mon-6975 is another essentially anonymous star that had not even been
considered an NGC~2264 member until now.   Our attention was drawn to
this star by its 2011 {\em CoRoT} light curve (it was not observed
by {\em CoRoT} in 2008).   We have subsequently obtained optical
and IR spectra of it from SOAR (see Appendix) which indicate a
spectral  type of M2.5 and moderately strong H$\alpha$ emission
(H$\alpha$ equivalent width = 15 \AA).  Good seeing images show it to
be a nearly equal brightness optical binary with a separation of order
2.4$\arcsec$ oriented NE-SW on the sky;  the NE component has a nearly
featureless spectrum that is much redder than that of the M2.5 star. 
Despite the very red color, the NE component  has H$\alpha$ weakly in
absorption, and no other spectral features that are detected with
certainty.  We therefore assume it is not a YSO and is only a
line-of-sight companion to the SW component. Both objects contribute
to the {\em CoRoT} light curve, though we assume that the flux dips
arise only from the young star.  We do not have a measure of its UV
excess, nor a complete set of IRAC photometry in order to classify
its disk.

\begin{figure*}
\begin{center}
\epsscale{1.0}
\plottwo{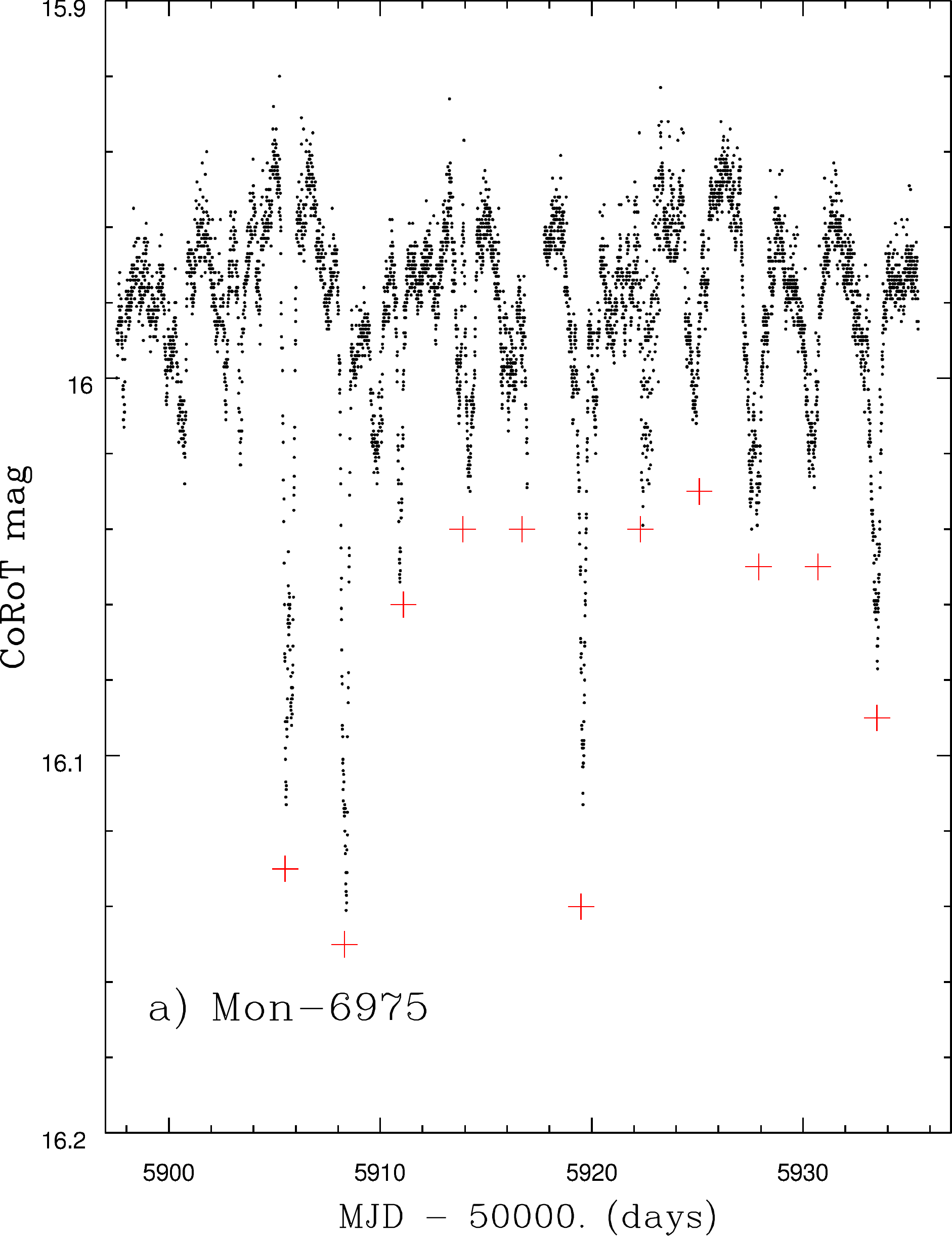}{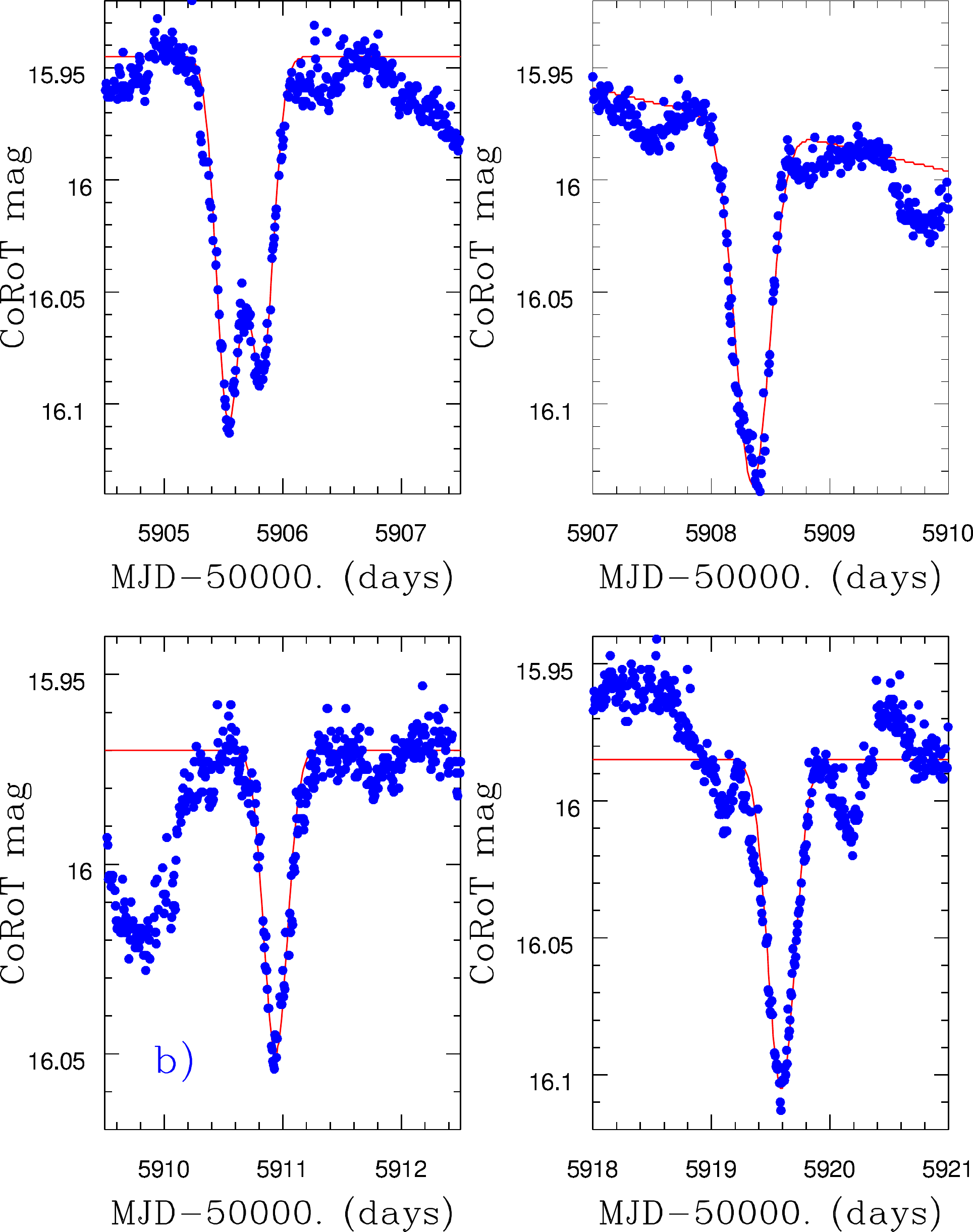}
\end{center}
\caption{a) {\em CoRoT} light curve for Mon-6975.  The plus sign under
the flux dip at MJD 55906 marks the centroid of that dip.  All of the
other plus signs are simply placed at exactly 2.8 day intervals with
respect to the first one; b) Gaussian fits to four of the narrow flux
dips for Mon-6975 marked in the previous figure.   All are well
approximated by either a single Gaussian shape or the blend of two
Gaussians.
\label{fig:mon6975_corot}}
\end{figure*}

The {\em CoRoT} light curve for Mon-6975 shows a nearly constant
``continuum" level and a large number of narrow flux dips with a
relatively wide range in both depth and width
(Figure~\ref{fig:mon6975_corot}a).   The periodicity of the flux dips
for this star is not as obvious as for Mon-21 and Mon-1165, but an
auto-correlation function yields a period of 2.8 days.  To guide the
eye, we have placed a cross below the first strong dip in the light
curve, and then successive crosses spaced 2.8 days apart for the
remainder of the light curve.   Some of the flux dips for Mon-6975 are
very narrow, with Gaussian $\sigma$ values ranging from 0.095 to 0.13
days. Figure~\ref{fig:mon6975_corot}b shows several of these dips, and
the Gaussian fits we have derived.

\begin{figure*}
\begin{center}
\epsfxsize=.99\columnwidth
\epsfbox{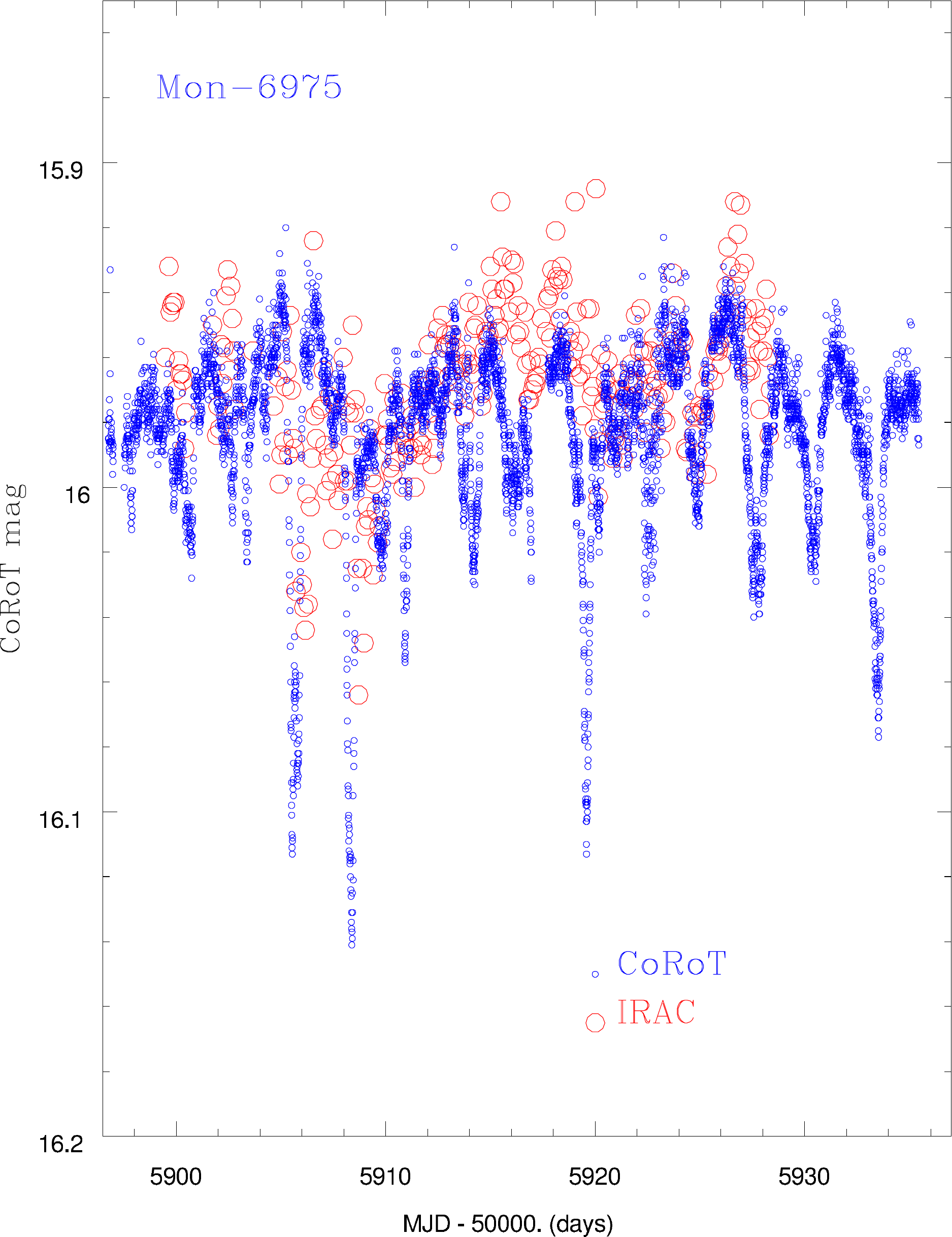}
\end{center}
\caption{Overlay of the {\em CoRoT} and IRAC light curves for
Mon-6975; a zero-point shift has been added to the IRAC magnitudes so
they approximately align with the {\em CoRoT} data. The two strongest
{\em CoRoT} flux dips have counterparts in the IRAC light curve, and
there is moderate correlation in the structure of the rest of the two
light curves.
\label{fig:mon6975_irac_corot}}
\end{figure*}

We have an IRAC 3.6 $\mu$m\ light curve for Mon-6975 -- it is overlaid on the
{\em CoRoT} light curve in Figure~\ref{fig:mon6975_irac_corot}.   The
IRAC light curve is similar in character to the {\em CoRoT} light
curve in the sense that it shows a relatively flat continuum level and
two probable flux dips at the location of two of the strongest flux
dips in the {\em CoRoT} light curve.   The IRAC light curve is,
however, much noisier, making an accurate assessment of the depth of
the two flux dips at IRAC wavelengths difficult. However, it is clear
that the derived ratio, after correction for the  contribution to the
IRAC flux from the disk, would be greater than 0.5 and possibly
consistent with 1.0.   Therefore, as for Mon 1165, the dip depth ratio
is either consistent with optically thick occulting structures or with
optically thin ``clouds" whose dust mixture is extremely depleted in
small grains.

\subsection{Mon-1131}

Mon-1131, also known as QT Mon, 
has {\em CoRoT} data from both 2008 and 2011.  The 2008 light
curve for this star is very different from the 2011 light curve, and
is probably dominated by variable accretion -- this light curve is
shown and discussed in the Appendix.  The 2011 light curve
(Figure~\ref{fig:mon1131_corot}a)  fits within the short-duration flux
dip class, although compared to the other prominent members of the
class, its flux dips are relatively broad.  The dip period is 5.08
days. The two narrowest dips have Gaussian $\sigma$ values of 0.27 and
0.30 days -- their profiles are shown in 
Figure~\ref{fig:mon1131_corot}b.    The IRAC light curve for Mon-1131
is essentially uncorrelated with the {\em CoRoT} light curve.   We
also obtained a SOAR optical spectrum of Mon-1131.   Based on measures
of the depths of the TiO bands in that spectrum, we estimate its
spectral type at M1.5. The Furusz et al.\ (2006) Multiple-Mirror
Telescope (MMT)  high resolution spectrum for Mon-1131  shows it to
have a broad and structured H$\alpha$\ profile.    Both its H$\alpha$\
equivalent width ($>$30\AA) and its UV-excess
(Figure~\ref{fig:colorcolor_IR}) identify it as the most strongly
accreting star amongst the short-duration flux dip group.

\begin{figure*}
\begin{center}
\epsscale{1.0}
\plottwo{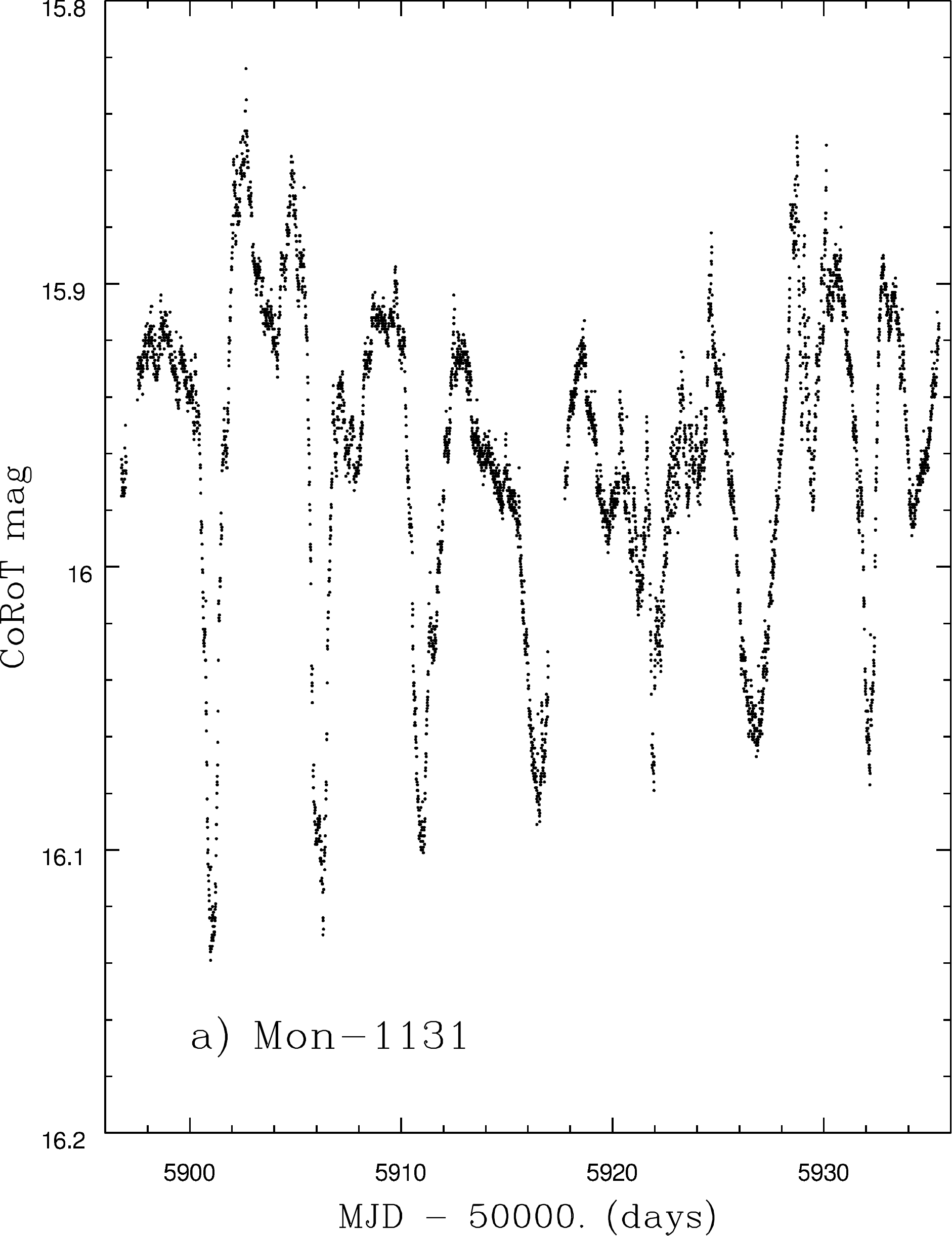}{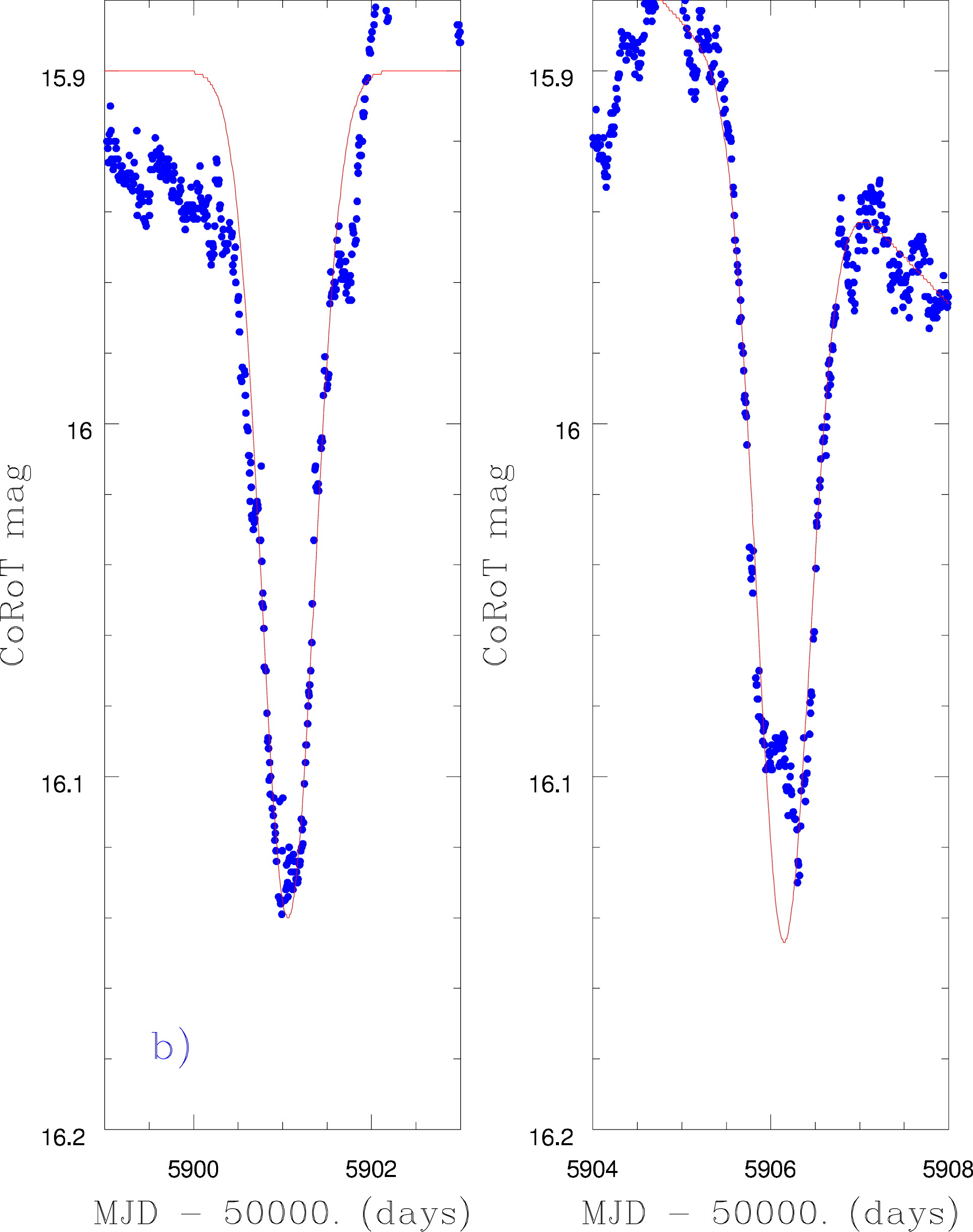}
\end{center}
\caption{a) {\em CoRoT} light curve for Mon-1131; b) Illustration that
some  of the flux dips for Mon-1131 are reasonably Gaussian in shape. 
\label{fig:mon1131_corot}}
\end{figure*}

\subsection{Mon-1580}

\begin{figure*}
\begin{center}
\epsscale{1.0}
\plottwo{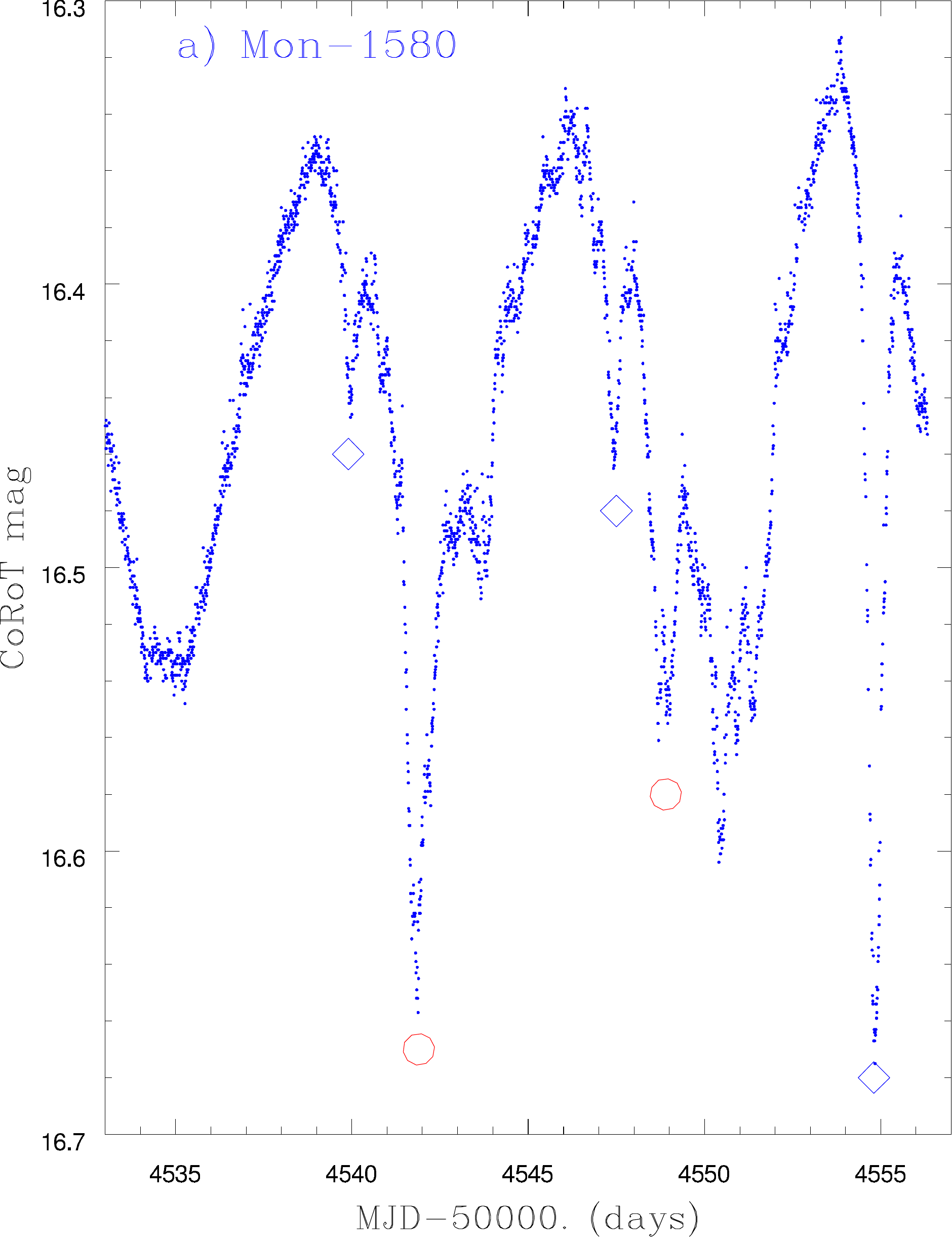}{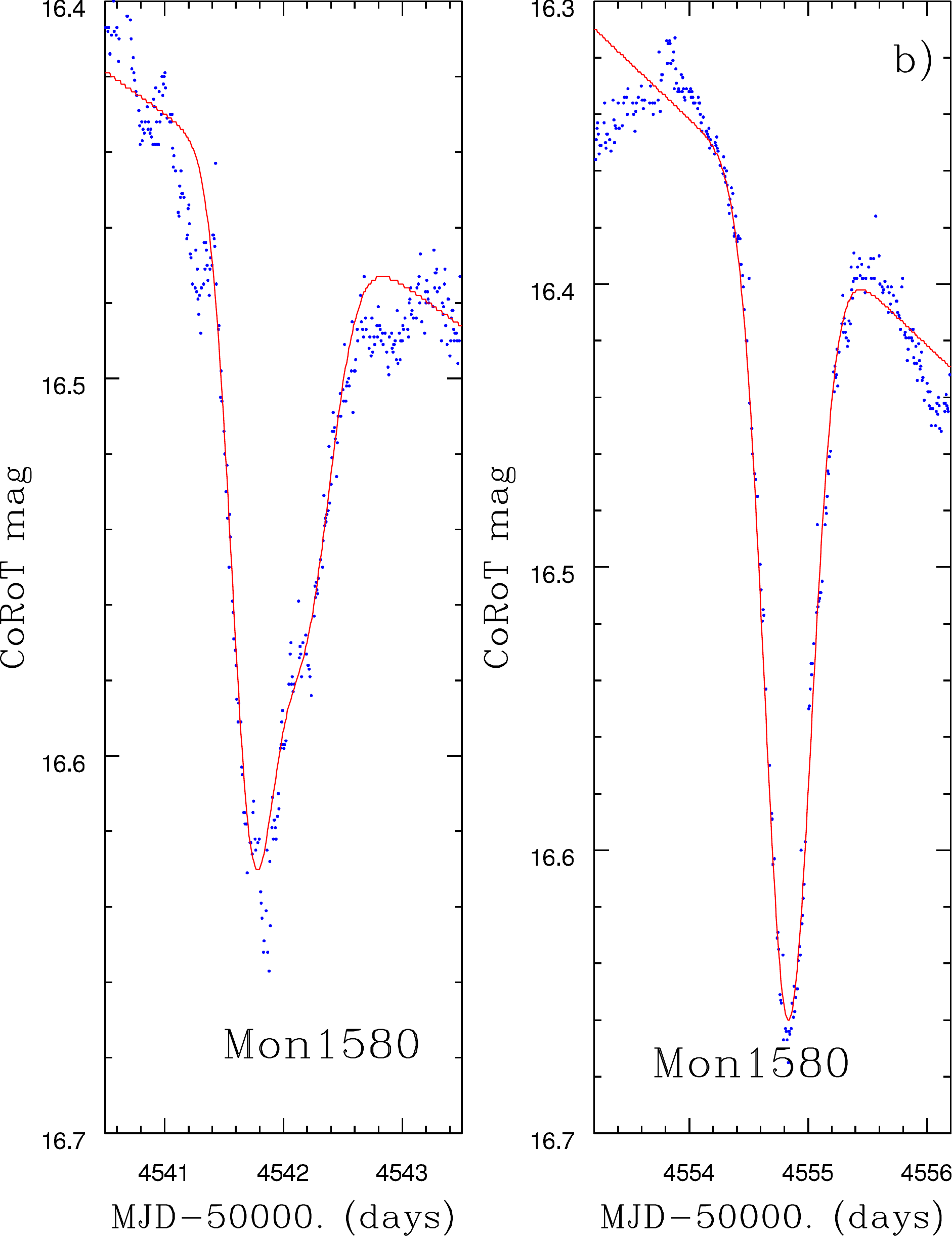}
\end{center}
\caption{a) {\em CoRoT} light curve for Mon-1580. This is another star where
there appears to be a sine-wave-like, spotted-star light curve,
superposed on which are narrow flux dips which recur for at least two
or three epochs.   Two such recurring dips are marked with diamonds and
circles in the plot; b) Illustration that some of the narrow flux dips 
for Mon-1580 can be well fit by single Gaussians or blends of 
two Gaussians (dips A and C in Table~\ref{tab:quant_information}).  Though
not well-determined due to the relative shortness of the observing campaign,
the period for the semi-sinusoidal (spotted-star) waveform and the average
interval between dip systems (blue diamonds or red circles) is about 7.1 days.
\label{fig:mon1580}}
\end{figure*}

Mon-1580 is another relatively little-studied member of NGC~2264. 
There is no published spectral type, however it is a known H$\alpha$\
emission line YSO (Reipurth et al.\ 1996; Furesz et al.\ 2006).  Based
on the Furesz et al.\ spectrum, we estimate the spectral type as M1.  
We have a 2008 {\em CoRoT} light curve, but no corresponding light
curve from the 2011 campaign.  The 2008 light curve
(Figure~\ref{fig:mon1580}a) is reminiscent of the 2008  Mon-21 light
curve, in the sense that it appears to be the superposition of a
spotted star light curve and a number of short duration flux dips. 
The spot light curve period and the flux dip period are the same ($P
\sim$7.1 days) to within our ability to measure the periods.  
Figure~\ref{fig:mon1580}b shows the Gaussian fits to the two strongest
flux dips.

\subsection{Mon-378}

\begin{figure*}
\begin{center}
\epsscale{1.0}
\plottwo{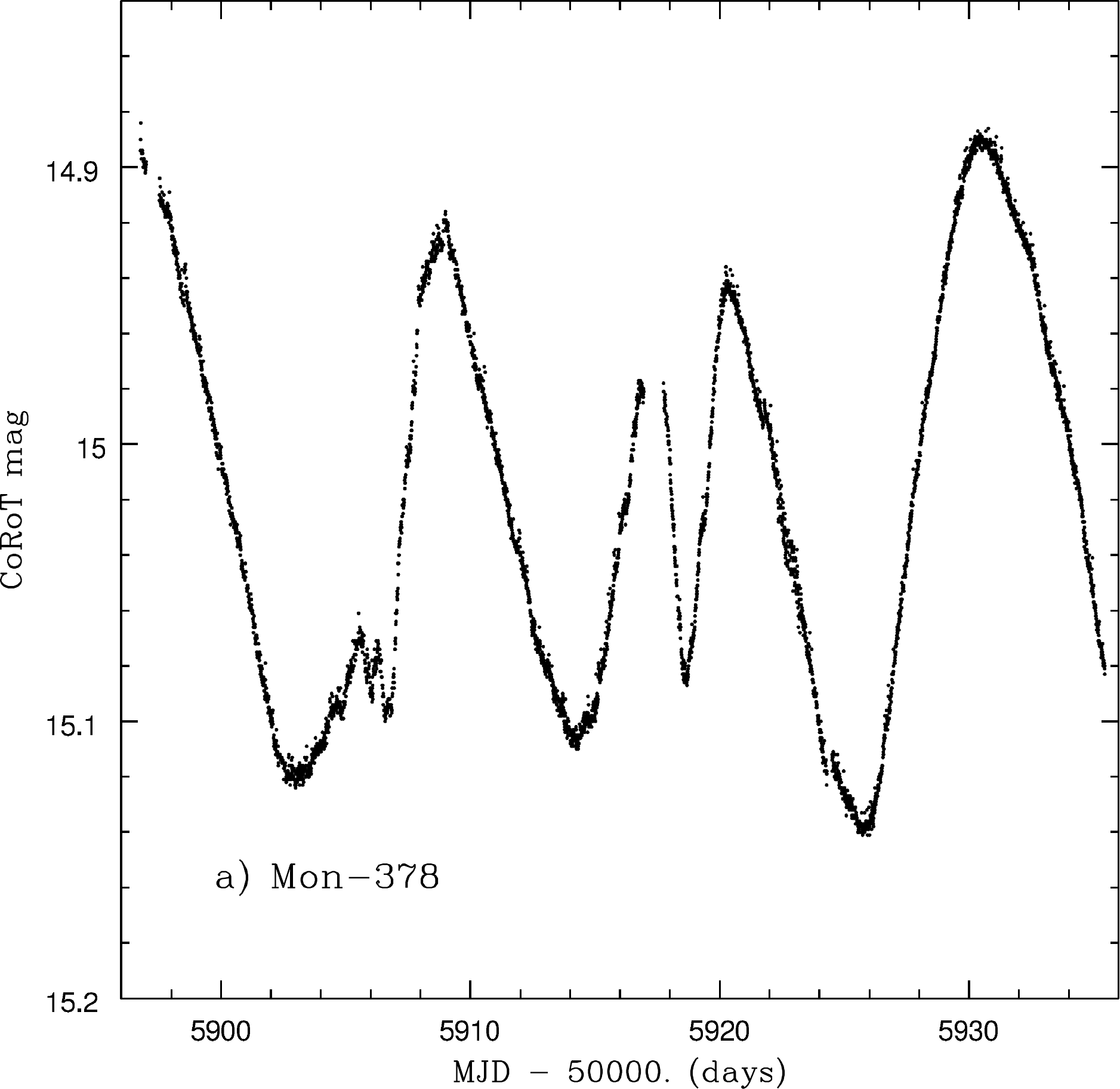}{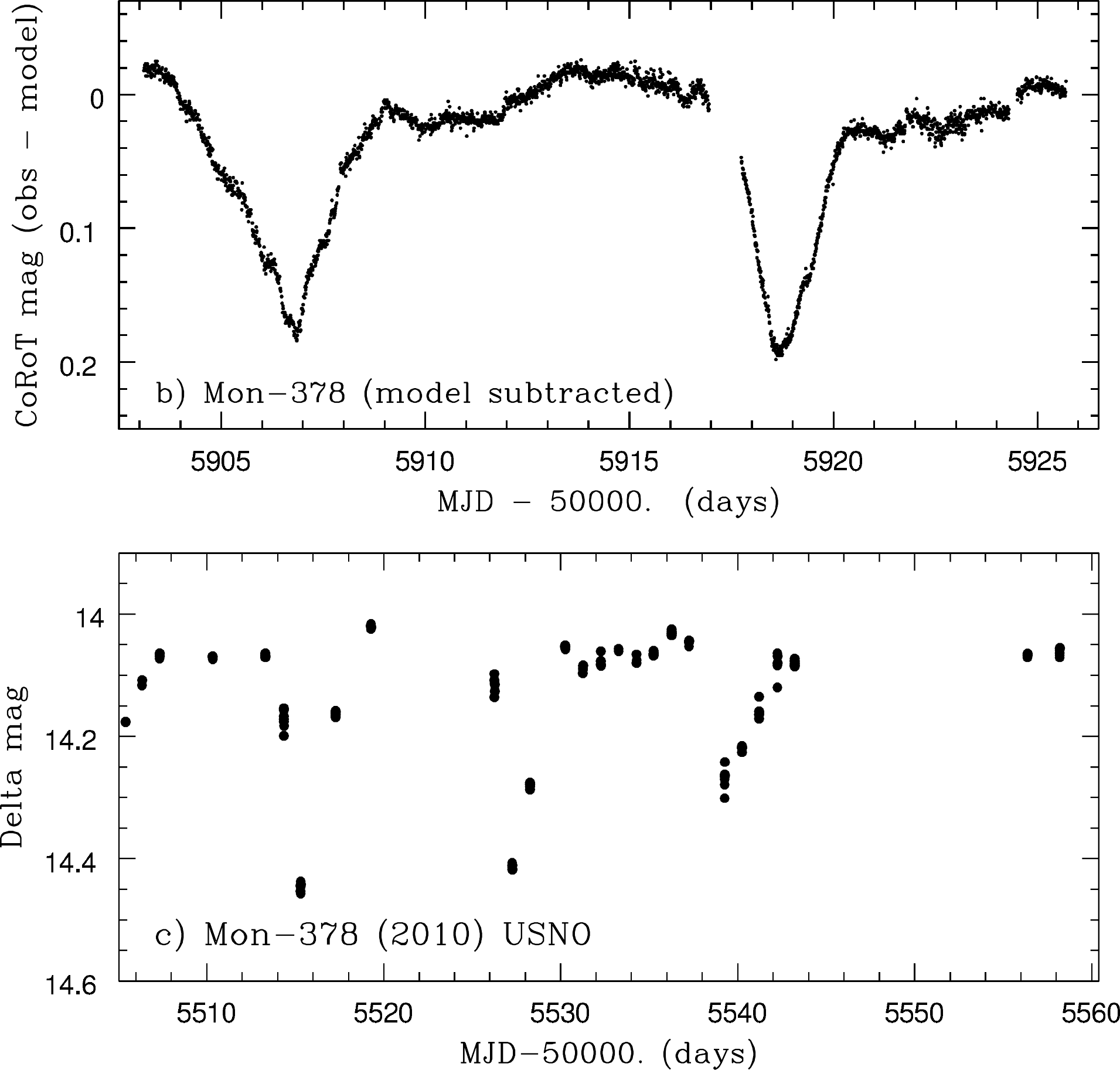}
\end{center}
\caption{a) Observed {\em CoRoT} 2011 light curve for Mon-378; b)
Result of subtracting a model spotted star light curve, generated as
in Sec. 3.1.1 for Mon-21, from the as-observed light curve; c) USNO
light curve for Mon-378 from observations in 2010.    The period for
both the spotted-star waveform and the short-duration flux dips is
about 11 days.
\label{fig:Mon378_obs_mod}}
\end{figure*}

Mon-378 is another system where the light curve is a superposition of
a spotted star light curve and narrow  flux dips
(Figure~\ref{fig:Mon378_obs_mod}a).  Mon-378 is a CTTS with an inverse
P Cygni H$\alpha$\ profile (Furesz et al.\ 2006),  spectral type K5.5
(Dahm \& Simon 2005), and location corresponding to a Class~II YSO in
the IRAC color-color diagram.  The {\em CoRoT} light curve is from
2011 -- it was not observed in the 2008 {\em CoRoT} campaign.  The IRAC
light curve shows no flux dips and little correlation with the {\em
CoRoT} light curve.  The {\em CoRoT} light curve shows a large
amplitude spotted star waveform with a period of about 11 days and 
with a single prominent, narrow flux dip at MJD $\sim$ 55919.   
However, after creating a model of the spotted star light curve as
done for Mon-21, and subtracting it from the original light curve, a
second prominent, narrow flux dip becomes evident
(Figure~\ref{fig:Mon378_obs_mod}b).   This by itself would be a weak
case that the dips are periodic. However, we also have a 2010
ground-based optical light curve, which shows (albeit poorly sampled)
three apparently narrow flux dips with the same $\sim$11 day period
(Figure~\ref{fig:Mon378_obs_mod}c).

\subsection{Mon-1076}

\begin{figure*}
\begin{center}
\epsscale{1.0}
\plottwo{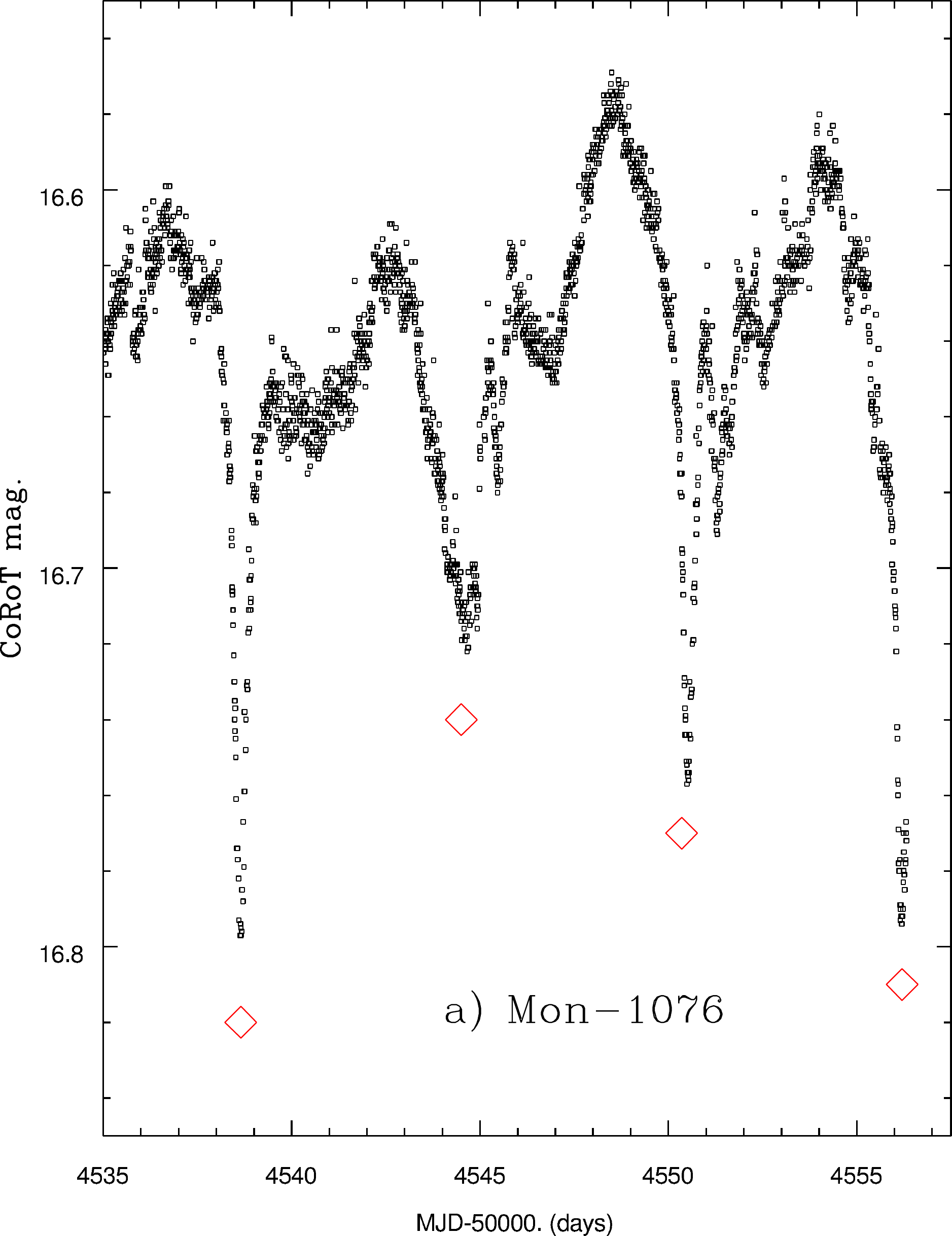}{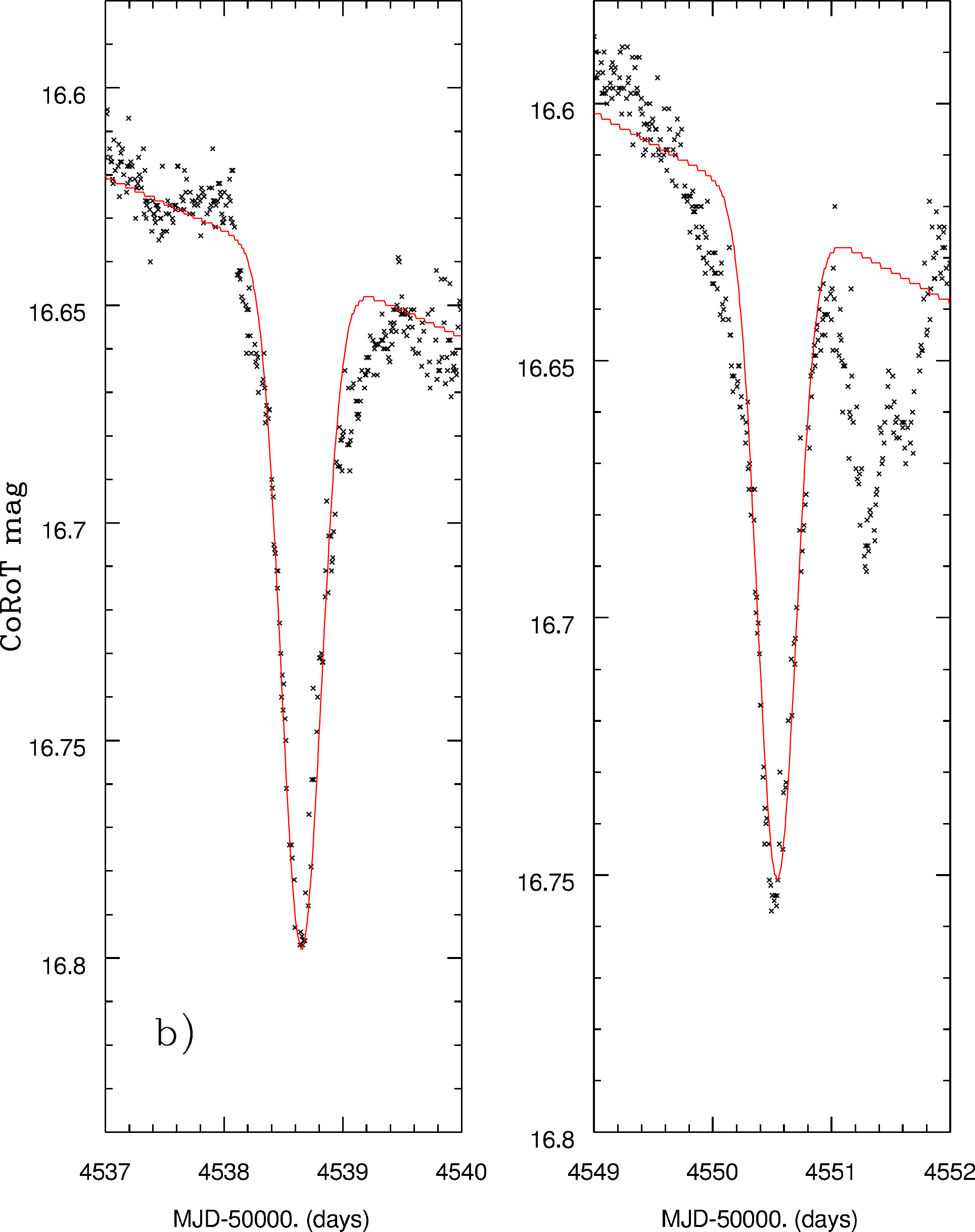}
\end{center}
\caption{a) {\em CoRoT} light curve for Mon-1076.   The first diamond is
centered at the measured location of the first flux dip; the other
diamonds are centered at the time predicted assuming a period
of 5.80 days; b) The two most prominent flux dips and our fits 
for Mon-1076 (dips A and B in Table~\ref{tab:quant_information}).
\label{fig:mon1076}}
\end{figure*}

Mon-1076 (aka V338 Mon) is somewhat similar to Mon-378 and Mon-1580.  
It has a 2008 {\em CoRoT} light curve with low amplitude variations
which may be due to star-spots but which are not stable enough to
yield an obviously periodic (sinusoidal) signature. The  {\em CoRoT}
light curve shows three prominent narrow flux dips at MJD 54539, 54551
and 54556.5, consistent with a period of $\sim$5.8 days
(Figure~\ref{fig:mon1076}a).  Fits to the two best observed  flux dips
are shown in Figure~\ref{fig:mon1076}b.  The published spectral type
for Mon-1076 is M1 (Dahm \& Simon 2005).  Mon-1076 was observed by
{\em CoRoT} in 2011, but bad systematics in the light curve preclude
using it for any analysis.

\subsection{Mon-566}

Mon-566 is a relatively late-type CTTS member of NGC~2264, with a
published spectral type of M3.5 (Dahm \& Simon 2005) and an H$\alpha$\
equivalent width of 19.4 \AA (Dahm \& Simon 2005).  
Its location in the IRAC color-color
diagram places it in the Class~II YSO category.  It was observed
by {\em CoRoT} in both 2008 and 2011.   Mon-566 is quite faint
in the optical, and therefore the {\em CoRoT} light curves are very
noisy which makes interpretation of the light curves difficult.
Nevertheless, both the 2008 and 2011 light curves appear to show
periodic, narrow, shallow flux dips.  The flux dip period is 4.5 days.

\begin{figure*}
\begin{center}
\epsscale{1.0}
\plottwo{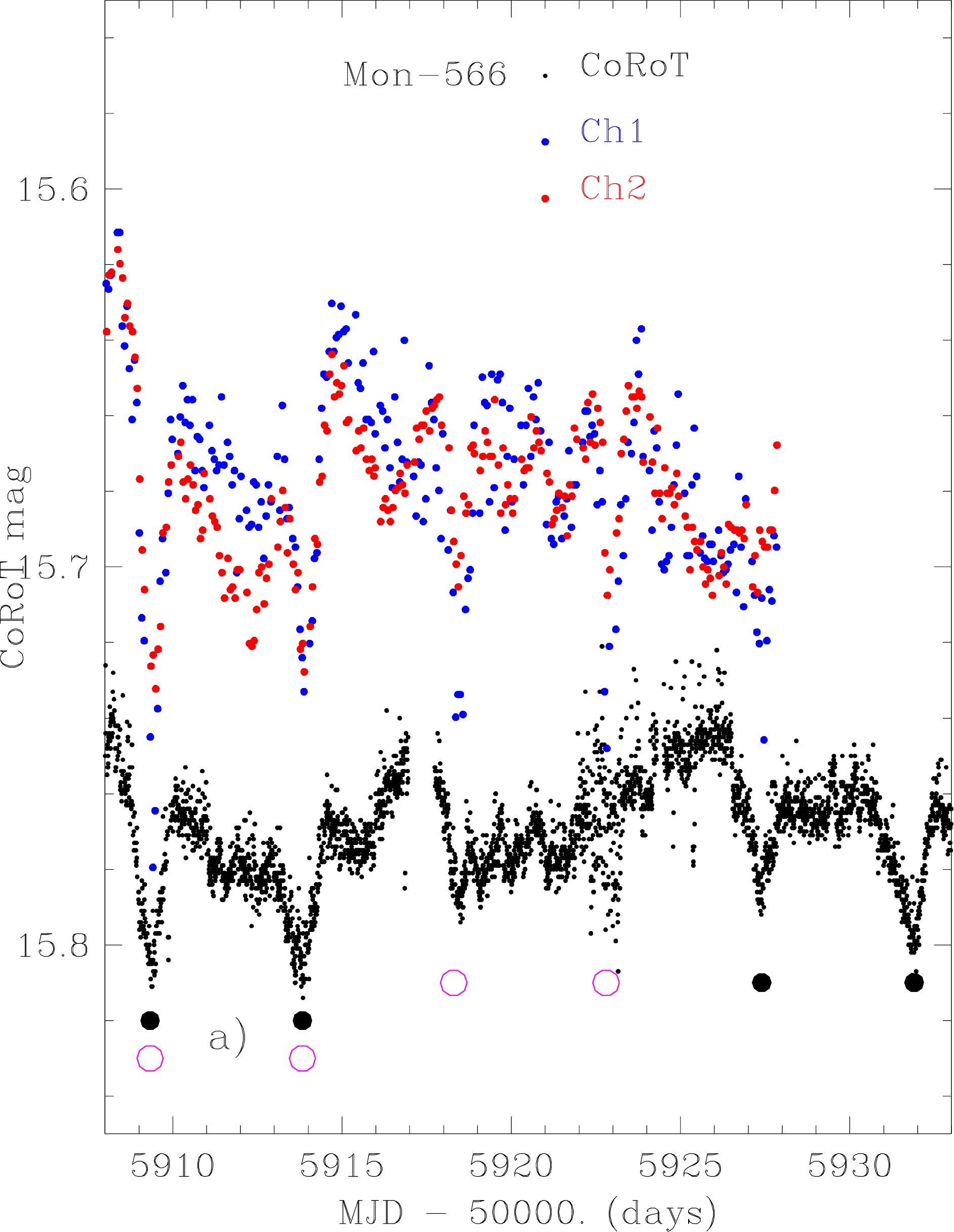}{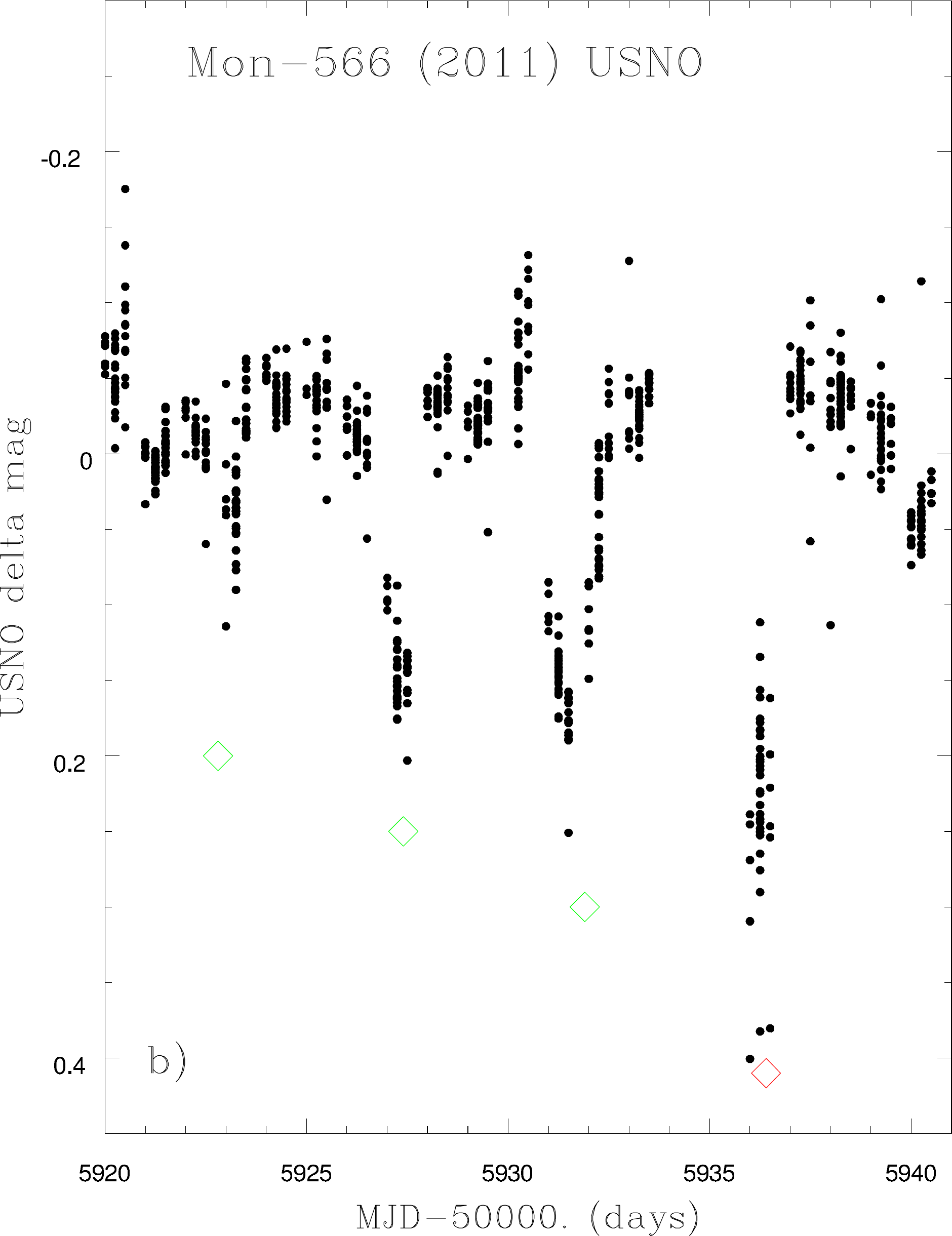}
\end{center}
\caption{ a) Overlay of the IRAC and {\em CoRoT} light curve for
Mon-566.   The magnitude zero points are shifted in order to align the
two IRAC light curves and displace them above the {\em CoRoT} light
curve for ease of comparison.  The large black filled dots mark flux dips
present in the {\em CoRoT} light curve; the magenta open circles mark
flux dips that are present in the IRAC light curves.  
b) {\em USNO} $I$-band light curve for
Mon-566; flux dips with depths up to about 0.15 mag are present,
coincident in time with some of the flux dips shown in the {\em
CoRoT}/{\em Spitzer} plot. The green diamonds are placed at the same
times as for panel (a); the red diamond marks a dip beyond the end of
the {\em CoRoT} campaign, where it would appear given the  measured
4.5 day period of the other flux dips.
\label{fig:mon566_spitzer_corot_usno}}
\end{figure*}

Two other sets of data give us confidence that the {\em CoRoT} flux
dips are real.  First, a {\em Spitzer} 2011 light curve also shows low
S/N flux dips of similar width and depth at the  location of some of
the {\em CoRoT} flux dips.  Second, we have a USNO light curve for
Mon-566 in 2011, and that light curve also shows flux dips coincident
with those present in the {\em CoRoT} and {\em Spitzer} light curves. 
All three light curves are shown in
Figure~\ref{fig:mon566_spitzer_corot_usno}.

The flux dips in the USNO light curve seem to be significantly deeper
than those in the {\em CoRoT} light curves.  We suspect this is
because Mon-566 is very faint for {\em CoRoT} ($r$ = 17.7),  and in
some cases the background subtraction routine that is used in the {\em
CoRoT}  pipeline results in too small (or large) a correction.  For
Mon-566, the correction is probably too small and the apparent flux
dip depths are therefore also too small.  Most of the stars in our
sample are considerably brighter in the optical, so we do not believe
their flux dips are significantly affected by  imprecision in the
background correction for their {\em CoRoT} light curves.

\section{YSOs Sharing Some Traits with the
  Short-Duration Flux Dip Stars}

Are there stars with light curves exhibiting superposed narrow {\it
and} broad, periodic flux dips?  Are there stars which have just one
or a couple narrow-flux dips over the timescale of the {\em CoRoT}
short runs?  Do Weak-lined T~Tauri stars (WTTS) ever show narrow flux
dips in their {\em CoRoT} light curves? These questions could shed
light on both the physical mechanism for the phenomena and the
requirements on our line-of-sight to the circumstellar disks in these
system.

\begin{figure*}
\begin{center}
\epsscale{1.0}
\plottwo{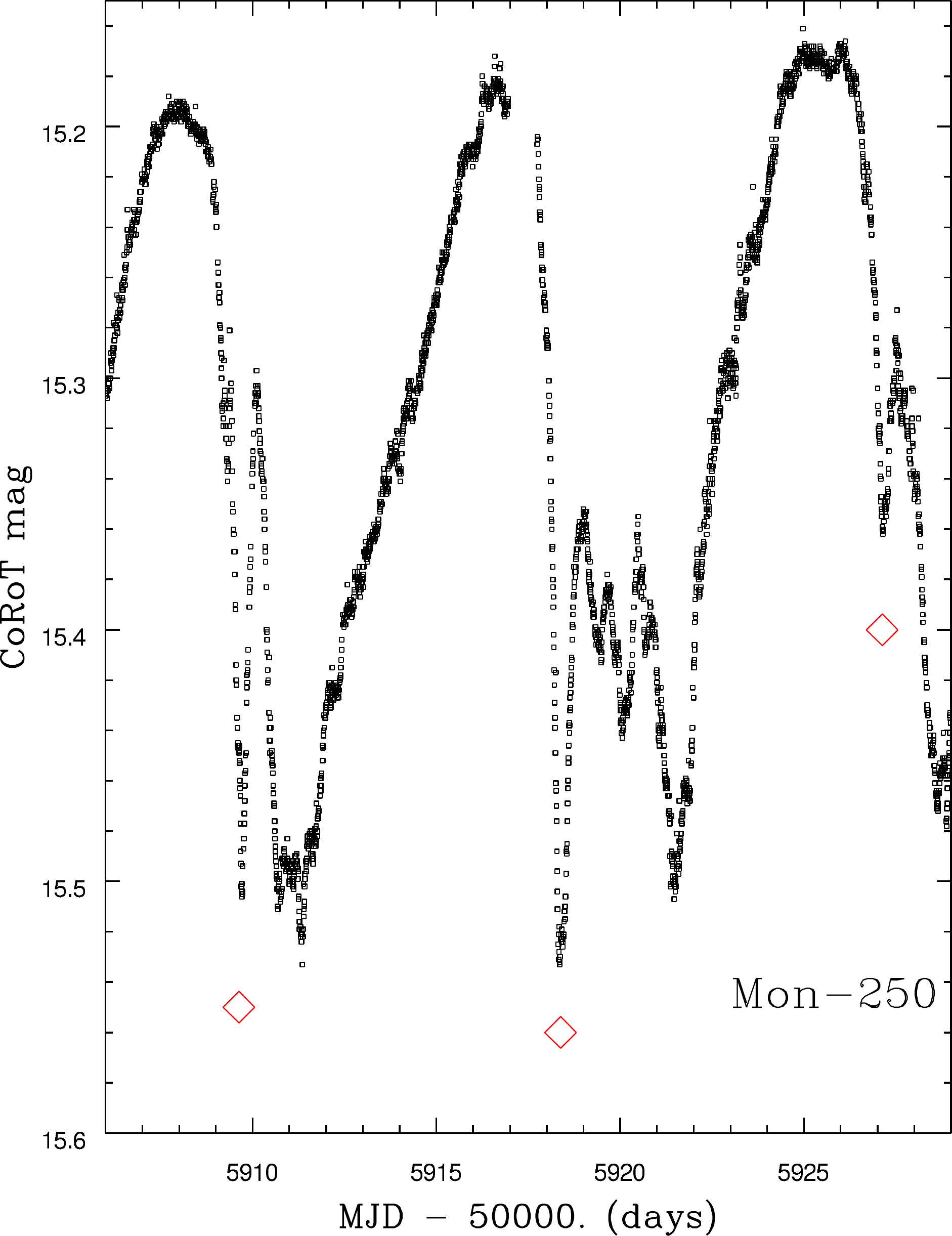}{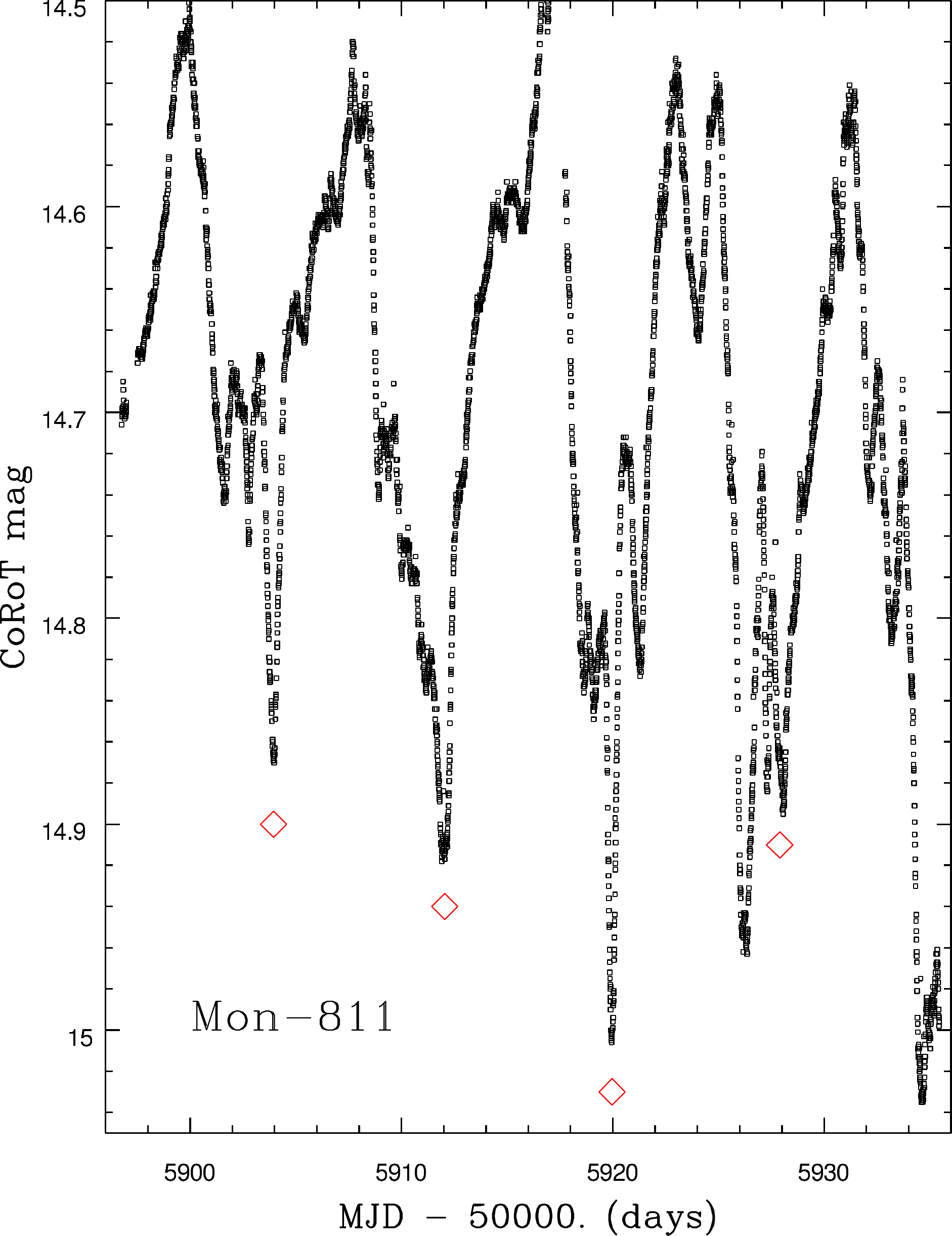}
\end{center}
\caption{View of a portion of the {\em CoRoT} light curve for two AA
Tau analogs, Mon-250 and Mon-811, highlighting in each light curve
several narrow flux dips that are spaced in time at the period defined
by the broad-flux dips.  The red diamonds mark the location of the
narrow flux dips.  The periods are $P$ = 8.75 day and 8.0 day, 
respectively.
\label{fig:mixed_dips}}
\end{figure*}

To answer the above questions, we critically examined the {\em CoRoT}
light curves for all $\sim$600 NGC~2264 members (CTTS, WTTS and unknown
classification)  in our catalog for which we have {\em CoRoT} light
curves as well as for 600 field stars with {\em CoRoT} light curves
obtained simultaneously with the NGC~2264 members.    The member stars
included all of the YSOs identified in McGinnis et al.\ (2015) as
AA~Tau analogs. We find that two of the stars with AA~Tau-type light
curves  do exhibit periodic, narrow flux dips in addition to the broad
flux dips required to be members of the AA~Tau class.  
Figure~\ref{fig:mixed_dips} shows these light curves, with the red
diamonds marking the location of the periodic, narrow flux dips.

In each case, the period inferred from the narrow flux dips is equal
(within the uncertainties) with that derived from the broad flux dips.
Gaussian fits to the narrow flux dips yield widths that are in the
same range as those measured for the narrow flux dip stars of
Table~\ref{tab:basicinformation}.   For these composite light curves,
the narrow flux dips always occur during the portion of the light
curve near the minimum in light for the broad-flux dips.  The
amplitudes for the narrow flux dips vary considerably from epoch to
epoch, but do not seem to correlate with the amplitudes of the broad
flux dips.

\begin{figure*}
\begin{center}
\epsscale{0.80}
\plottwo{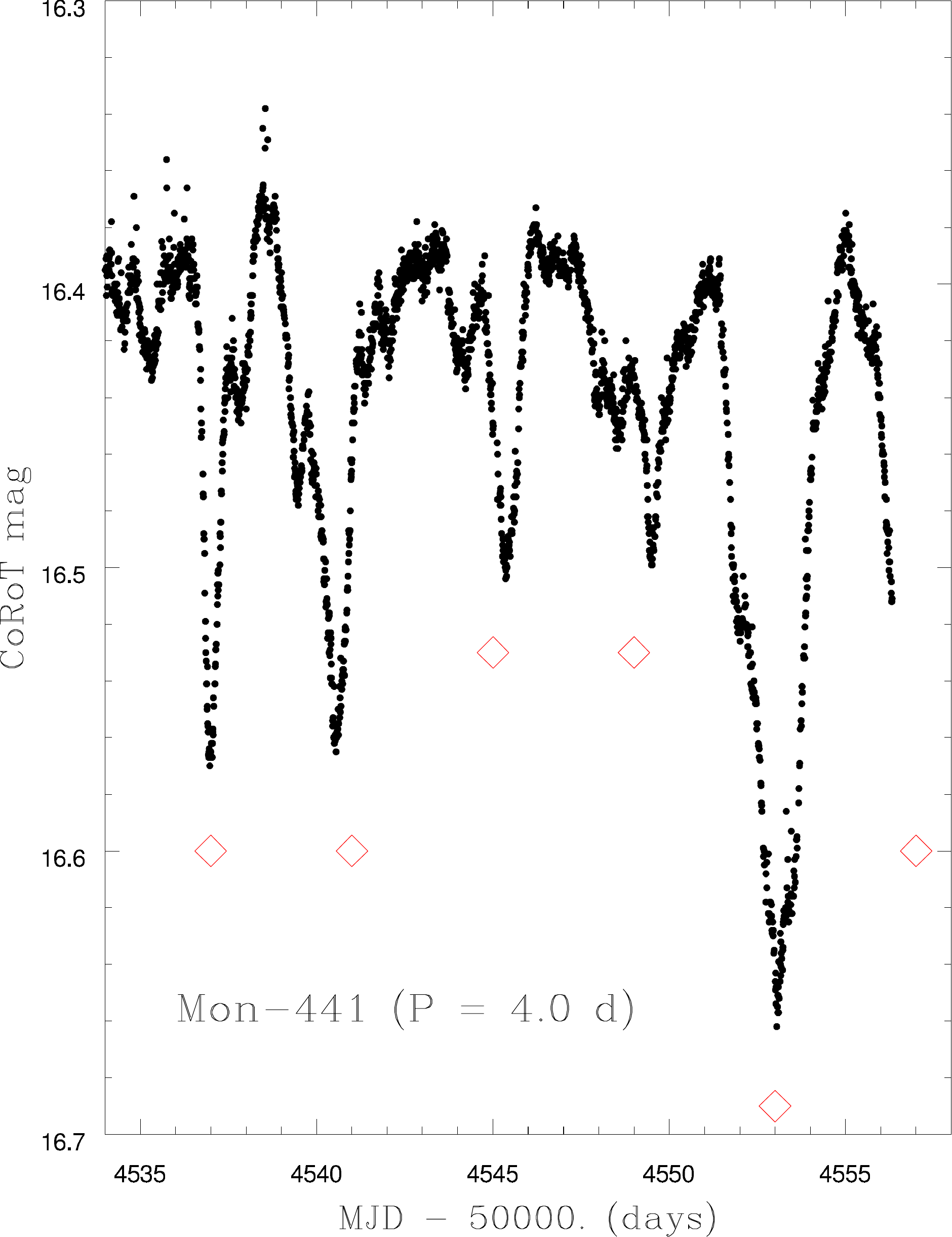}{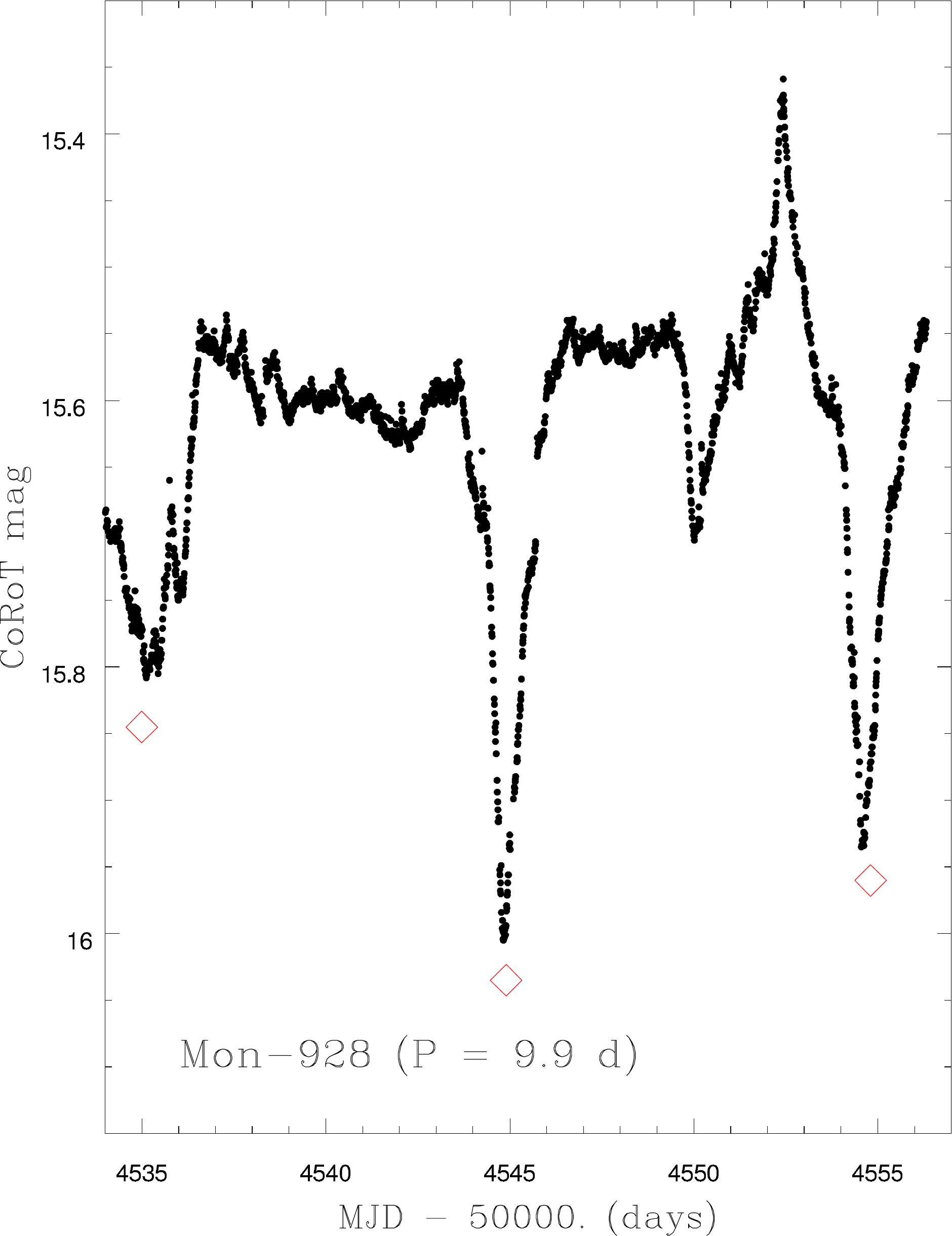}
\vspace{-0.cm}
\plottwo{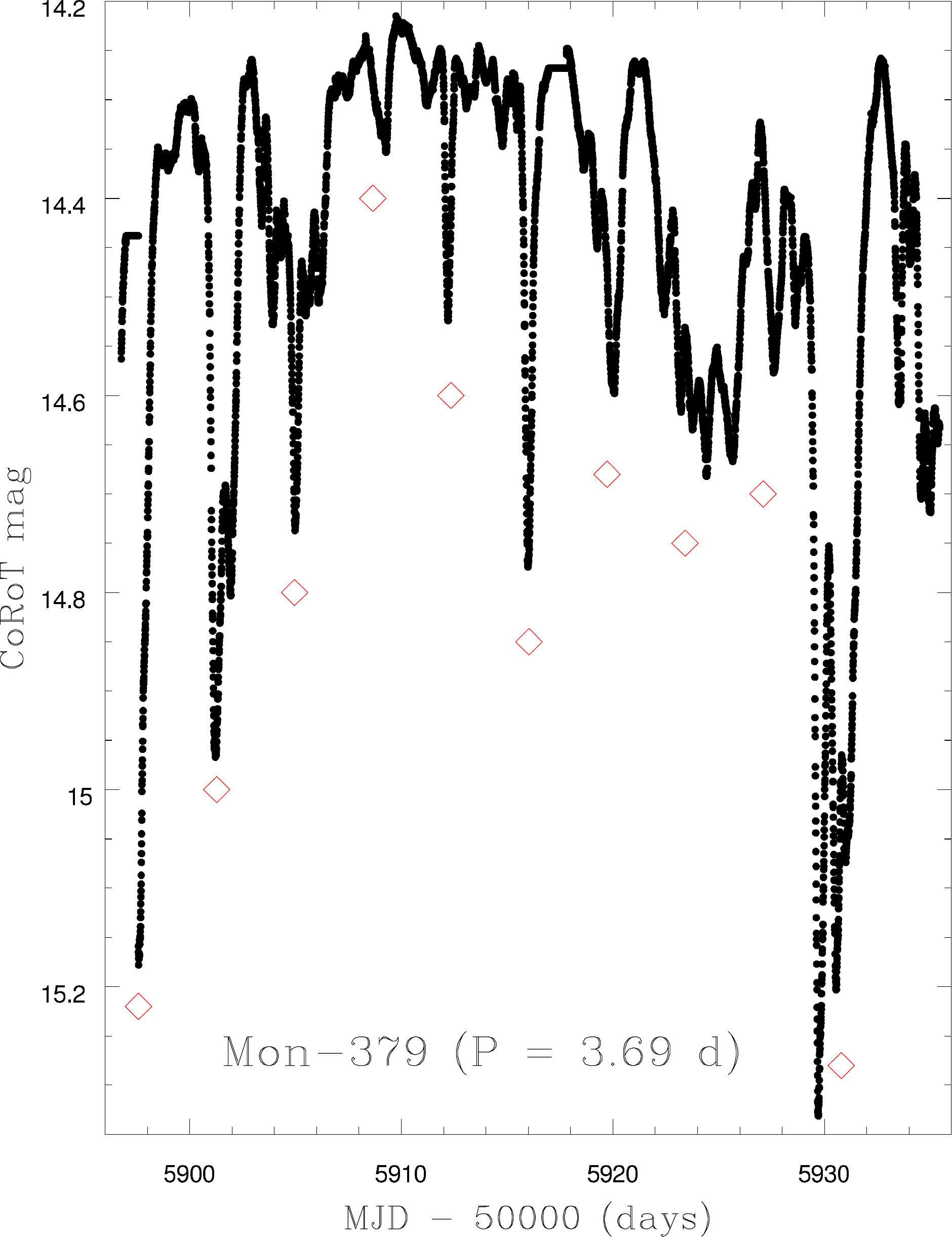}{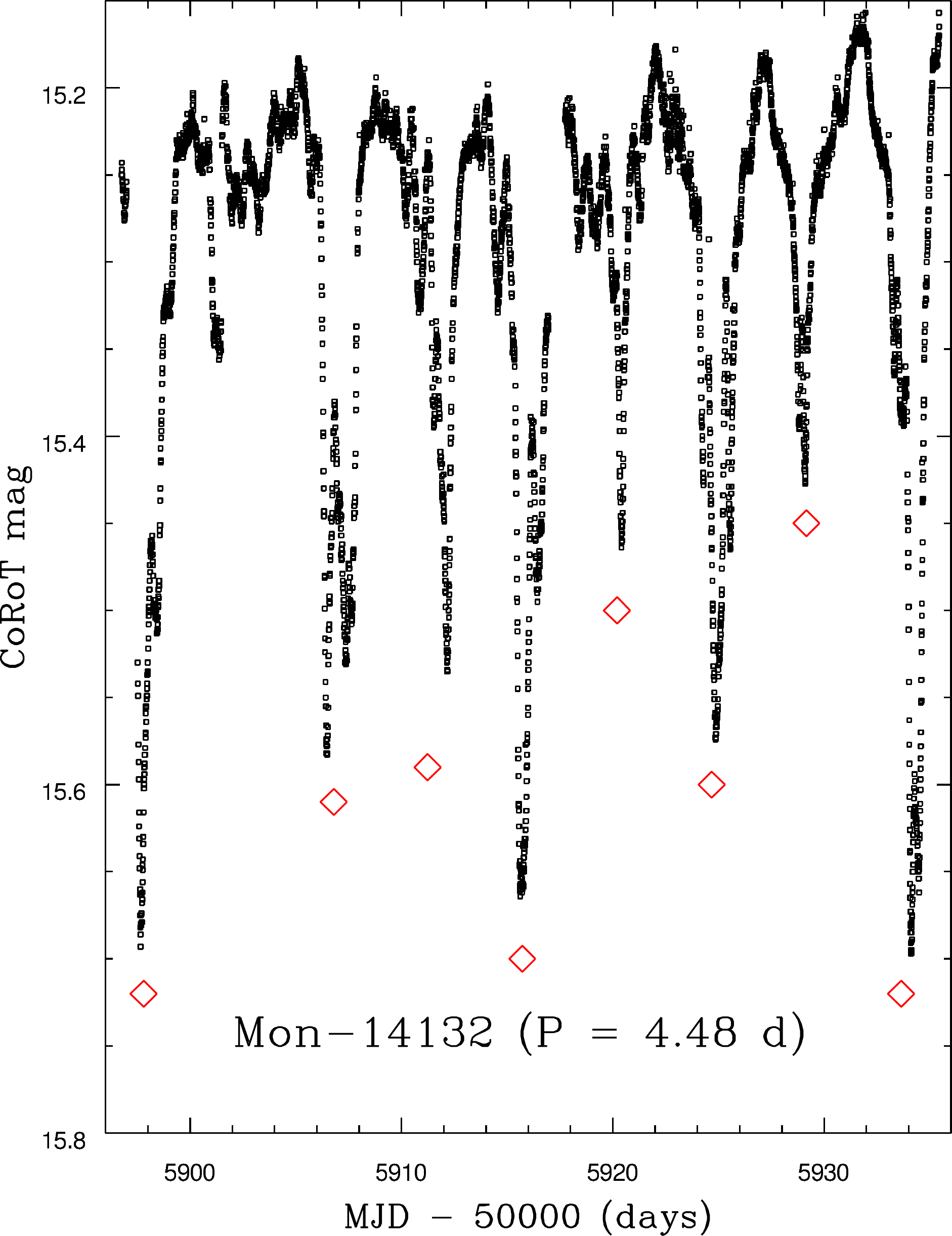}
\end{center}
\caption{Light curves for four NGC~2264 CTTS whose light curves show
periodic flux dips that are sometimes broad and sometimes narrow. The
red diamonds mark the location of the periodic, usually narrow flux
dips - in each case, placing the first diamond near the bottom of the
first dip, then spacing the other  diamonds by multiples of the period
we have determined.  We, and McGinnis et al. (2015) include these stars
with the classic AA Tau group, and they are shown as red circles with
inscribed red plus signs in Figure~\ref{fig:compare_dippers}, 
though their sometimes narrow flux dips suggests a link
with the stars of Table~\ref{tab:basicinformation}. \label{fig:composite_dips}}
\end{figure*}

We have also identified another type of composite light curve -- light
curves with periodic features that are sometimes narrow and sometimes
broad (defined arbitrarily as FWHM/Period $>$ 0.3).   There are at
least four such objects in our NGC~2264 dataset - Mon-441, Mon-928,
Mon-379, and Mon-14132;  all are included in McGinnis et al.\ (2015) as
AA Tau analogs.  Their light curves are shown in
Figure~\ref{fig:composite_dips}.  It could be argued that these four
stars should be added to our group of short-duration, periodic flux
dip stars.  We prefer not to do that because we prefer to restrict the
group to stars where all the dips in a {\em CoRoT} campaign are short
duration.  However, including these stars in the group would not
change any of our conclusions.

Based on the fact that the periodic narrow and broad flux dips can be
present in the same star and during the same campaign, the most
significant deduction one can make  is that the requirements for
producing these two types of dips overlap.  Because the AA Tau, warped
disk model requires a specific viewing angle to the disk, that same
viewing angle must also be compatible with production of the
short-duration, periodic flux dips.

\begin{figure*}
\begin{center}
\epsfxsize=.99\columnwidth
\epsfbox{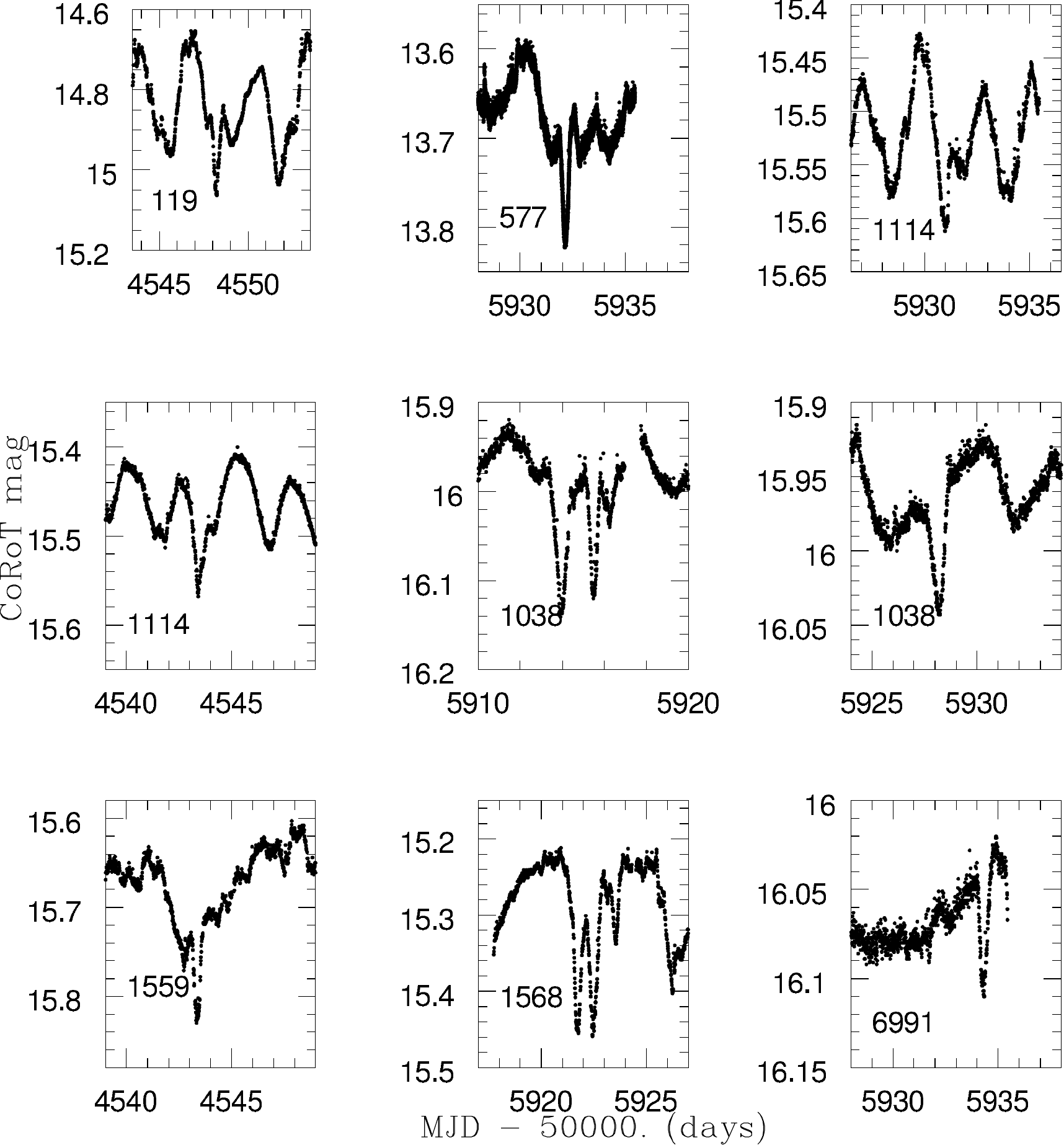}
\end{center}
\caption{Isolated, short-duration flux dips found in the light curves
of NGC~2264 CTTS, for stars with only one or a few such events per
{\em CoRoT} campaign -- and hence for stars not previously included in
our variable extinction classes. Biographical data for these stars are
provided in the Appendix (Table~\ref{tab:singledippers}). \label{fig:singledip_dips}}
\end{figure*}

Our visual re-examination of the other {\em CoRoT} light curves
yielded two additional results.   First, none of the stars  identified
as WTTS (nor any of the non-member stars) showed any narrow or broad
flux dips whose morphology mimicked those of the AA Tau light curves
or those of the periodic, short-duration flux dip stars.  This is, of
course, as one would expect if such dips are due to close-in
circumstellar dust, which WTTS and non-members should lack.  Second,
we identified a set of nine stars -- all CTTS -- whose light curves
showed only a single, or in some cases a few, short-duration  flux
dips (but with no hint of periodicity).  Light-curve snippets for a
sample of these flux dips are shown in
Figure~\ref{fig:singledip_dips}.  The widths and depths of these dips
fall within the range found for the stars of
Table~\ref{tab:basicinformation}.  The physical properties (UV
excesses; IR excesses; spectral types) of the stars with these single
dips are not obviously different from the other types of variable
extinction stars.  We assume that the same physical mechanism(s) for
producing the flux dips applies to this set of stars, but in their
case other dust raising events either do not lift the dust high enough
to occult our line of sight or yield dust opacities insufficient to
yield a dip detectable in our data. However, when considering the
statistics of what fraction of CTTS show any evidence of line-of-sight
extinction events, these objects should also be included in the
calculation.  A table of basic information for these stars
(Table~\ref{tab:singledippers}) is provided in the Appendix.

\section{Physical Properties of the Short Duration Flux Dip Class}

Why do the CTTS in Table~\ref{tab:basicinformation} show periodic,
short-duration, often Gaussian shaped flux dips?  If there are common
physical characteristics that are shared by this set of nine YSOs,
perhaps that would help allow us to assign a physical mechanism to the
flux dips.  We begin by examining the inferences we can make regarding
the orientation of the circumstellar disks to our line of sight.  We
then discuss a variety of color-color and color-magnitude diagrams
(CMDs) which sort the NGC~2264 YSOs by mass or accretion rate or IR
excess.  For the inferred properties in
Table~\ref{tab:basicinformation} and for the calculations in this
section, we adopt a distance of 760 pc for NGC~2264 (Sung et al.\
1997).  The Sung et al.\ distance is in good agreement with a recent
VLBI distance for a maser source in the cluster ($d$ =
738$^{+57}_{-50}$ pc; Kamezaki et al.\ 2014) and with a recent
analysis of data for a newly-identified NGC~2264 pre-main sequence
(PMS) eclipsing binary ($d$ = 756$\pm$96 pc; Gillen et al.\ 2014).

\subsection{Our View Angle to The Disks of the Narrow Flux-Dip Stars}

Practically the simplest, most straight-forward, test we can make to
constrain the origin of the material causing the flux dips is to
determine the inclination of our line of sight to the circumstellar
disk for these stars. In order for a YSO to show flux dips from disk
material passing through our line of sight, most models predict that
the inclination of the disk to our line of sight must be
$i>65\arcdeg$.  To increase the sample size for this test, we include
not only the stars in Table~\ref{tab:basicinformation} but also the
other stars discussed in \S 5 that have both narrow and broad flux
dips.  In total, we have eleven stars for which we have all of the
relevant  information (luminosity, stellar \teff, \vsini, and rotation
period). The input data, and our estimates of the line of sight
angles are provided in Table~\ref{tab:incl-estim}.

The periods listed in Table~\ref{tab:incl-estim} are those derived
from the flux dips.   In several cases (Mon-21, Mon-56, Mon-378,
Mon-1580),  the light curves also show semi-sinusoidal variations
plausibly due to photospheric spots, and in those cases the spot
periods are indeed essentially the same as the periods derived from
the dips (and thus consistent with ``disk locking").   We assume for
the other stars that their flux dip periods are also equal to their
photospheric rotation periods.   The \vsini\ values  are either from
VLT-FLAMES or Keck-HIRES spectra we have obtained for this program, or
from reanalysis of MMT-Hectochelle spectra obtained by Furesz et al.\
(2006).   A description of the spectral analysis is provided in
McGinnis et al.\ (2015). The stellar luminosities are derived from the
measured J-band magnitudes, extinctions based on fits to multicolor
photometry, the adopted distance of 760 pc, and bolometric corrections
from Pecaut \& Mamajek (2013); the derivation of these luminosities is
described in Venuti et al. (2014). Effective temperatures are derived
from spectral types -- either from the literature or from our own
spectra as described in the text and the appendix -- and a conversion
from spectral type to \teff\ appropriate for YSOs.  In order to
illustrate the impact of the temperature scale, we show results for
two representative temperature scales  in Table~\ref{tab:incl-estim},
one from Table 6 of Pecaut \& Mamajek (2013; PM13)  and one from
Cohen \& Kuhi (1979; CK79).

\begin{figure*}
\begin{center}
\epsfxsize=.99\columnwidth
\epsfbox{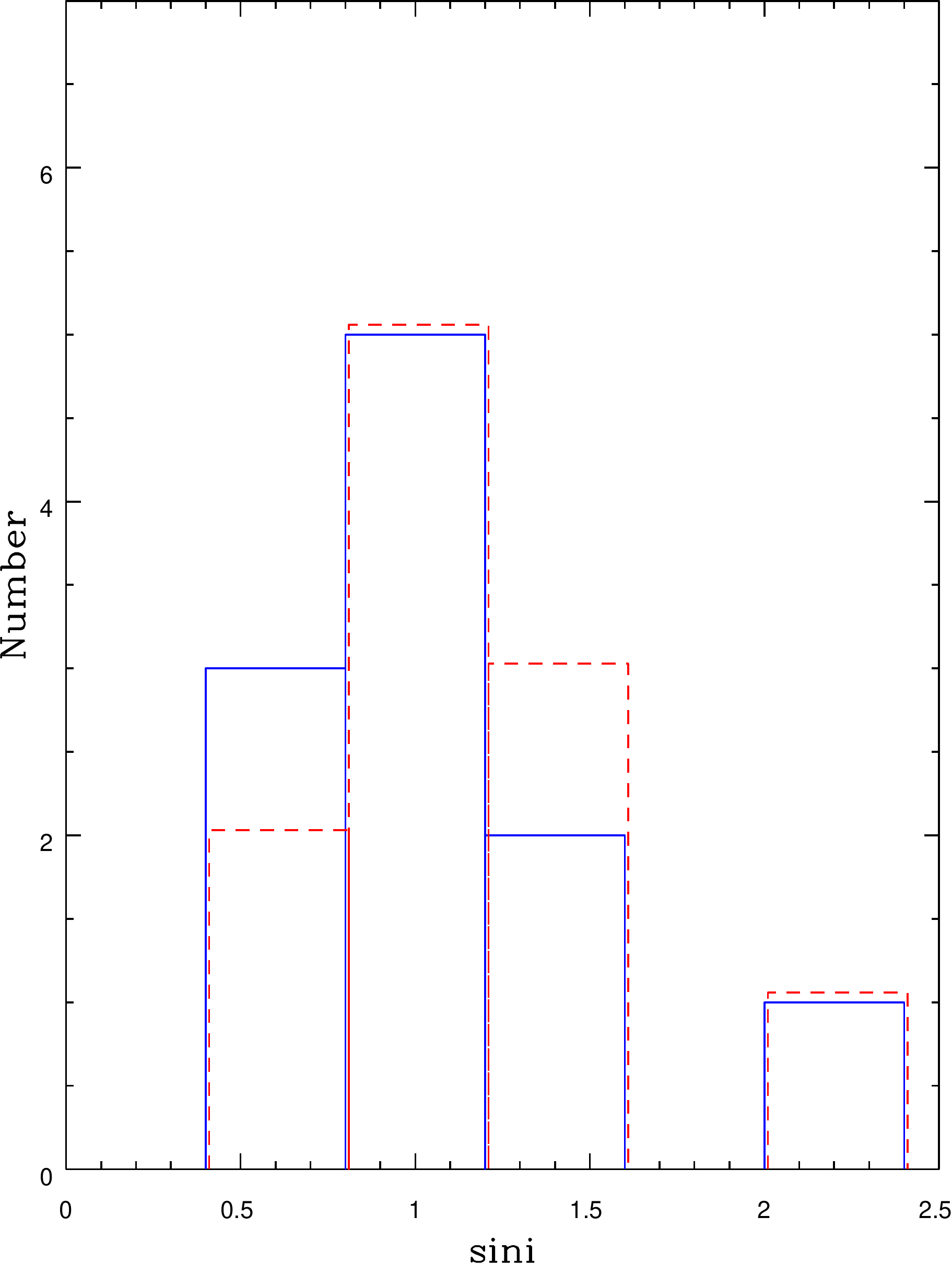}
\end{center}
\caption{Histograms of the derived \sini\ estimates for the stars
exhibiting periodic short-duration flux dips.  The red dashed
histogram corresponds to the estimates made using the CK79 temperature
scale; the blue solid histogram corresponds to the calculations made
with the PM13 young star temperature scale (their Table 6).
\label{fig:inclination_hist}}
\end{figure*}

For many of the stars, the calculation returns a value of \sini\ $>$ 1.0,
and in these cases we tabulate an inclination of 90$\arcdeg$.  A visual
summary of the results is provided in  Figure~\ref{fig:inclination_hist}.
The mean values and uncertainties are: $< \sin i >$ = 1.16 $\pm$ 0.55 (CK79) and
$< \sin i >$ = 1.08 $\pm$ 0.53 (PM13) when including all eleven stars; and
$< \sin i >$ = 1.03 $\pm$ 0.35 (CK79) and \sini\ = 0.96 $\pm$ 0.35 (PM13)
if one excludes the highly discrepant values for Mon-928.   These estimates
support the hypothesis that we are viewing these systems nearly edge on
to their disks, and that the material passing through our line of sight
to produce the flux dips plausibly lies within the disk or near the disk surface.
It is clear, however, that individual \sini\ estimates have large 
uncertainties, making this result less robust than we had hoped.  Assuming
that \teff\ and \vsini\ uncertainties dominate, and that they 
contribute approximately equally, the \sini\ estimates suggest the true
uncertainties in \teff\ are of order 10\% and in \vsini\ are of
order 20\%.   

\begin{deluxetable*}{lccccccccccc}
\tabletypesize{\scriptsize}
\tablecolumns{12}
\tablewidth{0pt}
\tablecaption{View Angles to the Disk Axis\label{tab:incl-estim}}
\tablehead{
\colhead{Mon ID}  & \colhead{Period} &
\colhead{\vsini} & \colhead{Luminosity} &
\colhead{Spectral} & \colhead{Ref}\tablenotemark{a} &
\colhead{\teff} & \colhead{\teff} &
\colhead{R$_{star}$(CK79)} &
\colhead{\sini (CK79)} &
\colhead{i-PM13} & \colhead{i-CK79}  \\
\colhead{} & \colhead{(days)} & \colhead{(km/s)} & \colhead{(L$_{\odot}$)} &
\colhead{Type} & \colhead{} &
\colhead{(PM13)} & \colhead{(CK79)} & \colhead{(R$_{\odot}$)} & \colhead{(R$_{\odot}$)} &
\colhead{(degrees)} & \colhead{(degrees)}
}     
\startdata
CSIMon-000021 & 3.20 & 19.2 $\pm$ 1.5 & 0.957 & K5 & R02    & 4140.0 & 4395.0 & 1.68 & 0.719 & 39.53 & 45.97 \\
CSIMon-000056 & 5.83 & 11.1 $\pm$ 2.2 & 0.774 & K5 & DS05   & 4140.0 & 4395.0 & 1.52 & 0.842 & 48.20 & 57.35 \\
CSIMon-000378 & 11.1 & 10.4 $\pm$ 0.4 & 0.896 & K5.5 & DS05 & 4080.0 & 4295.0 & 1.71 & 1.333 & 90.00 & 90.00 \\
CSIMon-001076 & 5.85 &  9.8 $\pm$ 2.4 & 0.256 & M1 & DS05   & 3630.0 & 3681.0 & 1.23 & 0.908 & 61.98 & 65.20 \\
CSIMon-001131 & 5.18 &  8.8 $\pm$ 1.7 & 0.545 & M1.5: & S14 & 3560.0 & 3590.0 & 1.91 & 0.470 & 27.60 & 28.00 \\
CSIMon-001165 & 5.50 & 16.8 $\pm$ 2.3 & 0.130 & M3: & S14   & 3360.0 & 3360.0 & 1.06 & 1.711 & 90.00 & 90.00 \\
CSIMon-001580 & 7.10 & 11.4 $\pm$ 2.0 & 0.265 & M1: & S14   & 3630.0 & 3681.0 & 1.26 & 1.260 & 90.00 & 90.00 \\
CSIMon-000250 & 8.70 &  9.8 $\pm$ 2.2 & 1.229 & K3 & DS05   & 4550.0 & 4775.0 & 1.62 & 1.037 & 70.29 & 90.00 \\
CSIMon-000379 & 3.69 & 25.0 $\pm$ 1.2 & 2.121 & K2 & DS05   & 4760.0 & 4955.0 & 1.97 & 0.920 & 58.06 & 66.86 \\
CSIMon-000811 & 7.88 & 13.5 $\pm$ 1.7 & 1.076 & K6 & DS05   & 4020.0 & 4200.0 & 1.96 & 1.070 & 78.50 & 90.00 \\
CSIMon-000928 & 9.92 & 20.1 $\pm$ 2.8 & 0.576 & M0 & DS05   & 3770.0 & 3915.0 & 1.65 & 2.490 & 90.00 & 90.00 \\
\enddata
\tablenotetext{a} {Spectral type references: R02 - Rebull et al.\ (2002); DS05 = Dahm \& Simon (2005);
S14 - this paper}
\end{deluxetable*}

\subsection{Stellar and Disk Properties of the Narrow Flux-Dip Stars}

Color-magnitude or color-color diagrams could provide clues to the
physical mechanisms driving the photometric variability in our three
light curve classes (AA Tau analogs, stars with aperiodic flux dips,
and the new short-duration flux dip class).  If a mechanism depended
on the stellar mass or luminosity, this should be reflected by the
location of the class members in a CMD. If the mechanisms depended
sensitively on our view angle to the circumstellar disk, then the mean
$A_V$ to the system as estimated from a $J - H$ vs. $H - K_s$ diagram
could differ amongst the three classes.  If the mechanism(s) depend
sensitively on the  radius of the inner disk wall or the degree of
disk flaring, that could be reflected in the IRAC [3.6]$-$[4.5] vs.
[5.8]$-$[8.0] diagram.

\begin{figure*}
\begin{center}
\epsscale{1.0}
\plottwo{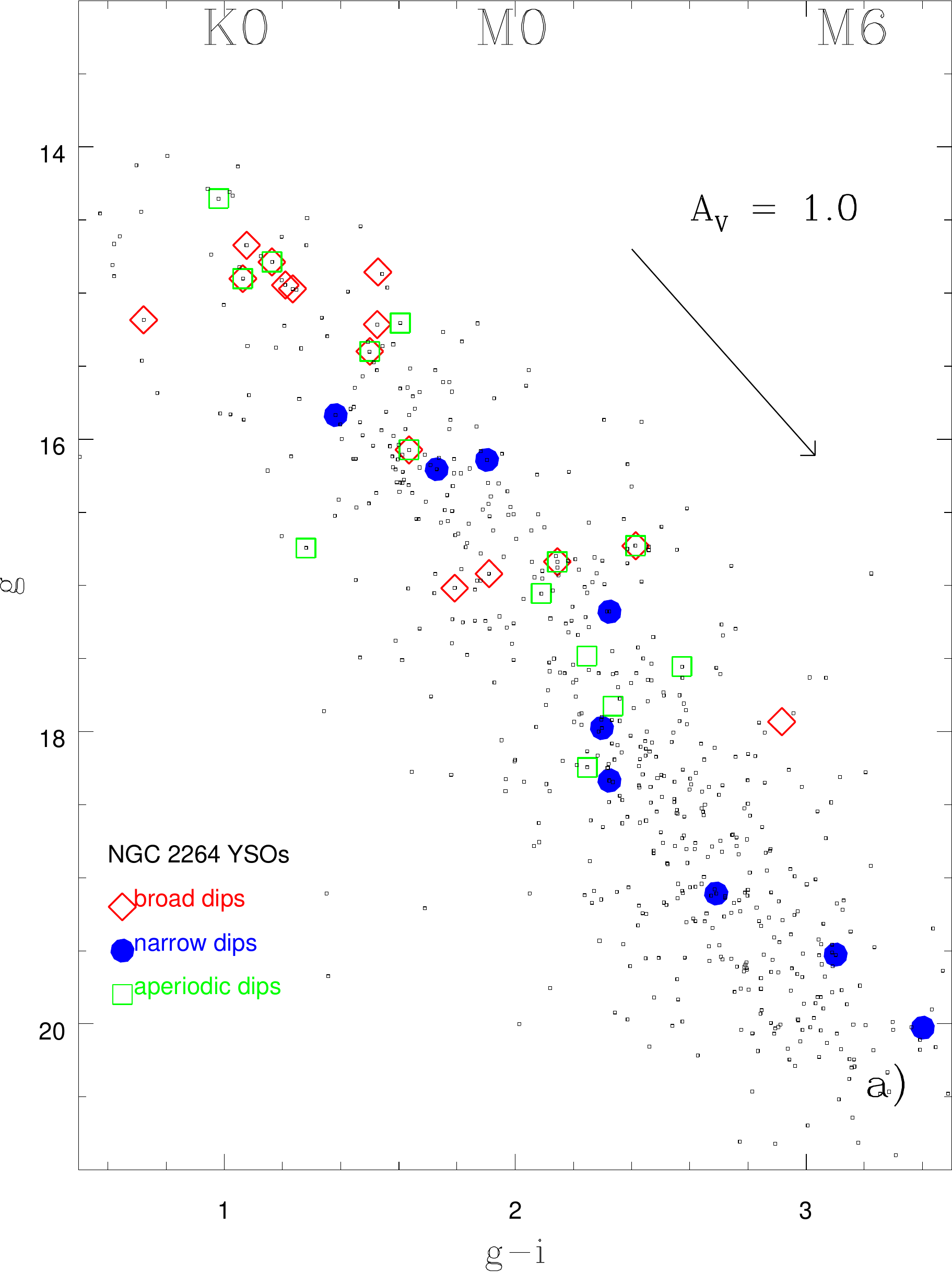}{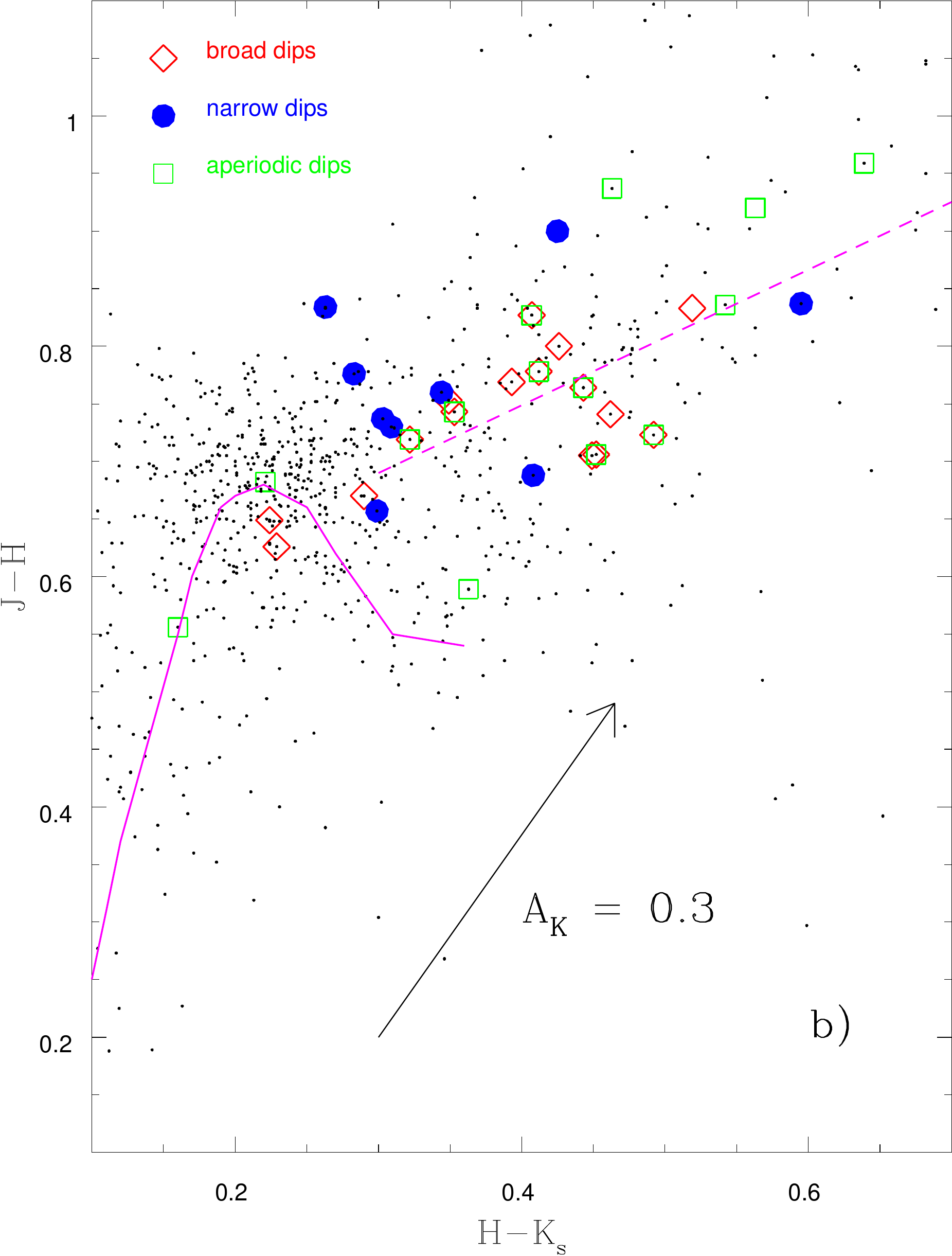}
\end{center}
\caption{Comparison of the location of members of the three different
classes of variable extinction YSOs in an optical CMD (panel a) and in
the near-IR $J-H$ vs. $H-K_s$ diagram (panel b).  The magenta curve
in  panel (b) is the 2MASS main-sequence locus (Pecaut \& Mamajek
2013); the  magenta dashed line is the Meyer et al.\ (1997) CTTS
extension to the main sequence locus (transformed to the 2MASS filter
system). The small black dots are all probable NGC 2264 members (Cody et al.\ 2014), 
regardless of whether we have {\em CoRoT} light curves or
not.  A number of objects are marked both as broad dip and aperiodic
dip members because their light curve in 2008 falls in one class and
their 2011 light curve falls in the other class.
\label{fig:cmdandJHKdiagrams}}
\end{figure*}

\begin{figure*}
\begin{center}
\epsfxsize=.99\columnwidth
\epsfbox{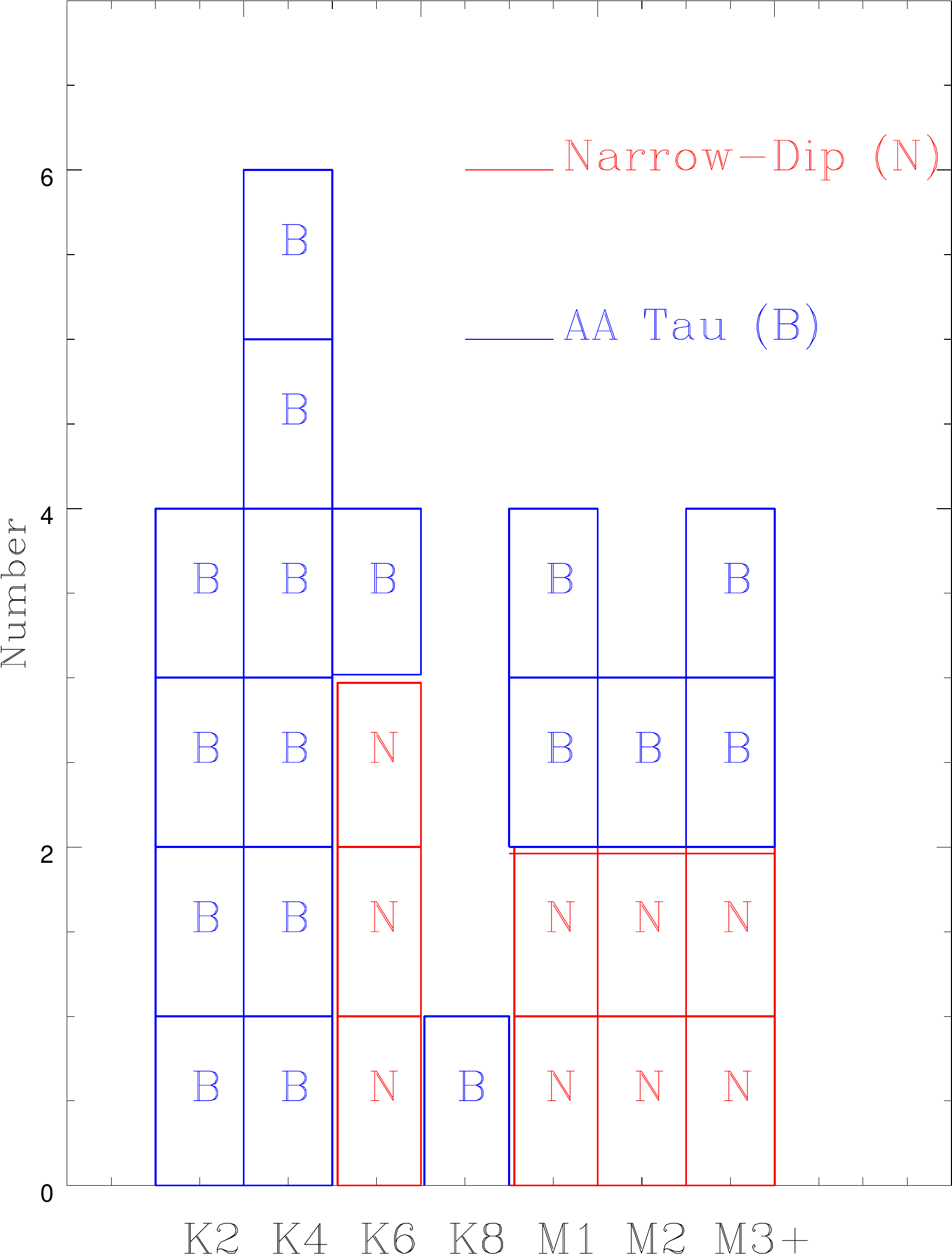}
\end{center}
\caption{Spectral types for the stars in Table~\ref{tab:basicinformation}
compared to those for the classical AA Tau stars in NGC~2264 (same stars
as in Figure~\ref{fig:compare_dippers}).  For one of the narrow-dip stars
and six of the AA Tau analogs, there is no published spectral type and 
we have adopted the \teff\ values from Venuti et al.\ (2014) to estimate
spectral types.  The spectral bins used were K1-K3; K3-K5; K5-K7; K7-M0;
M0-M1.5; M1.5-M3; and M3-M4.5.
\label{fig:sptype.histo}}
\end{figure*}

Figure~\ref{fig:cmdandJHKdiagrams}a compares the location in an
optical CMD of the short duration flux dip stars to the set of
classical AA Tau analogs (except as noted in footnote 22) and the
aperiodic extinctor candidates in NGC~2264 as identified in McGinnis
et al.\ (2015).  No reddening correction has been attempted.  The
primary conclusion we draw from this diagram is that the AA Tau
analogs and stars with aperiodic flux dips appear to be bluer and
brighter (higher mass), on average, than the narrow dip stars, with
only a couple having $g-i$ colors corresponding to M dwarfs. 
Secondarily, we see no obvious age difference between the three groups
of stars or of these stars relative to the entire set of NGC~2264
members. As illustrated in Figure~\ref{fig:sptype.histo}, the spectral
types for the short-duration flux dip stars are also significantly
later, on average, than those for the  NGC~2264 stars we classify as
``classic" AA Tau analogs, lending support to the belief that they are
generally lower in mass.  Based on a Monte  Carlo simulation, there is
less than a 2\% chance the two groups are drawn  from the same parent
spectral type distribution.

Figure~\ref{fig:cmdandJHKdiagrams}b provides the $J-H$ vs.\ $H-K_s$
diagram for the same set of stars, as a means to assess their
reddening. Since all of the variable extinction stars are CTTS, we
assume that the Meyer et al.\ (1997) CTTS locus represents their
intrinsic (unreddened) colors, at least for low mass CTTS.  The
aperiodic dip star located near $H - K_s$ = 0.4, $J - H$ = 0.6
(Mon-325) is a high mass star (spectral type G6), and presumably
dereddens back to the ascending portion of the main-sequence locus. 
Most of the variable extinction stars lie near the Meyer et al.\ CTTS
locus, and therefore have little additional reddening as inferred from
their mean 2MASS photometry.   One exception is Mon-6975, whose IR
colors of $H - K_s$ = 0.42, $J - H$ = 0.90 indicates relatively large
inferred reddening. Mon-6975 is also the reddest variable extinction
star in the optical CMD (Figure~\ref{fig:cmdandJHKdiagrams}a).  We
conclude that outside of the flux dips, the other YSOs with narrow and
broad flux dips  have at most modest reddening, and therefore that
reddening corrections would not significantly alter their locations in
optical CMDs such as Figure~\ref{fig:cmdandJHKdiagrams}a.

\begin{figure*}
\begin{center}
\epsscale{1.0}
\plottwo{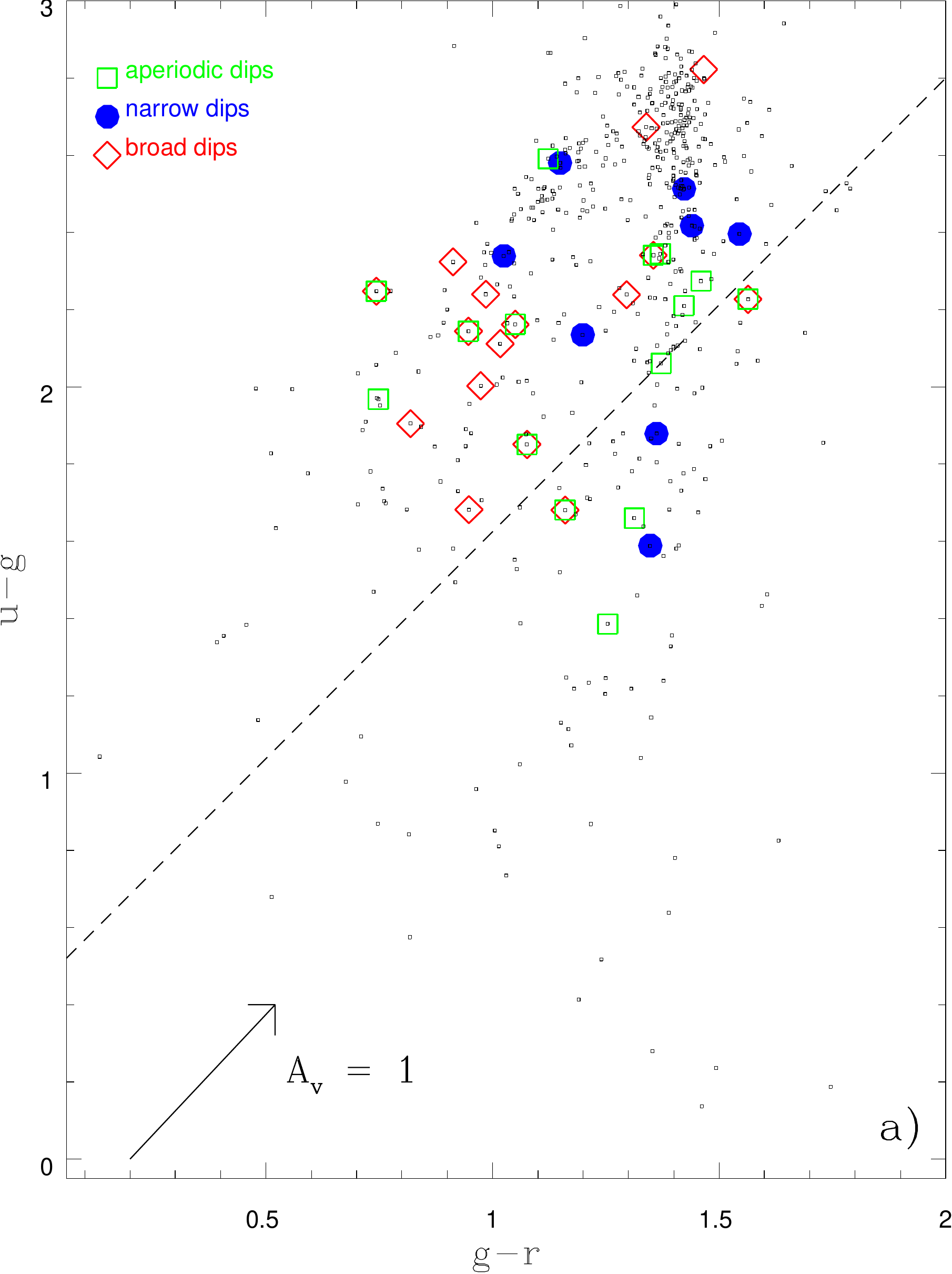}{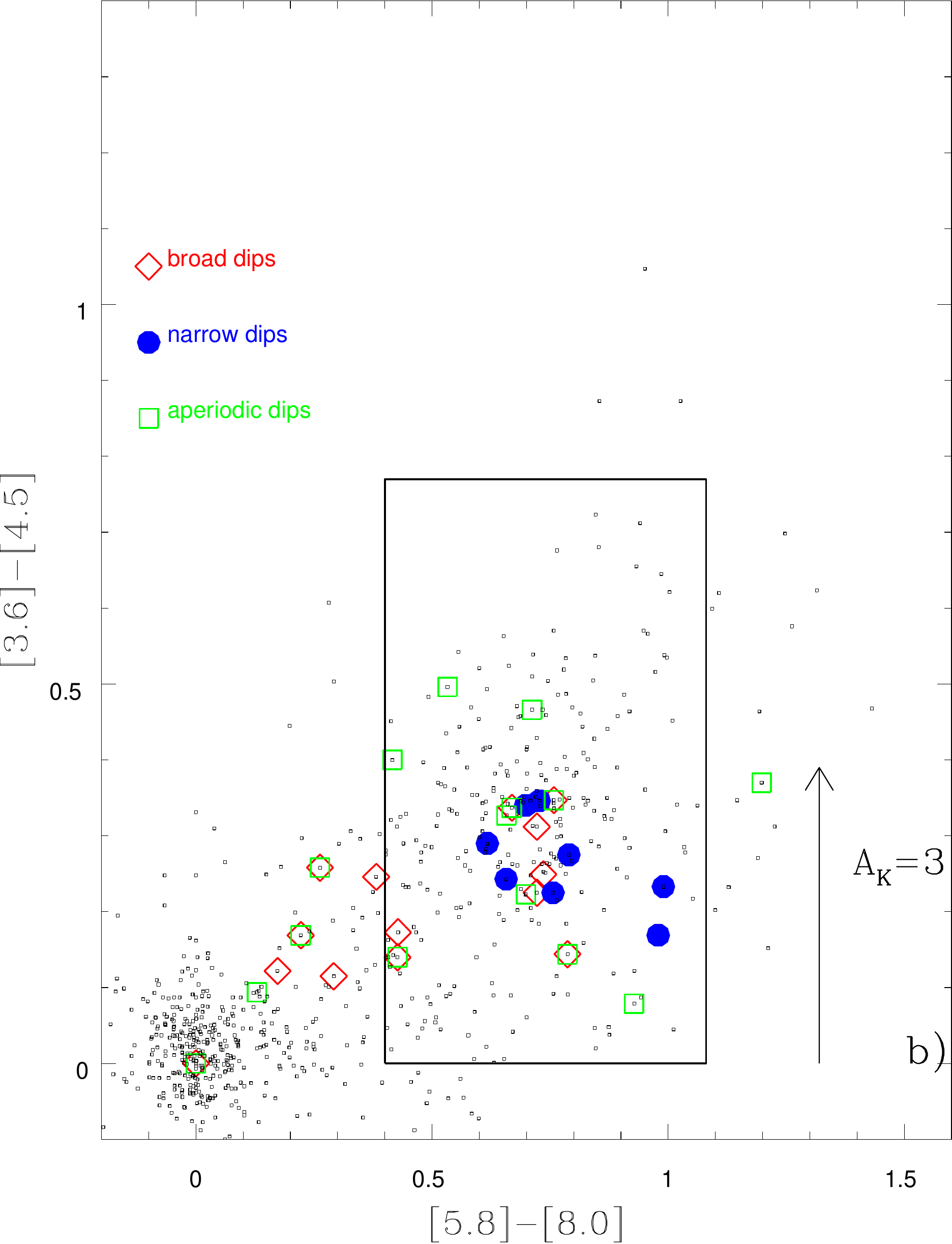}
\end{center}
\caption{Location of the short-duration flux dip class members in two
diagrams sensitive to on-going accretion and warm dust.  The dashed
line in the left-hand plot separates stars that are strongly accreting
(those below the line) from those with weak or no on-going accretion
(Venuti et al.\ 2014).  The box in  the right-hand plot is the locus of
Class~II YSOs from Allen et al.\ (2004).    Class III YSOs have colors
near 0,0 in this diagram. The small black dots are all NGC 2264
members and candidate members, regardless of whether we have {\em
CoRoT} light curves or not.   \label{fig:colorcolor_IR}}
\end{figure*}

Figure~\ref{fig:colorcolor_IR}a shows the $u-g$ vs. $g-r$ color-color
diagram (Venuti et al.\ 2014; Stauffer et al.\ 2014), discussed
extensively in Venuti et al.\ (2014) as an accretion diagnostic.  YSOs
lacking active accretion (WTTS) form a roughly inverted V-shaped locus
in this plane, with late F stars located near $g-r$ = 0.6, $u-g$ = 1.6
and later spectral types becoming redder until about M0 at $g-r$ =
1.4, $u-g$ = 2.8. Later type M dwarfs maintain $g-r$ $\sim$ 1.4 but
have bluer $u-g$ colors for later M subtypes, presumably due to active
chromospheres.   Accretion hot spots primarily affect the $u-g$
colors, displacing the star towards the bottom of the plot (bluer
$u-g$ colors).  The diagram shows that most of the variable extinction
stars have weak UV excesses.  The short duration flux dip stars have
locations in this diagram indistinguishable from the members of the
two other variable extinction light curve classes.

Figure~\ref{fig:colorcolor_IR}b provides the IRAC color-color diagram,
which allows one to separate stars with IR SEDs characteristic of
Class I, II and III YSOs (Allen et al.\ 2004).  The diagram shows that
a significant fraction of the AA Tau analogs and aperiodic flux-dip
stars have IR colors intermediate between Class~II and III, whereas
the short duration flux dip sample  all have relatively red colors
consistent with the Class~II locus.  The red IR colors imply that warm
dust near the inner disk wall often provides the majority of the flux
in the IRAC bands (see column 7 of Table~\ref{tab:basicinformation}). 
That, plus the expected lower selective extinction at 4 $\mu$m
compared to the optical, may help explain why, when we do have
simultaneous optical and IR light curves, we usually  see only hints
of the flux dips in the IRAC data.

\section{Towards A Physical Explanation for the Short-Duration Flux Dips}

In the previous sections, we have attempted to establish quantitative
data characterizing the short duration flux dips and the stars whose
light curves exhibit these features.  We summarize these properties:

\begin{itemize}

\item  About 5\% (8 of 159) of the CTTS in NGC~2264 show this type of
light curve (periodic, short-duration flux dips) at least some of the
time.  After removing two objects we classify here as short-duration
flux dip stars, McGinnis et al.\ (2015) consider 31 of 159 of the CTTS
with CoRoT light curves as either AA Tau analogs or aperiodic
extinctors, corresponding to a 19\% occurrence rate.

\item Another nine CTTS show one or a few short-duration flux dips in
their CoRoT light curves.   Therefore, the total fraction of CTTS with
some evidence for flux dips due to dust structures intersecting our
line of sight is 48/159 or 30\%.

\item For the stars of Table ~\ref{tab:basicinformation},  the flux
dip durations are typically  $\sigma<$ 0.25 days, with a range
from about 0.10 to 0.4 days, based on Gaussian fits.

\item Many of the flux dips are well-fit by single Gaussian profiles,
 or in some cases, blends of 2 or 3 Gaussians.

\item  The depths are never large -- the largest we have measured is
      of order 25\%; most are only of order 10\%.

\item  The periods exhibited range from about 3 to about 10 days.  In
some cases, we also can derive stellar rotation periods from
superposed semi-sinusoidal flux variations in the light curves, and
when we do so, the periods are equal or nearly equal to the dip
periods.

\item  There are significant changes in both depth and shape of the
dips from epoch to epoch; in most cases the widths are more constant
than the depths.

\item  The flux dips persist for typically many orbital periods.  In
  at least two cases, periodic narrow dips are present in both 2008
  and 2011.

\item  In three cases (Mon-21 in 2008, first and last dips; Mon-56,
several dips; and Mon-1165, all dips), there are paired flux dips with
the dips separated well enough so as not to be ``blended". When this
occurs, the separation in time between the two dip centroids stays
approximately constant.

\item  Where we have estimates of the inclination of the disk to our
line of sight, that inclination indicates that we usually view the
disk close to edge on.

\item  Compared to the other CTTS in NGC~2264 with variable extinction
light curves, those with short duration flux dips are, on average,
later in spectral type, with the majority being early to mid M dwarfs.

\item  In the IRAC color-color diagram, the short-duration flux dip
stars have colors which place them in the middle of the Class II
distribution, whereas the classic AA Tau and aperiodic flux dip stars
often have mid-IR colors in the transition region between Class II and
Class III.

\item  A small number of stars show both broad (AA Tau-like) and
narrow periodic flux dips; another small set show one or two narrow
flux dips, with no evidence of periodicity.

\item  For the very small number of events where we have sufficient
coverage at both optical and IR wavelengths, the ratio of the flux dip
depth in the optical to that at 4 $\mu$m indicates that either the
occulting structures are optically thick, or if the ``cloud" is
optically thin, then the dust grain mixture must be very depleted of
small grains compared to the standard ISM composition.

\end{itemize}

\subsection{Inferred Characteristics Which Any Model Must Incorporate}

We assume that any physical model to explain the short duration flux
dips will have as its starting point that the dips are due to our line
of sight being intercepted by structures in or near the inner
circumstellar disk that rotate through and occult our line of sight.  
A fundamental parameter of such structures is their physical size.  We
can estimate the average size scale involved from the parameters in
Table~\ref{tab:basicinformation} and
Figure~\ref{fig:compare_dippers}.   From
Figure~\ref{fig:compare_dippers}, the average value of the ratio of
the dip FWHM and the period for the short-duration flux dips is about
0.15.  Assuming the dips are due to material orbiting the star, that
ratio should correspond approximately to the angular size of the
structure as seen from the star, in this case indicating an angular
size of about 0.95 radians.  The average co-rotation radius for the
stars in Table~\ref{tab:basicinformation} is 11.3 R$_{\odot}$.  
Assuming that the dust structures are located at or near the
co-rotation radius, the average physical size of the structures would
then be about 11 R$_{\odot}$.  For our stars in
Table~\ref{tab:basicinformation}, the stellar radius R$_{star}$ $\sim$
1.5 R$_{\odot}$, and hence the physical size of the structure
corresponds to about 7 R$_{star}$, or to about 0.05 AU.

Further, the shallowness of the dips -- usually just 5-10\% --
implies  either that only a small part of the star is being occulted,
or that the average optical depth through the dust structure is
small.  The lack of a flat bottom to the flux dips requires that the
dust structure not be very small relative to the star, nor much bigger
than the star (completely occulting it) and homogenous.

Finally, we emphasize again the frequent occurrence of flux dips whose
shapes are approximately Gaussian.   In combination with the preceding
two points, this characteristic places significant constraints on the
geometric structure of the occulting object.

With these considerations in mind, we offer four possible physical
models which could produce short duration, periodic flux dips in the
light curves of CTTS.

\subsection{Model 1: The Short-Duration Periodic Flux Dips Fall within the Standard
   Warped Disk Model for AA Tau Light Curves}

Classical AA Tau light curves usually show considerable structure
within their broad flux dips, which is often highly variable from
epoch to epoch. These characteristics presumably attest to physical
structures in or near the inner disk wall whose size, shape, and/or
optical depth also change considerably on the  timescales of the
Keplerian orbital period at that radius.  The duration of the AA Tau
flux dips is set by the fraction of the orbital longitude range where
the upper edge of the inner wall intersects our line of sight.  Even
if the height of the upper lip of the disk wall varies smoothly with
longitude, there will be certain lines of sight where the longitudinal
range which intersects our line of sight is small, resulting in a
short-duration, periodic flux dip.  More probably, the height of the
inner disk wall will vary irregularly both in time and in longitude,
with the highest points being  potential sources for  short-duration
flux dips given an appropriate view angle to the disk.  

\begin{figure*}
\begin{center}
\epsfxsize=.99\columnwidth
\epsfbox{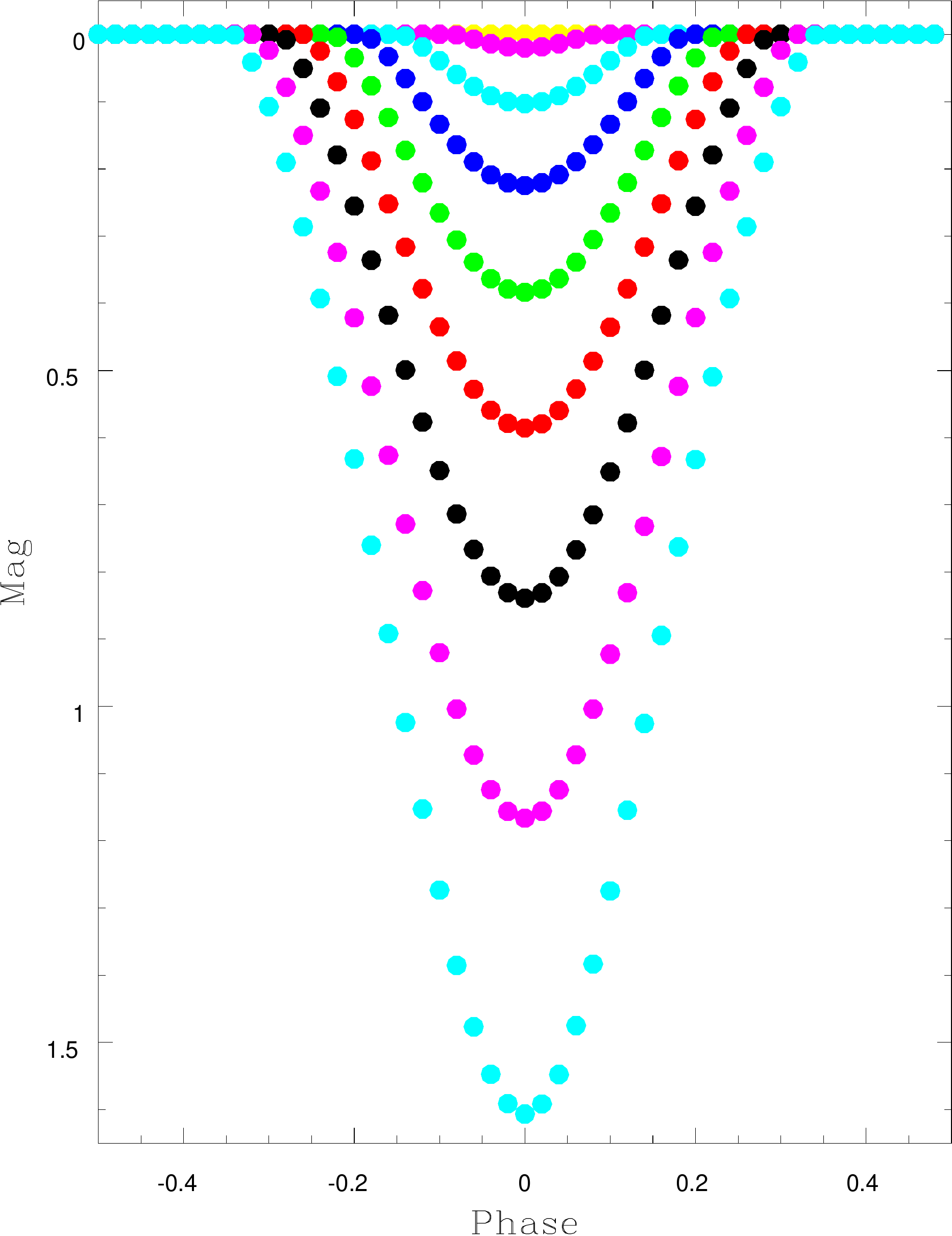}
\end{center}
\caption{Model flux dips for a disk warp of fixed height and extent,
where the inclination of our line of sight is varied from 65 (yellow) 
to 75 (light blue, at bottom)
degrees in steps of 1.25 degrees.  The model used is that from Bouvier
et al.\ (1999), with the disk parameters that were appropriate to AA
Tau -- $h_{max}$ = 2.7 $R_{\rm star}$, inner disk radius $R_i$ = 8.8
$R_{\rm star}$, and $\phi = \pm 180 \arcdeg$.
\label{fig:model_width_depth}}
\end{figure*}

Arguments in favor of this interpretation of the data are that it is
the simplest explanation, and it can fairly naturally explain stars
with periodic flux dips having a range of (dip FWHM)/period. The
simple mathematical formula proposed by Bouvier et al.\ (1999) results
naturally in there being a width-depth correlation for flux dips
produced by disk warps of varying inclination to our line of sight --
as illustrated in Figure~\ref{fig:model_width_depth}.   Given the
Bouvier et al.\ formula to describe the disk warp for AA Tau, flux dips
with maximum depths $>$ 0.3 magnitudes and FWHM/Period $>$ 0.25
(classic AA Taus)  would be predicted to be somewhat more common than
those with shallow but detectable depths, again as observed.   The
simple model also yields dip shapes that are reasonably close to
Gaussian.

In detail, however, some aspects of the data do not seem to fit with
this simple model.  The distribution of variable extinction stars in 
Figure~\ref{fig:compare_dippers}  is somewhat bimodal, with a deficit
of objects with FWHM/Period $\sim$ 0.3 and Depth $\sim$ 0.3, perhaps
more suggestive of two populations rather than a single population.  
The tendency for the classic AA Taus to have earlier spectral types
and to be bluer and brighter in the optical CMD argue against there
being just a geometrical (view-angle) difference between the classic
AA Taus and the short-duration flux dip sample.  Instead, there could
be a mass or temperature effect.  For example, this could be explained
by differing magnetic field geometries as a function of mass, over the
K1 to M4 range in spectral types for our stars. Also, while the simple
geometrical model can produce Gaussian flux dip shapes, the actual
flux dip shapes for the classic AA Taus are highly structured,
requiring the material in the disk warp to also have a highly
non-uniform and variable distribution.   If the disk warp model were
adjusted to yield structured dips to match the classic AA Taus, it is
not obvious it would yield predicted narrow flux dips that would often
still be well-approximated as Gaussian in shape.

\subsection{Model 2: Flux Dips Produced by the Footpoint of A Funnel-Flow Accretion
   Column or by Inhomogeneities in a Dusty Disk Wind}

Romanova et al.\ (2004; 2008) predict that for YSOs with relatively
low  accretion rates, the accretion may occur in stable, relatively
long-lived ``funnel flows".  Typically, two tongues of gas arise from
the inner disk, separated by about 180$\arcdeg$.  The gas originates
from a relatively broad range in longitude and then gathers together
into a narrower stream (it ``funnels").  Some empirical evidence for
these types of accretion curtains has been provided through analysis
of CTTS emission line profiles and their variability (Symington et
al.\ 2005; Bouvier et al.\ 2006).     If the gas flow originates from
the region near the dust sublimation radius, and that radius is beyond
the co-rotation radius, then it will start with a lower angular
rotation rate than the star and a trailing spiral pattern will result,
as illustrated in  Figure 1 of Romanova \& Kurosawa (2014).   Dust
could be entrained in this flow and survive briefly as the gas/dust
mixture rises above the disk plane (though the dust will soon
sublimate as it flows towards the star).    Alternatively, Bans \&
K\"onigl (2012) have advocated that disk winds originating from beyond
the dust sublimation radius may levitate sufficient dust to contribute
a large fraction of the 3 to 4 micron flux in YSOs, and if that wind
were significantly non-axisymmetric it could be the source of
potentially short-duration flux dips.

\begin{figure*}
\begin{center}
\epsscale{1.0}
\plottwo{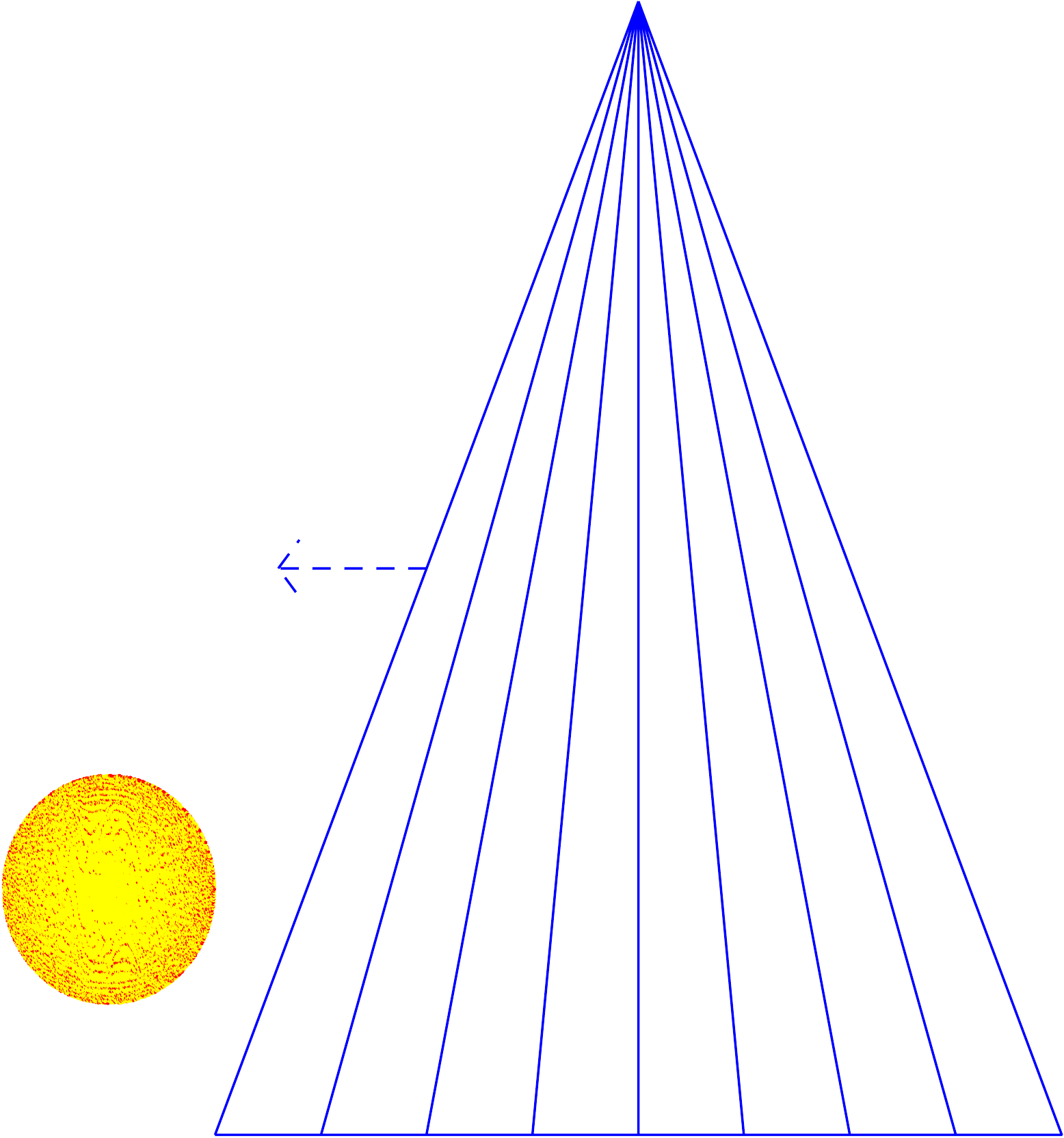}{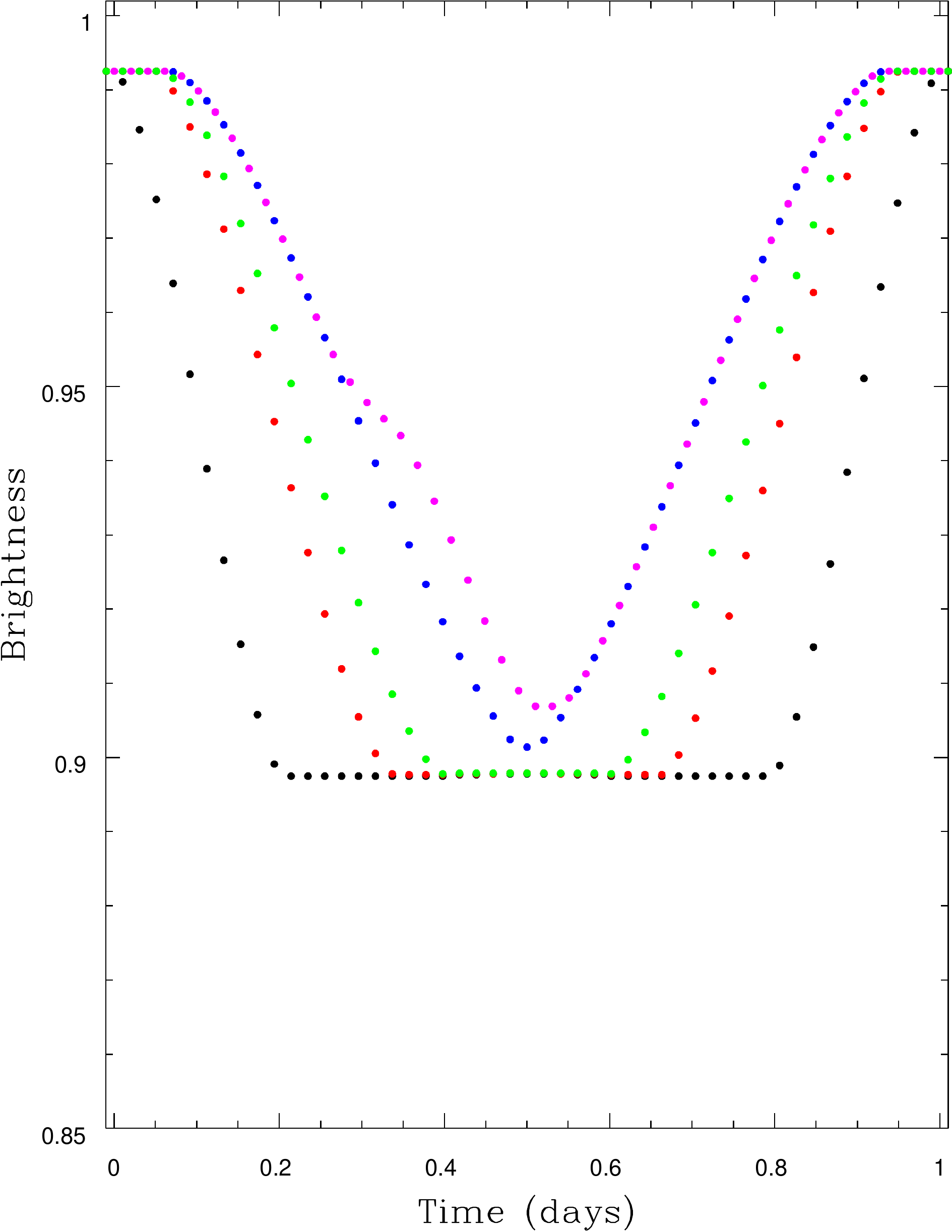}
\end{center}
\caption{ Illustration of the geometry used by the toy model (left panel), and
light curves produced by allowing orbital motion to pass the model
accretion curtain through our line of sight to the star (right panel).  
The curves
plotted are for convergent heights of 20 $R_{\rm star}$, 2 $R_{\rm
star}$, 1.5 $R_{\rm star}$, and 1 $R_{\rm star}$.   The top (magenta)
curve is also for a convergent height of 1 $R_{\rm star}$, but with
one panel of the curtain set to half the opacity of the other panels.
\label{fig:toy_funnel}}
\end{figure*}

We have constructed a toy model of an accretion curtain in order to
illustrate constraints that the narrow dip profiles place on such a
curtain.  We tune the model to the flux dips in Mon-21.  Specifically,
the model must produce approximately Gaussian shaped flux dips with
$\sigma \sim$ 0.2 days (FWHM $\sim$ 0.45 days) for a system with
$R_{\rm star}$\ = 1.7 $R_{\odot}$, $P$ = 3.2 days, and $R_{co}$ = 9.7
$R_{\odot}$. We place the curtain at the co-rotation radius.  Based on
the geometrical arguments from \S  7.1, we adopt a curtain width of 8
$R_{\rm star}$.   We divide the curtain into 8 vertical panels, each
of base width 1 $R_{\rm star}$ (the left panel of Figure~\ref{fig:toy_funnel}
illustrates the geometry of the toy model).   The panels are actually
triangles, all meeting at a common apex at height $y_{\rm convergent}$
above the midpoint of the curtain.  For very large $y_{\rm
convergent}$, the panel is essentially a vertical wall of width 8
$R_{\rm star}$.   For $y_{\rm convergent}$ = 8 $R_{\rm star}$, the
curtain becomes an equilateral triangle.  We adopt a uniform optical
depth of 0.1 for dust within the curtain.   Finally, we place the
bottom of the curtain so that it is aligned with the bottom of the
star as viewed from our vantage point.  We place the edge of the
curtain near the leading limb of the star, and let orbital motion
(with the Mon-21 period) carry the curtain in front of the star.  

With the above idealized geometry, we wish to determine if flux dips
like that observed can be reproduced - and if so, what range of
convergent point heights is allowed?  The right panel of Figure~\ref{fig:toy_funnel}
shows the flux dips produced for values of the convergent height from
20 $R_{\rm star}$ (bottom) to 1 $R_{\rm star}$ (upper blue curve).  
The model with a convergent height of 1 $R_{\rm star}$ produces an
approximately Gaussian shaped profile whose FWHM $\sim$ 0.4 days,
within the range of dip widths observed for Mon-21.  These results
emphasize two things: (a) in order to produce a flux dip as broad as
observed, an accretion curtain must be quite wide (several times the
stellar diameter) at its base;  and (b) in order to produce the
Gaussian shape, assuming a uniform dust opacity, the curtain must
``funnel" down to a much smaller region over a very small scale
height.  The alternative to this would be to assume instead that the
curtain has an  opacity distribution smoothly varying with longitude,
peaking near its midpoint.

What types of limits on opacity variations within the curtain could we
detect? We constructed one model, for convergent height of 1
$R_{\rm star}$, where we set the opacity of one of the panels to half the
normal value.  The top model flux dip in Figure~\ref{fig:toy_funnel}b
shows this result.  For many of the flux dips illustrated in this
paper, the observed dip shape is more symmetric than this model and we
could exclude an opacity fluctuation of this order.  However, some of
the flux dips we have shown do have asymmetries which could perhaps be
indicative of opacity fluctuations (rather than as a blend of two
Gaussian waveforms).

One criticism of this model is that it could simply be regarded as a
reformulation of the first model.  That is, the tip of a warped disk
is the place where one might expect an accretion curtain to form, and
whether one labels the dust structure there as part of the disk or the
base of the accretion flow is, perhaps, only semantics.  One reason to
keep the two ideas separate is that if the line of sight inclinations
to Mon-21 and especially Mon-1131 (Table~\ref{tab:incl-estim}) are
confirmed  at a value near our current estimates ($i \sim$ 40 and 27
degrees, respectively), this could not be accomodated within a warped
disk model but could be compatible with the accretion curtain model if
dust could survive in the accretion flow to heights of several stellar
radii above the disk plane. In this case, the duration of the flux dip
would be set by the inclination of the accretion flow relative to the
orbital axis.

The, on average, later spectral types and dustier disks for the
short-duration flux dip stars are not an obvious feature of the model
but could possibly be made to fit.   The lower luminosity of the star
and the dustier disk could correspond to allowing the base of the
accretion column to be located beyond the dust sublimation radius,
whereas the higher luminosity, less dusty disks of the classical AA
Taus might have their accretion columns be dust free.  Alternatively,
the less dusty nature of the AA Tau group's disks may simply reflect
their higher stellar mass and the expected dependence of disk
evolution timescales on stellar mass (longer timescales for lower mass
stars).

\subsection{Model 3: The Short-Duration Periodic Flux Dips Originate
  from Dust Associated with Vortex Instabilities Located Near the
  Inner Disk Rim}

The short durations and small amplitudes of the flux dips could arise
from mountains on the surface of the disk -- that is, features
standing up above their surroundings.  Such features would either be
low-relief, sweeping across only the bottom of the star's face but
optically thick, or taller, but diffuse and of moderate optical depth.
Mountains arising from the disk would naturally cross the star once
per local orbit, while varying from one crossing to the next due to
the internal turbulence.  The issue is how to lift the mountains above
the plains.

One way to produce vertical relief on the disk surface is through
vortices driven by Rossby wave or baroclinic instabilities.
Sub-millimeter and millimeter interferometric imaging of nearby
transition-disk YSOs has revealed several stars with large azimuthal
asymmetries in surface brightness at separations from the star of tens
of AU (van der Marel et al.\ 2013; Isella et al.\ 2013; Perez et al.\
2014).  The asymmetries are interpreted as millimeter-sized particles
concentrated in long-lived hydrodynamical vortices rotating in the
disk plane (Lyra \& Mac Low 2012; Lyra \& Lin 2013).  Similar vortices
occurring within a fraction of an AU of the star might account for the
periodic flux dips we observe.  The vortices have higher surface
densities than their surroundings and thus could increase the column
along lines of sight to the star in some range of viewing angles. 
However to account for all the narrow dips, the vortices would need to
obscure the star in 5-10\% of all systems, and thus to increase the
height of the starlight unit-optical-depth surface near the dust
sublimation front by 5-10\% of the radius -- or more if not every disk
has a vortex near the relevant radius.

In the existing hydrodynamical models, the surface density enhancements
at the vortices are about a factor of two (Lyra et al.\ 2009; Meheut et
al.\ 2010; Lyra \& Mac Low 2012; Meheut et al.\ 2012).  We construct a
model consisting of a star of 0.5 $M_\odot$ and 2 $R_\odot$, with
effective temperature 4000~K.  Orbiting the star is a disk accreting
at $10^{-8} M_\odot$~yr$^{-1}$.  The disk's surface density is
1000~g~cm$^{-2}$ at 0.1~AU and falls off inversely with the radius.
An inner edge is created by rolling off the surface density inside
0.1~AU as a Gaussian with half-width 0.015~AU.  The disk's outer edge
is defined by an exponential cutoff with radial scale 40~AU.  To
prevent the outer disk from obscuring the star before the inner edge
as the system is tilted over to higher inclinations, we give the outer
disk a lower dust abundance.  The dust-to-gas mass ratio is 1\% inside
0.2~AU and falls off as a Gaussian with half-width 0.2~AU toward
dust-to-gas mass ratio $10^{-4}$.  The disk material's opacity comes
from well-mixed compact spherical grains chosen to match the observed
spectra of protostellar sources (Preibisch et al.\ 1993).  We place a
vortex at 0.1~AU, giving it a surface density contrast of two and
profiles in the radial and azimuthal directions that are Gaussians
with full-widths at half maximum of one-quarter the radius and
$60\arcdeg$, respectively.

We translate the disk's surface densities into volume densities using
a temperature profile taken from the midplane of a Monte Carlo
radiative transfer calculation in a similar but axially-symmetric
density distribution, lacking the vortex.  This places the model disk
in approximate hydrostatic equilibrium.  We then carry out a 3-D
transfer calculation to obtain the distributions of temperature and
radiation intensity in the disk with vortex, using the Bjorkman \&
Wood (2001) algorithm and Lucy's (1999) method as implemented by
Turner et al.\ (2012).  Isotropic scattering is included.  We solve the
transfer equation on a grid of rays (Yorke 1986) including the
scattering and thermal emission source terms, to construct synthetic
images of the system as viewed at different points around the vortex's
orbit.  We conduct photometry on the images and assemble them into a
lightcurve.  The resulting V-band photometric amplitude is 1\% or less
at all viewing angles.  The model vortex thus has too little vertical
relief to explain the dips' amplitudes.

To account for the possibility that vortices can eventually reach
contrasts up to an order of magnitude relative to their surroundings,
as suggested by long-duration 2-D hydrodynamical calculations (Regaly
et al. 2012), we carry out a further run in which the surface density
contrast is increased to ten.  The V-band amplitude exceeds 3\% when
the system is inclined between 81 and 84$\arcdeg$, peaking at 7\% for
inclination 83$\arcdeg$.  Therefore, even this seemingly extreme
vortex  cannot easily explain either the full range of the narrow dip
amplitudes or the number of objects in which they occur.  However,
vortices concentrate  pebble- to boulder-sized rubble (Barge \&
Sommeria 1995), and collisions between the rubble particles could
yield fragments that raise the abundance of sub-micron dust even more
than by a factor of 10. Determining whether the increase makes
vortices a viable explanation for the narrow flux dips will require
better models of the dust sources, sinks and transport.

\subsection{Model 4: Flux Dips Produced by A Spiral Arm in the
  Inner Disk}

A planet embedded in a disk exerts a gravitational force that raises a
spiral arm in the surrounding material.  The pattern co-rotates with
the planet and is strongest near the planet's location (Rafikov 2002).
In particular, a planet orbiting interior to the disk's inner rim
could raise a bump on the rim that rotates with the planet's orbital
period, which is shorter than the period at the rim.  Hot Jupiters
orbiting mature stars cluster near periods 3-5 days and eccentricity
zero, while less massive planets show no such clustering.  Hot gas
giants cannot easily grow in situ, as reviewed by Helled et al.\
(2013).  Instead they probably formed outside the snow line and
reached their present orbits either by migrating through the disks
that gave them birth, or by later dynamical interactions with third
bodies (Baruteau et al.\ 2013).  Detecting planets interacting with
the disks would help separate these possibilities.

\begin{figure*}
\begin{center}
\scalebox{.20}{\includegraphics{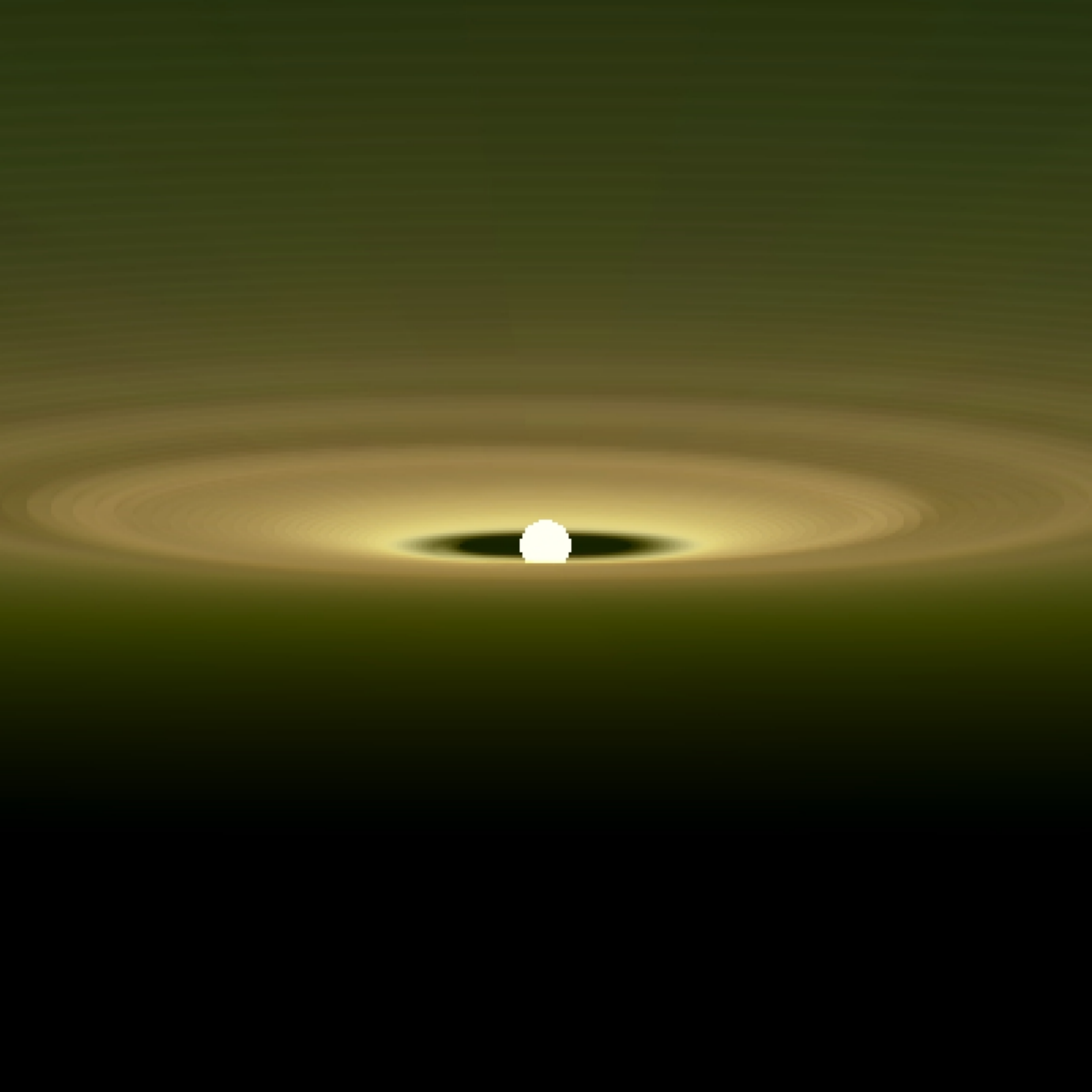}}
\scalebox{.20}{\includegraphics{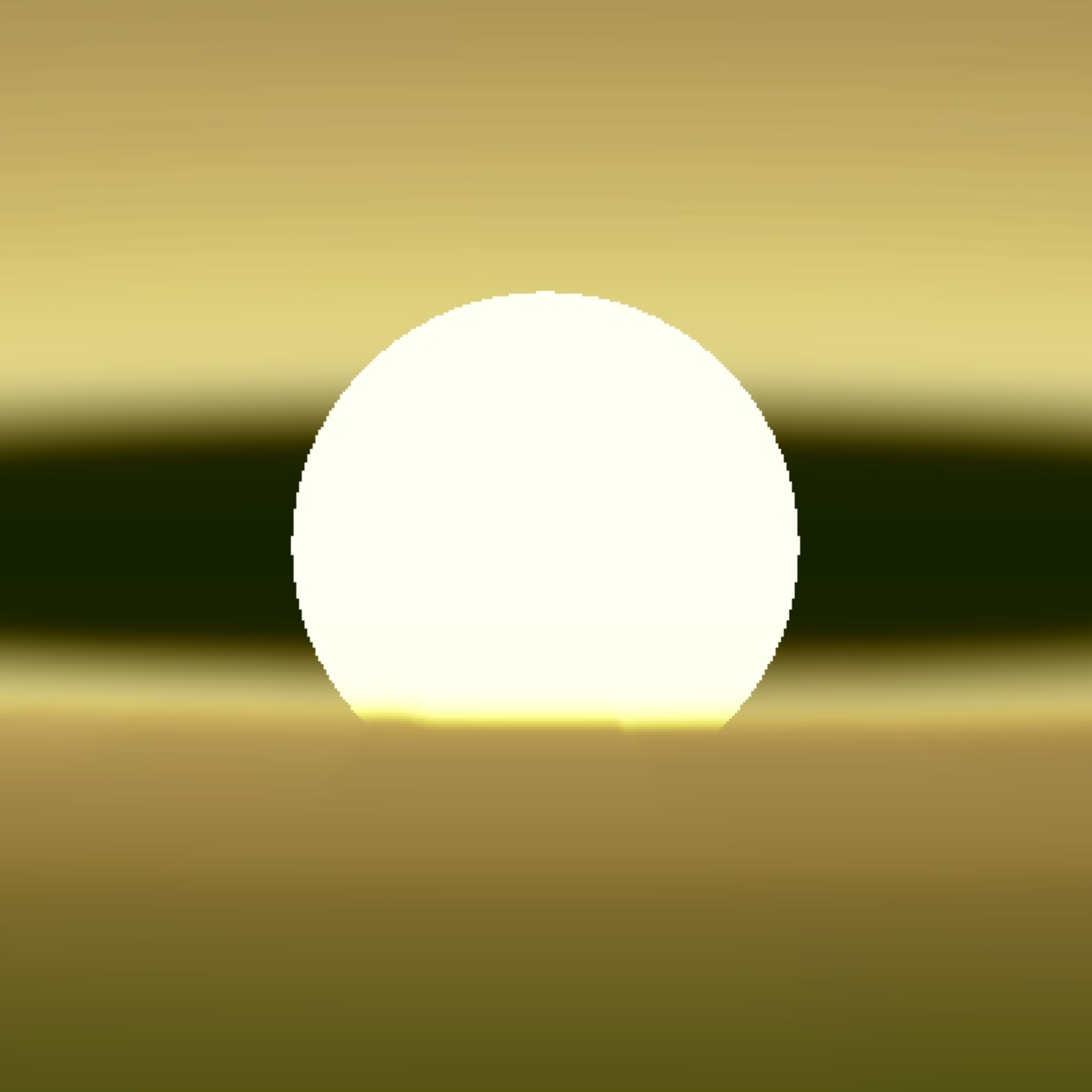}}\\
\scalebox{.44}{\includegraphics[angle=-90,origin=c]{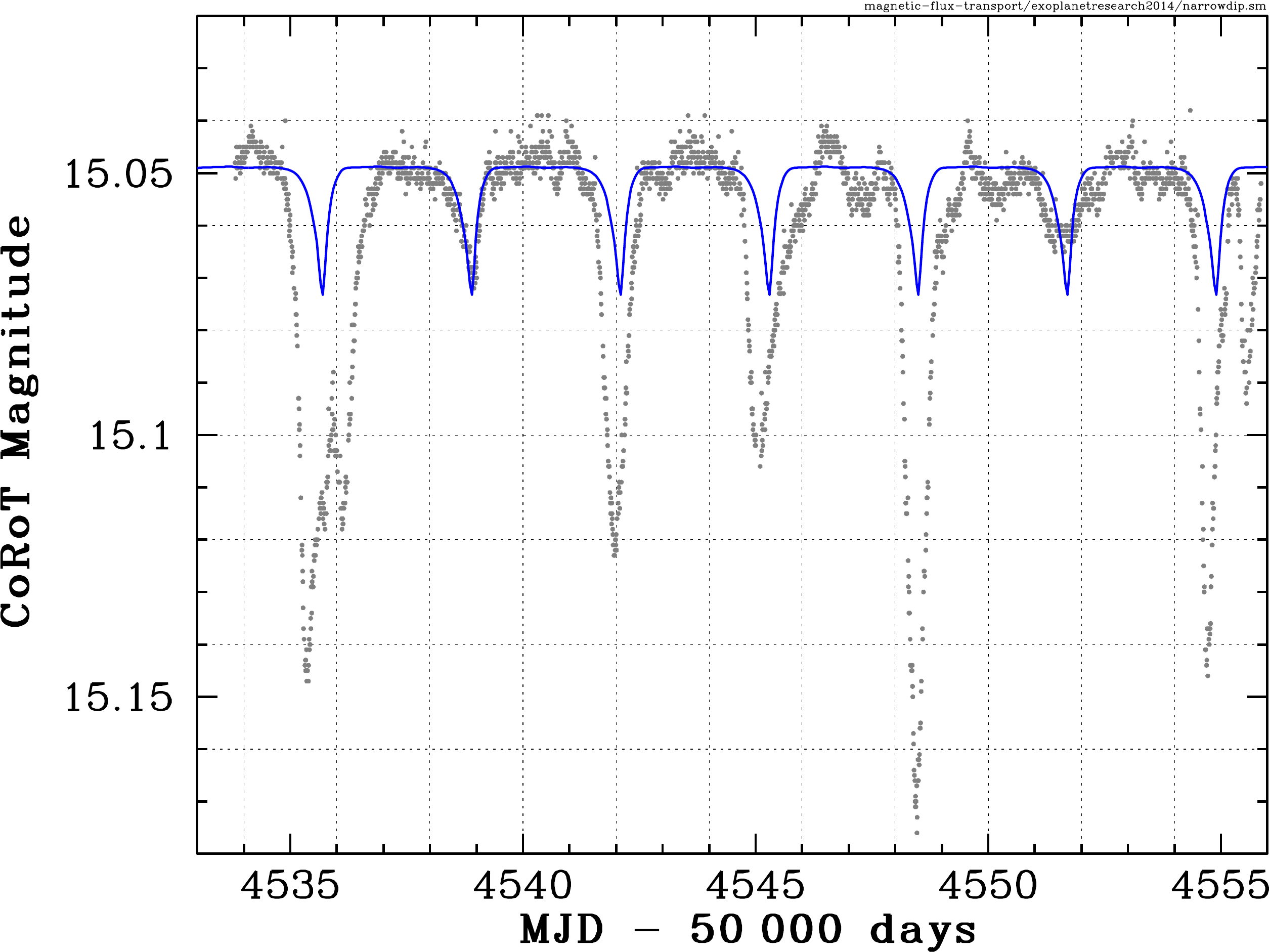}}
\end{center}
\caption{ \label{fig:spiral} Photometric signature of the spiral
  density wave launched in a T~Tauri disk by a planet orbiting at
  0.07~AU near the disk's inner edge, according to a simple
  near-hydrostatic model.  The synthetic images at top have red, green
  and blue channels corresponding to the $V$, $K$ and 3.6 $\mu$m
  bands.  The image at left is 0.4~AU across and the close-up at right
  is magnified ten times.  Both show the spiral wave's peak about to
  cross the star's face.  The system is $82\arcdeg$ from face-on.  At
  bottom is the $V$-band lightcurve (blue curve) overlaid on the
  starspot-subtracted {\em CoRoT} lightcurve of Mon-21 (gray points).  The
  synthetic lightcurve's period has been adjusted to match the
  observed 3.2-day dip interval, but the amplitude has been left
  unscaled.  }
\end{figure*}

To test whether planet-disk interactions could cause the narrow dips,
we construct a model using the same 0.5 $M_\odot$ star and disk as in
the vortex picture above.  The vortex is replaced by a 5 $M_{\earth}$
planet orbiting at 0.07~AU, which raises a spiral overdensity in the
disk following the shape of Muto et al.\ (2012) eq.~1, which is
derived from a linear analysis by Rafikov (2002) and therefore is
strictly valid only for planets with masses less than about 1 $M_\earth$.
We build the density and temperature distribution using the same
procedure as for the vortex model, and again conduct synthetic
observations using the radiative transfer code.  The results are shown
in Figure~\ref{fig:spiral}, where the head of the spiral arm passing
in front of the star produces a narrow dip with optical amplitude 2.3\%,
in the range observed among the stars in NGC~2264.  A more massive
disk or higher dust abundance would likely yield larger amplitudes.
Also, the spiral arm's star-facing side is hotter than the
axisymmetric model used to set up the approximate hydrostatic
equilibrium, so that restoring the balance would make the atmosphere
expand, increasing the dip amplitude.  However whether there is time
to reach hydrostatic equilibrium is unclear, since the spiral arm
propagates around the disk at the planet's orbital frequency.  A more
serious difficulty with this picture is that it yields dips in less
than 5\% of all randomly-oriented systems, even if every star has a
similar hot planet.  The planet-driven bump simply does not extend
high enough above the surrounding disk to obscure the star over a wide
range of viewing angles.  A more massive planet could yield a taller
bump, but accurate modeling would require non-linear numerical
hydrodynamical calculations.  If the amplitude can be made large
enough, then among the four pictures we have considered, the
planet-raised spiral arms could best explain any individual
system in which the dips are regularly-spaced and repeat more often
than the orbital period at the dust sublimation radius.  However, unless
hot Jupiters are much more frequent in NGC~2264 than amongst field
stars, this mechanism could not be invoked for the majority of the
periodic, narrow, flux dip stars.

\section{Conclusions} 

This paper, and the others in this series (Alencar et al.\ 2010; Cody
et al.\ 2014; Stauffer et al\. 2014; McGinnis et al.\ 2015) demonstrate
how high cadence, long duration, sensitive, photometrically stable
light curves can be used to infer physical properties of YSOs that are
otherwise impossible to investigate.  In this paper, we have
identified a new YSO light curve class characterized by
short-duration, shallow, periodic flux dips. Depending on which
physical mechanism is responsible for producing these short-duration
flux dips, detailed study of this set of stars has the potential to
significantly improve our understanding of the structure of the inner
disks and accretion flows in YSOs and  the role of instabilities and
migrating planets in shaping the inner disks and their dust content.

Regardless of the precise physical mechanism for producing the
short-duration flux dips, if we are correct in assigning the
semi-sinusoidal variations to photospheric spots and the
short-duration flux dips to dust in or near the co-rotation radius,
then our observations for at least Mon-21 (2008), Mon-378, and
Mon-1580  confirm that their stellar rotation rate is locked to the 
Keplerian rotation period at the inner disk.  This has been a 
long-standing prediction of some models for the angular momentum
evolution of YSOs (e.g., K\"onigl 1991; Collier-Cameron and Campbell
1993)  but with generally only indirect evidence to support it.

We offer here one possible interpretation of our data, which attempts
to incorporate as many as possible of the constraints on the physical
mechanisms we have described in Sec.~7.  Assume that extinction events
can be produced both by a warped inner disk and by dust entrained
within an accretion flow of the sort envisioned in the Romanova et
al.\ models.  Assume further that for the youngest, least evolved
disks, our line of sight to the inner disk may be obscured by
optically thick dust in the outer, flared disk.  Adopt the idea
proposed in Keane et al.\ (2014) that transition disks, as well as
having partially cleared inner disks,  also have less flared (or less
dusty) outer disks.  Given those thoughts, the stars with isolated,
narrow flux dips (Figure~\ref{fig:singledip_dips}) could be systems
where dust in an accretion flow has survived to an unusually high
distance above the disk plane.  We see only one or two of these dips
per 40 day light curve because only rarely does the dust survive long
enough to reach a height to intersect our line of sight.  These
systems are, therefore, viewed at somewhat smaller inclinations than
the other variable extinction stars.  The periodic, narrow flux dip
stars are systems viewed at larger inclinations where it is more
likely the dust will generally be present in the accretion flows. 
Finally, the classic AA Tau analogs  are the systems with more evolved
-- less dusty -- outer disks where  we see into the inner disk and
where a warped inner disk can then occult our line of sight.   The
systems with both broad and narrow flux dips have extinction events
produced by dust in both the  warped disk and in the accretion
funnels.  We offer this scenario primarily as a way to focus future
studies.  The main results of this paper are the empirical constraints
on the narrow flux-dip light curves and their host stars provided in
\S 6 and \S 7, which any model of these events must explain.

A general concern with any model for the variable extinction stars is
simply that there seems to be too many of them in NGC~2264. There are
roughly equal numbers of stars with each of the three light curve
types (AA Tau, aperiodic dip, and short-duration dip), with about 39
such stars in all, corresponding to about 25\% of the total number of
CTTS for which we have {\em CoRoT} light curves.  If one includes the
single-dip stars from Figure~\ref{fig:singledip_dips}, then about 30\%
of the CTTS have some evidence for dust clumps within 0.1 AU of the
star intersecting our line of sight.   That seems high, given an
expectation that the inner disk wall height should be of order 15\% of
the disk radius at that location (Dullemond et al.\ 2001; Muzerolle et
al.\ 2003),  even allowing for some YSOs having warped disks.  A
provocative solution to this concern is suggested by the claim from
Klagyivik et al.\ (2013) that there are many more eclipsing binary
members of NGC~2264 than should occur by chance if their axes were
randomly oriented, from which they conclude that the EB axes are
preferentially aligned and that we are near the plane of their orbital
motions.   If this is also true for the angular momentum vectors of
the single stars in NGC~2264, then we would also detect a higher
fraction of variable extinction stars than would be true for a random
orientation of axes.   Other authors have concluded that the
rotational velocities (\vsini\ and periods) of stars in open clusters
are compatible with a random orientation of axes and incompatible with
any strong alignment of axes (e.g., Kraft 1970; Jackson \& Jeffries
2013), so this either argues against such an effect in NGC~2264 or
that NGC~2264 is unusual in this respect.

\acknowledgements{This work is based on observations made with the
{\em Spitzer} Space Telescope, which is operated by the Jet Propulsion
Laboratory, California Institute of Technology, under a contract with
NASA. Support for this work was provided by NASA through an award
issued by JPL/Caltech. This research was carried out in part at the
Jet Propulsion Laboratory, California Institute of Technology, under a
contract with the National Aeronautics and Space Administration and
with the support of the NASA Origins of Solar Systems program via
grant 11-OSS11-0074.  RG gratefully acknowledges funding support from
NASA ADAP grants  NNX11AD14G and NNX13AF08G and Caltech/JPL awards
1373081, 1424329, and  1440160 in support of Spitzer Space Telescope
observing programs. SHPA and PTM acknowledge support from CNPq, CAPES
and Fapemig.  }

{\it{Facility:} \facility{Spitzer (IRAC)}, \facility{CoRoT}, 
\facility{CFHT (MegaCam)}, \facility{VLT (FLAMES)}.}

\section*{Appendix}

\subsection*{Separating the Spotted-Star and Flux Dip Signatures
   in the 2008 Light Curve for Mon-21}

In \S 4.1.1, we used the Mon-21 light curve itself to generate a model
for the portion of the light curve we attribute to spots, and
subtracted that model from the data in order to derive an estimate of
the light curve structure due to variable extinction.  In order to
demonstrate that this procedure is adequate to the task, we offer here
another means to estimate the flux dip light curve.

\begin{figure*}
\begin{center}
\epsfxsize=.99\columnwidth
\epsfbox{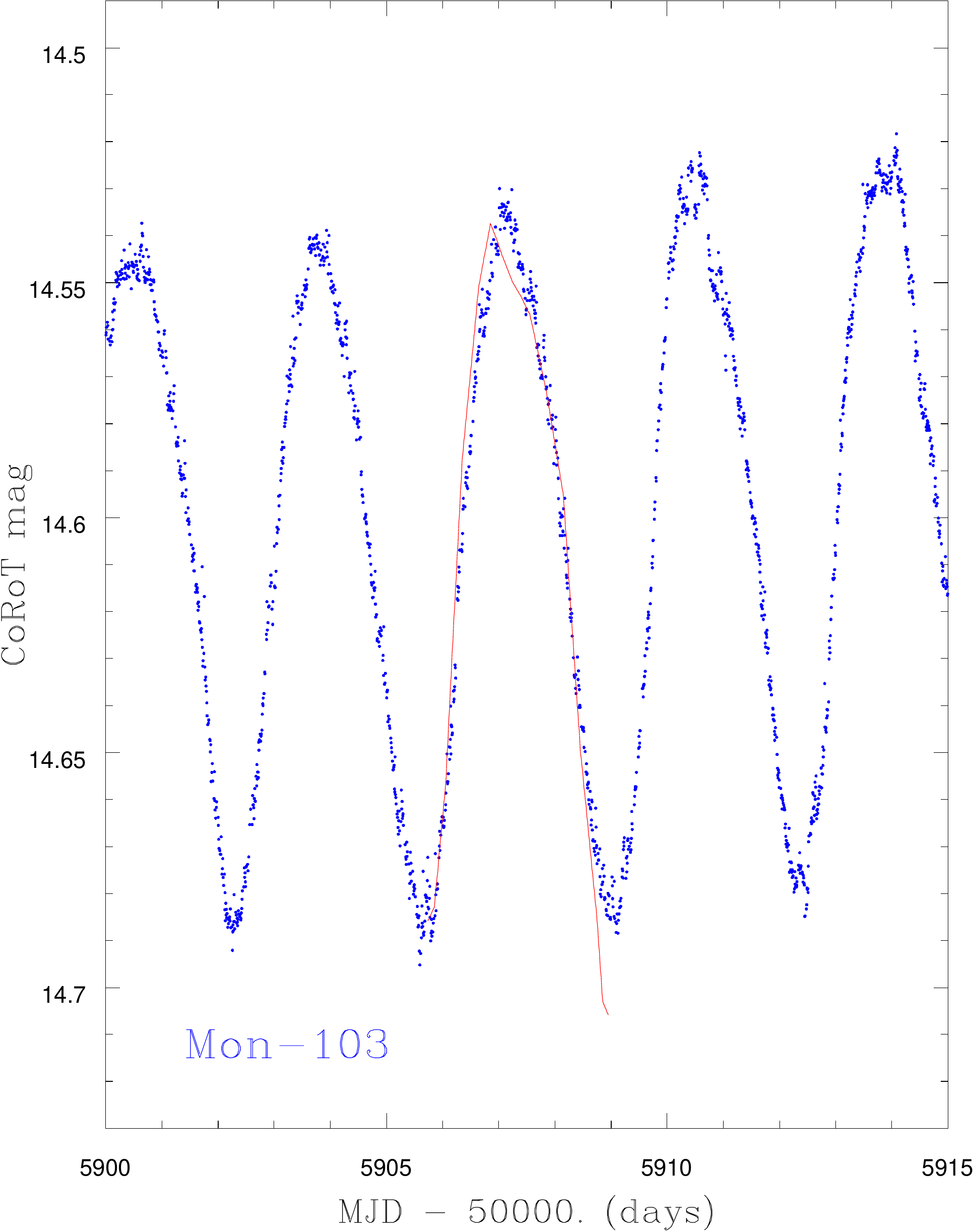}
\end{center}
\caption{{\em CoRoT} 2011 light curve for Mon-103 ({\em CoRoT} SRa01
223980447),  a YSO whose light curve shows a fairly stable, nearly
sinusoidal pattern likely attributable to a relatively high latitude
spot-group.  The red curve is the template waveform derived from the
Mon-21 2008 {\em CoRoT} light curve, as discussed in \S 4.1.1.
\label{fig:Mon103_observed}}
\end{figure*}

All modern studies agree that the dominant physical mechanism which
drives the visual appearance of optical light curves for most WTTS is
the rotational modulation of star-spots that are distributed
asymmetrically over the stellar surface.  When these spots are at
moderately high-latitude, which is typical, the light curve shapes can
be approximately sinusoidal.  Dozens of the NGC~2264 {\em CoRoT} light
curves fall into this category.  Given that large a library, it seemed
reasonable to us that one of those light curves might have a shape
that closely resembled the smoothly varying portion of the Mon-21
light curve.   We believe that is indeed the case.  
Figure~\ref{fig:Mon103_observed} shows the 2011 {\em CoRoT} light
curve of Mon-103, a YSO with spectral type very similar to that for
Mon-21 (K6 vs. K5 for Mon-21).  The period for Mon-103 is just very
slightly greater than for Mon-21 -- 3.3 days vs. 3.15 days.   The red
curve in Figure~\ref{fig:Mon103_observed}  is the template  waveform
for the Mon-21 model, simply shifted in zeropoint to match phase with
Mon-103 and multiplied in amplitude to match also; the waveform shapes
are indeed very similar.

\begin{figure*}
\begin{center}
\epsscale{1.0}
\plottwo{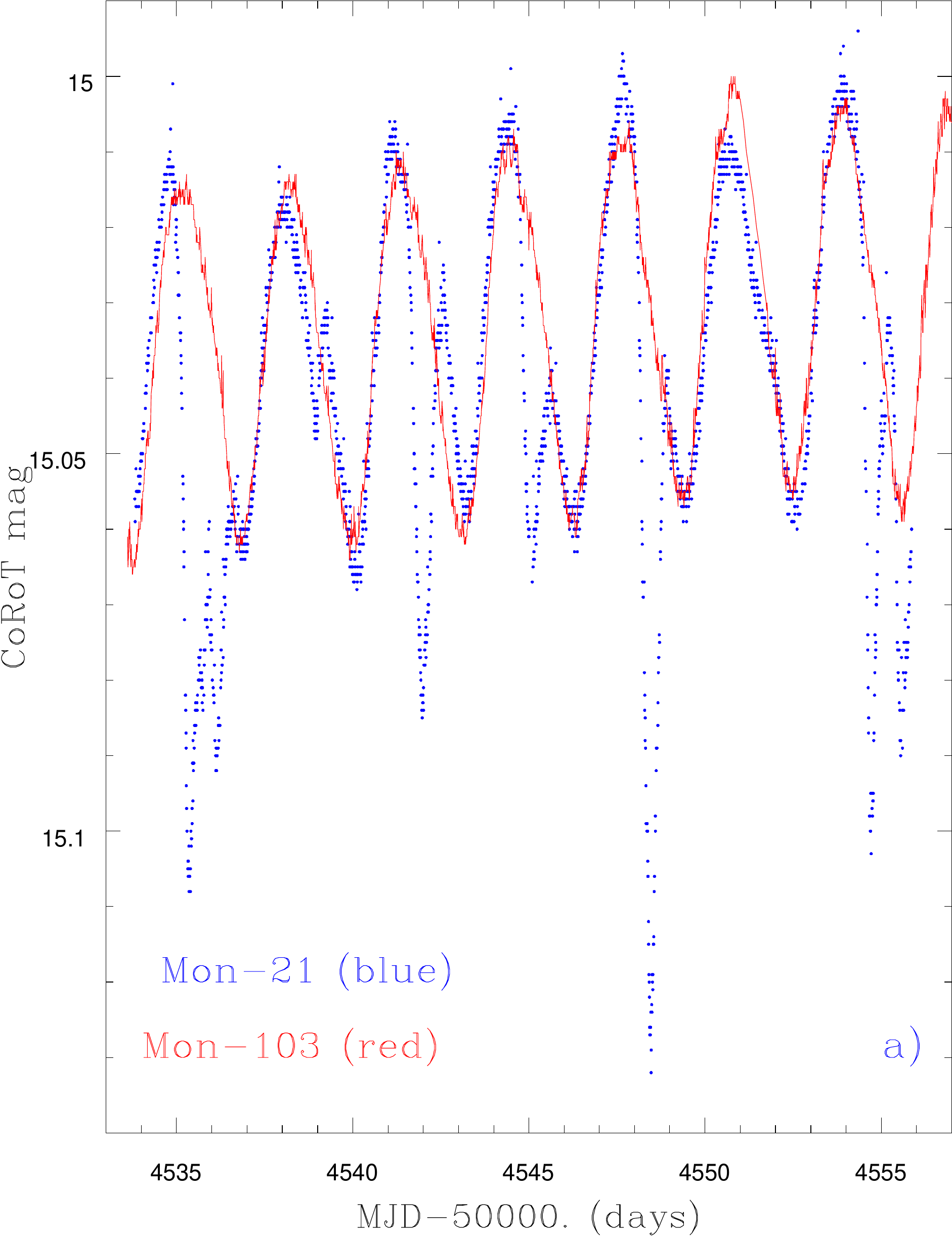}{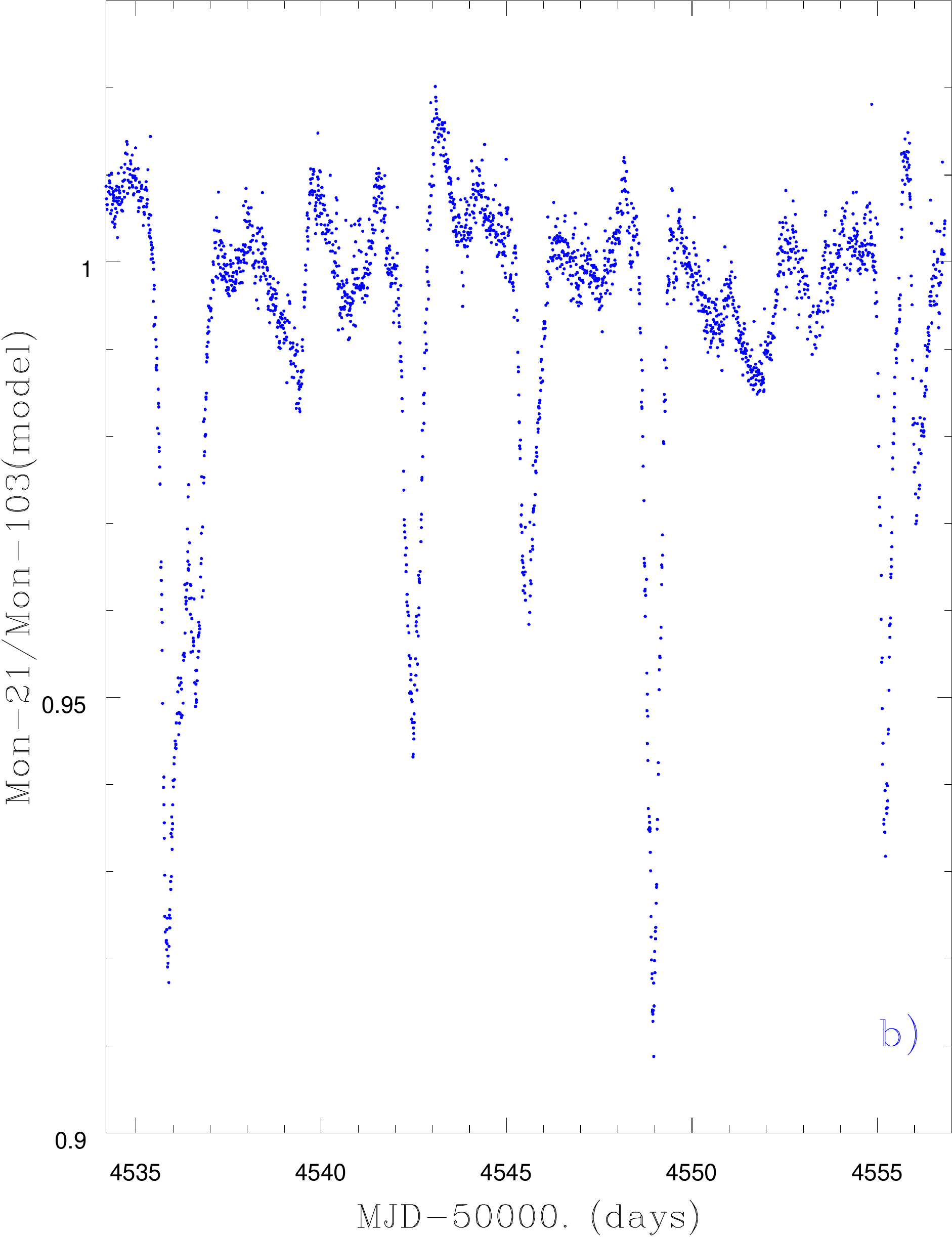}
\end{center}
\caption{a) The as-observed {\em CoRoT} light curve for Mon-21 overlaid by the
2011 light curve for Mon-103, after adjusting the latter light curve by
zero point shifts and linear stretches in both axes; b) Light curve resulting
from subtraction of the modified Mon-103 light curve from the Mon-21
light curve.
\label{fig:Mon21.div.103}}
\end{figure*}

We can use the Mon-103 light curve as a template to remove the
spotted-star component from the Mon-21 light curve.  In order to
provide as good a fit as possible, we make the following slight
modifications to the observed light curve: (a) we do a zero point
shift in time to align the two beginning points in phase; (b) we shift
the Mon-103 light curve counts by a multiplicative factor and a zero
point offset so that the mean magnitude and average amplitude match
that of Mon-21; and (c) we multiply the x-axis timescale by 0.935 in
order to correct for the slight period difference between the two
stars.  Figure~\ref{fig:Mon21.div.103}a overlays the two light curves,
and Figure~\ref{fig:Mon21.div.103}b  shows the Mon-21 flux dip light
curves that results from subtracting the modified Mon-103 light curve
from the observed data.  The main dips appear essentially the same as
in our original procedure (Figure~\ref{fig:Mon21_modela}c); the dip widths and depths
(Table~\ref{tab:quant_information}) differ by less than 10\% between
the two renditions.

\subsection*{Mon-1131 in 2008}

\begin{figure*}
\begin{center}
\epsfxsize=.99\columnwidth
\epsfbox{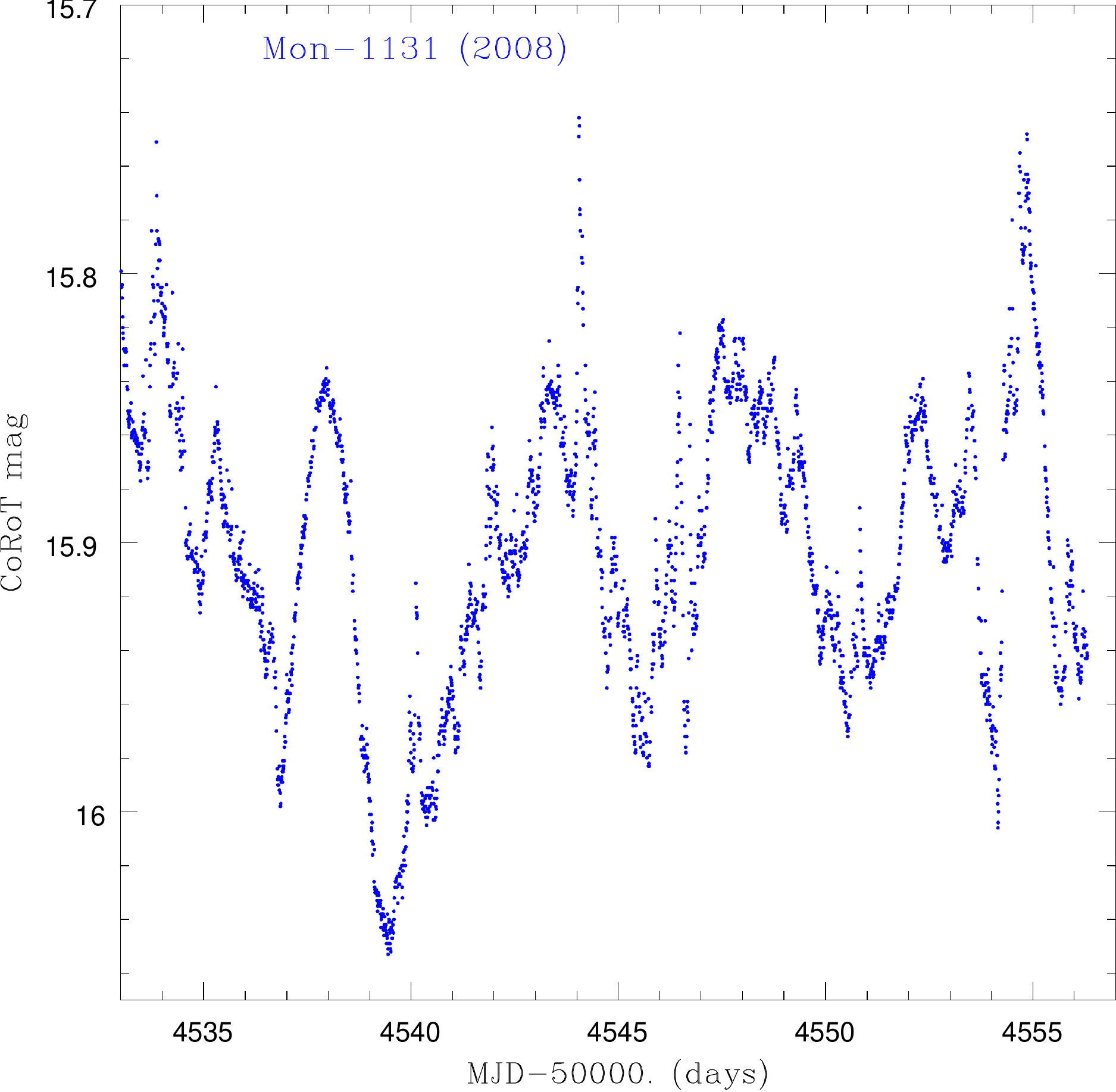}
\end{center}
\caption{{\em CoRoT} light curve for Mon-1131 in 2008.
\label{fig:mon1131_2008}}
\end{figure*}

In 2011, the {\em CoRoT} light curve for Mon-1131 shows flux dips that
recur periodically at P $\sim$ 5.1 day intervals.  The light curve
shows little other structure except the flux dips.  The 2008 light
curve shape is much more complex (see Figure~\ref{fig:mon1131_2008}) 
and probably owes its shape to several physical mechanisms whose
effects are superposed.  Yet, the mean magnitude is nearly the same
during the two epochs, and the full range of the variations is also
nearly the same (about 0.3 mag). The 2008 light curve simply has very
different character.  In terms of the quantitative classification
scheme of Cody et al.\ (2014), the 2008 light curve would be
classified as stochastic (with $M$ = 0.12 and $Q$ = 0.71), meaning the
variations are fairly symmetric about the mean and essentially
aperiodic.  There is still some power in the auto-correlation function
at $P$ = 5.1 days, but other variability dominates.

\subsection*{New Spectra and Spectral Types}

We obtained new spectra for several of the stars discussed in this
paper, primarily in order to provide spectral types or \vsini\ for
stars for which those data were lacking in the published
literature.    In particular, we obtained three Keck HIRES spectra
(all on the same night) for Mon-21, a single Keck HIRES spectrum for
Mon-1165, SOAR Goodman spectrograph medium  resolution spectra for
Mon-1131 and Mon-6975, and an INT low-resolution spectrum of
Mon-1165..  

\begin{figure*}
\begin{center}
\epsfxsize=.99\columnwidth
\epsfbox{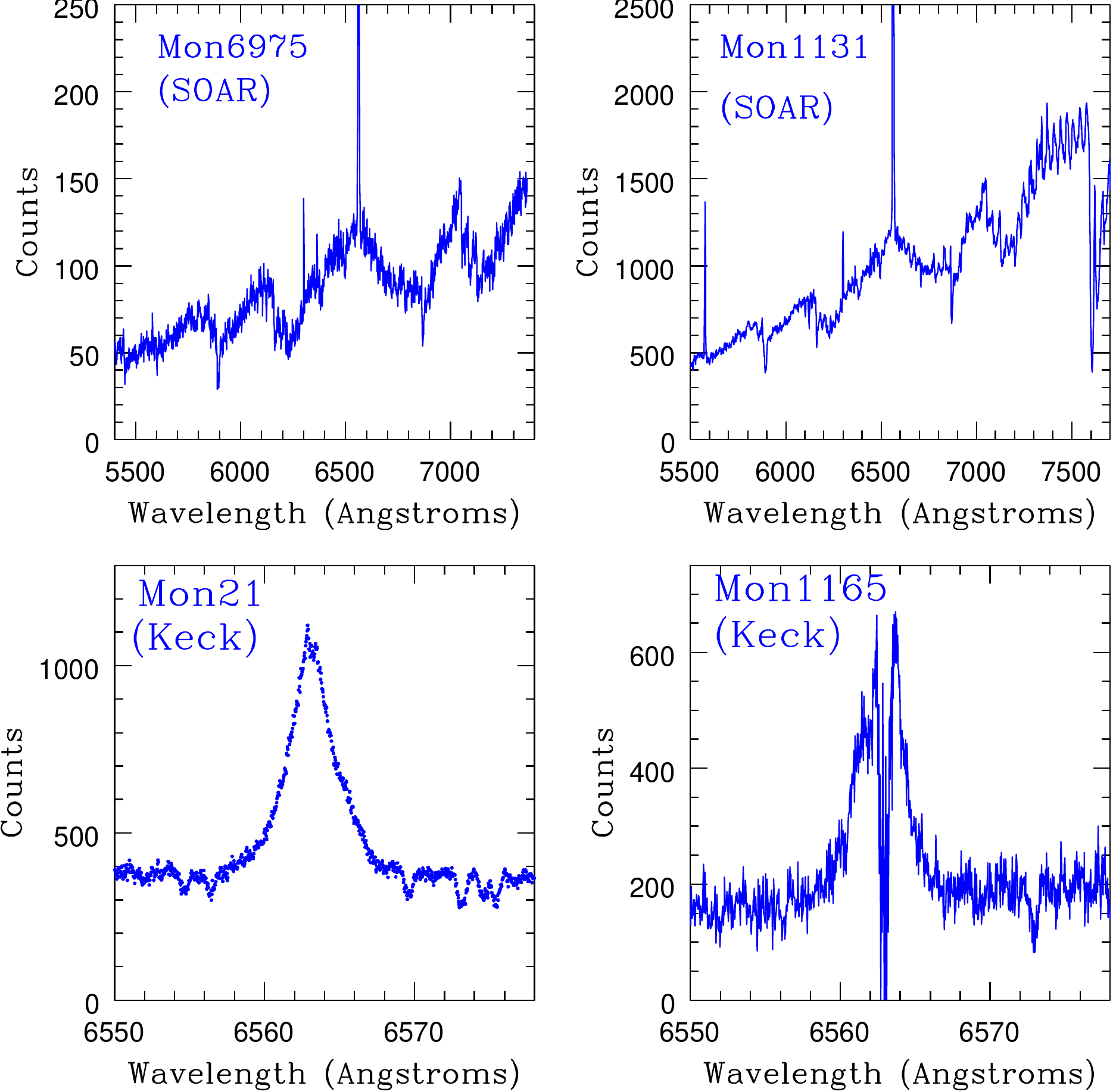}
\end{center}
\caption{Top panels show low-resolution spectra obtained with the
Goodman Spectrograph at SOAR for two of the short-duration flux-dip
stars.  Both are moderately early M dwarfs (M1.5 for Mon-1131 and M2.5
for Mon-6975).   Bottom panels show H$\alpha$\ emission profiles present
in Keck HIRES spectra for Mon-21 and Mon-1165.   The
H$\alpha$\ profiles are typical of modestly active T Tauri stars;
the central absorption feature for Mon-1165 may be partially or
wholly attributable to imperfect sky subtraction.
\label{fig:new_spectra}}
\end{figure*}

Snippets from these spectra are shown in
Figure~\ref{fig:new_spectra}.   For Mon-21 and  Mon-1165, we used the
Spectroscopy Made Easy (SME; Valenti \& Piskunov 1996) and  BINMAG3
(Oleg Kochukhov; \url{http://www.astro.uu.se/oleg})  packages to
derive  \vsini\ estimates,  as described more completely in the  
companion paper by McGinnis et al.\ (2015).

For the lower resolution, INT and SOAR spectra of Mon-1131, Mon-1165
and Mon-6975, visual inspection showed that all three stars were early
to mid M dwarfs.  We therefore used spectral indices based on the
depths of the observed TiO bands to infer spectral types; more
specifically, we used the relations in Stauffer (1982) and Stauffer \&
Hartmann (1986) for field M dwarfs and relations in Herczeg \&
Hillenbrand (2014) calibrated for YSOs.  The derived spectral types
are M1.5 for Mon-1131, M3 for Mon-1165, and M2.5 for Mon-6975.  For
Mon-1131, the field star relation yields M2 and the YSO relation
yields M1 -- we take the average; for Mon-6975, both relations yield
M2.5.   We consider these spectral types preliminary because the
relations are based on flux calibrated spectra, and we did not have
the data needed to flux calibrate the SOAR or INT spectra.

\subsection*{Coordinates and {\em CoRoT} ID's for Stars with Single or Just
  a Few Short-Duration Flux Dips}

As discussed in \S 5, a small number of the YSOs in NGC~2264 for which
we have {\em CoRoT} light curves have either just one or a couple
short-duration, shallow flux dips in their light curves.  They
therefore do not fall into any of our previously defined variable
extinction classes.  These stars are listed here in
Table~\ref{tab:singledippers}.  All of them are Class II systems based
on their IRAC photometry, though in some cases their IR excesses are
relatively small.

\begin{deluxetable*}{lccccc}
\tabletypesize{\scriptsize}
\tablecolumns{6}
\tablewidth{0pt}
\tablecaption{Basic Information for YSOs with Isolated Flux Dips\label{tab:singledippers}}
\tablehead{
\colhead{Mon ID\tablenotemark{a}}  & \colhead{2MASS ID} &
\colhead{{\em CoRoT}\tablenotemark{b}} & \colhead{{\em CoRoT}\tablenotemark{c}} &
\colhead{SpT\tablenotemark{d}} & \colhead{H$\alpha$ EW (\AA)}
}
\startdata
CSIMon-000119 & 06412100+0933361 & 223985987 & 223985987 & K6   & 10.6 \\ 
CSIMon-000342 & 06405573+0946456 &  \nodata  & 616872592 & M4   & 21.1 \\
CSIMon-000577 & 06414382+0940500 &  \nodata  & 616895846 & K1   &  5.9 \\
CSIMon-000650 & 06410098+0932444 & 223980807 & 223980807 & K1   &  6.4 \\
CSIMon-001038 & 06402262+0949462 &  \nodata  & 602095739 & M0   &  2.7 \\
CSIMon-001114 & 06393339+0952017 & 223957142 & 223957142 & M1.5 & 19.7 \\
CSIMon-001559 & 06393276+0904137 & 223956963 &  \nodata  & \nodata & \nodata \\
CSIMon-001568 & 06410684+0902161 & 223982299 & 223982299 & \nodata & \nodata \\
CSIMon-006991 & 06392200+1006233 & 223953770 & 223953770 & \nodata & \nodata \\
\enddata
\tablenotetext{a}{The Mon IDs are our internal naming scheme for stars in
the field of NGC~2264 -- see Cody et al.\ (2014).  }
\tablenotetext{b}{{\em CoRoT} identification number in 2008 SRa01 campaign.  }
\tablenotetext{c}{{\em CoRoT} identification number in 2011 SRa05 campaign. }
\tablenotetext{d}{See Cody et al.\ (2014) for the sources of the
spectral type and H$\alpha$ equivalent width (EW) data. All of these
are in emission.}
\end{deluxetable*}

\newpage
\newpage

\end{document}